\documentclass[conference]{IEEEtran}
\IEEEoverridecommandlockouts
\usepackage{cite}
\usepackage{amsmath,amssymb,amsfonts}
\usepackage{algorithm}
\usepackage{algorithmic}
\usepackage{graphicx}
\usepackage{textcomp}
\usepackage{comment}
\usepackage{xcolor}
\usepackage{subcaption}
\usepackage{amsmath}
\usepackage{amssymb}
\usepackage{gensymb}
\usepackage{multirow}
\usepackage{array}
\usepackage{bm}
\usepackage{subcaption}
\usepackage{stfloats}
\usepackage{tikz}
\usepackage[hyphens]{url}
\usepackage[hidelinks]{hyperref}
\usepackage{xurl}
\usepackage{nicematrix}
\usepackage{bm} 
\usepackage{amsmath,amssymb,booktabs,array}
\usepackage[left=1.62cm,right=1.62cm,top=1.9cm]{geometry}
\usepackage{makecell}
\usepackage{float}
\usepackage{amsmath}

\usepackage{array}
\usepackage{lipsum}
\usepackage{booktabs}
\usepackage[table]{xcolor}
\usepackage{balance}
\usepackage{tabularx}
\usepackage{soul}
\usepackage[normalem]{ulem}
\usepackage{tikz}
\usetikzlibrary{
    positioning,
    calc,
    arrows.meta,
    shapes.geometric,
    decorations.pathreplacing
}

\def\BibTeX{{\rm B\kern-.05em{\sc i\kern-.025em b}\kern-.08em
    T\kern-.1667em\lower.7ex\hbox{E}\kern-.125emX}}
\begin{document}
\title{Centralized PPO-Based DRL for Multi-UAV-BS Positioning and Trajectory Optimization in Disaster Response Networks}

\author{
\IEEEauthorblockN{Azim Akhtarshenas$^1$; Mario Rico Ibáñez$^2$; Matteo Bernabè$^1$; David L\'opez-P\'erez$^{1,3}$; M\'erouane Debbah$^{4}$}
\IEEEauthorblockA{\textit{$^1$Universitat Polit\`ecnica de Val\`encia (UPV), Valencia, Spain}\\
$^2$\textit{\'Ecole Polytechnique F\'ed\'erale de Lausanne (EPFL), Lausanne, Switzerland}\\
$^3$\textit{Beihang Valencia Polytechnic Institute (BVPI), Hangzhou, China}\\
$^4$\textit{Research Institute for Digital Future, Khalifa University, 127788 Abu Dhabi, UAE}}
\textit{Aakhtar@upv.edu.es}
\thanks{This research is supported by the Generalitat Valenciana
through the CIDEGENT PlaGenT, Grant CIDEXG/2022/17, Project iTENTE, and the action CNS2023-144333, financed by
MCIN/AEI/10.13039/501100011033 and the European Union
“NextGenerationEU”/PRTR.}}

\maketitle


\begin{abstract}
Unmanned aerial vehicle-mounted base stations (UAV-BSs) constitute a flexible and effective solution for global positioning system (GPS)-free emergency and disaster scenarios, where the rapid deployment of communication infrastructure is critical for maximizing life-saving operations. In this work, we extend a centralized learning framework to a multi-UAV-BS network architecture, in which a single centralized UAV-BS -as an intelligent agent- coordinates the three-dimensional positioning and navigation of multiple UAV-BSs, while the remaining UAV-BSs actively serve ground user equipments (UEs) with uncertain positions.
We formulate a fairness-aware sum-throughput maximization problem for UAV-BS coordination, which is inherently non-convex due to the non-linear and interference-coupled throughput expressions. 

To address this challenge, we cast the problem as a Markov Decision Process (MDP) and solve it using a deep reinforcement learning (DRL) framework based on Proximal Policy Optimization (PPO). The central agent interacts with the environment and learns optimal joint positioning policies that guide the serving UAV-BSs to provide efficient, adaptive, and resilient wireless coverage. The proposed approach exploits spatial configuration and radio signal sensing capabilities to dynamically adapt to heterogeneous UE mobility patterns.
Extensive simulations are conducted to evaluate the performance of the proposed method.
Numerical results demonstrate that PPO shows competitive performance during both training and evaluation phases. Furthermore, comparative analysis with state-of-the-art RL algorithms, namely Deep Deterministic Policy Gradient (DDPG) and Deep Q-Network (DQN), shows that PPO consistently outperforms these methods in terms of convergence stability, mean reward, and network throughput.\\
\end{abstract}


\section{Introduction} 
\textit{``We humans, in this modern age with advanced technologies, have established our dominance over the Earth and are even pursuing space exploration. Yet, despite all of this progress, we often find ourselves helpless in the face of devastating natural disasters that claim the lives of many on this very planet."}\\
DANA (Depresión Aislada en Niveles Altos) struck Valencia on October 29, 2024, transforming a severe weather event into a devastating humanitarian crisis. The floods claimed 224 lives, demolished infrastructure, and left deep physical and emotional scars across the region \cite{galvez2025dana}. Disasters like DANA \cite{galvez2025dana}, Typhoon Haiyan~\cite{aquino2013typhoon}, and 9/11~\cite{11s_comm} occur worldwide each year, disrupting lives, damaging infrastructure, and cutting off communication and internet services, making rescue efforts difficult and worsening human suffering.\\
%
%
%
In catastrophe situations, keeping consistent communication is crucial for organizing search and rescue activities. When conventional networks fail, user equipment (UEs) are disconnected. To solve this, unmanned aerial vehicle-mounted base stations (UAV-BSs) have developed as an effective method for quickly deploying emergency communication systems~\cite{uav_emergency}.
UAV-BSs are extremely useful in situations where terrestrial BSs are completely unavailable. In such high-stakes environments, the swift deployment of broadband and ultra-reliable low-latency communication (URLLC) networks becomes essential to assist first responders~\cite{do2021joint, kaleem2019uav}. For the sake of notational simplicity, throughout this paper, the term UE is used to refer to first responders.

\subsection{Related Works}

\par A substantial body of recent research on UAV-assisted emergency communication networks has centered on optimizing UAV-BS
positioning and trajectory to enhance coverage and throughput performance.
For example, Rahman~\textit{et al.}~\cite{rahman2018uav, ur2018positioning} propose a heuristic algorithm to determine the optimal UAV position for maximizing system throughput while also ensuring all UEs remain within transmission range. Their results indicate that optimal UAV placement can significantly enhance throughput, especially when UE locations or data demands are uneven.
However, both studies assume static UE positions and traffic demands, which may not reflect the dynamic nature of emergency scenarios. In contrast, Xie \emph{et al.} \cite{xie2018throughput} examine a UAV-enabled wireless powered communication network (WPCN) where a UAV serves as a mobile access point, wirelessly charges ground UEs through radio frequency energy transfer, and uses the energy it harvests to gather their uplink data. Under UAV speed and user energy limits, the system seeks to maximize the minimum uplink throughput by simultaneously optimizing the UAV trajectory and wireless resource allocation. For the ideal scenario, a multi-location hovering strategy is defined, and for actual speed constraints, a feasible consecutive hover-and-fly solution is suggested. Comparing numerical findings to traditional WPCNs with fixed access points, significant throughput increases are shown.
However, all the abovementioned studies ~\cite{rahman2018uav, ur2018positioning, xie2018throughput} 
 assume single-agent scenarios with a single UAV-BS that offers limited coverage and capabilities, making it difficult to support complex missions, while a multi-UAV scenario enables coordinated and scalable operations, combining individual UAV-BS autonomy with collective functionality to handle large-scale, multi-objective tasks more effectively.

To overcome the limitations of the single-agent scenario,
the authors in \cite{veeraswamy2025optimal} address multiple UAV-BSs' positioning with fixed altitude by formulating the UE coverage problem as a mixed-integer linear program. To manage computational complexity in large-scale scenarios, a clustering-based heuristic is introduced. The approach provides effective performance for rapid deployment in disaster response and offers both scalability and flexibility. Lyu~\textit{et al.}\cite{lyu2016placement} introduce a clever geometric approach to multiple UAV-BSs' placement. During their study, UAV-BSs are sequentially deployed along a spiral path from the perimeter inward, dynamically adapting to uncovered ground terminals while minimizing the number of UAV-BSs. 
The authors demonstrate that the geometric algorithm achieves better performance than existing schemes in terms of both the required number of UAV-BSs and computational complexity.
\par Nevertheless, traditional methods employed in ~\cite{rahman2018uav, ur2018positioning, xie2018throughput, veeraswamy2025optimal, lyu2016placement} lack adaptability and real-time decision-making due to their static nature and inability to model the complex, non-convex, non-linear, and stochastic nature of real-world environments. In contrast, AI methods-particularly, reinforcement learning (RL)-allow UAV-BSs to learn and adapt dynamically, making them more effective for positioning and trajectory problems in unpredictable emergency scenarios.
In this context, the study \cite{mandloi2023q} introduces a Q-learning-based approach for deploying UAV-BSs to restore wireless services in disaster-affected areas. The authors aim to develop a 3D range-based localization algorithm to estimate the positions of ground UEs. Then, a Q-learning-based UAV transmit power optimization approach is used to maximize the mean data rate of the UEs. Their simulation results show that the proposed approach outperforms two benchmark schemes in terms of UE mean data rate and outage probability.
 Parisotto~\textit{et al.} in \cite{8877247} propose a Q-learning-based approach to jointly optimize the 3D positioning and transmit power of multiple UAV-BSs, aiming to maximize network coverage while dynamically adapting to UE mobility. Their proposed algorithm achieves the same coverage with fewer drones, lower transmit power, and reduced outages.
\par Notwithstanding, Q-learning approach, used in~\cite{mandloi2023q, 8877247},  becomes inefficient in large-scale environments due to the curse of dimensionality, where the vast state-action space hinders scalability. To address this, deep reinforcement learning (DRL) emerges as a more practical and promising solution by leveraging deep neural networks (DNNs) to approximate the Q-values, enabling more effective learning in complex scenarios. 
For example, 
Zeyu~\textit{et al.} in~\cite{tian2026energy} propose an energy-efficient framework for joint communication and localization in UAV-assisted emergency systems. The authors consider a single UAV cooperating with a ground rescuer and employ an AoA-based localization method combined with a DRL algorithm to jointly optimize UAV trajectory, transmission power, and flight time. Simulation results show improvements in communication performance, localization accuracy, and system energy efficiency.  However, the authors assume a single UAV and rely on accurate location or geometric information for AoA-based localization~\cite{tian2026energy}. Tang \emph{et al.} \cite{tang2020minimum} study multi-UAV trajectory planning and resource allocation for throughput maximization using a deep Q-learning (DQL) approach. The problem was formulated as a non-convex optimization and addressed via successive convex approximation. During their simulation results, they demonstrate that the proposed DQL-based method achieves better performance than conventional WPCN schemes in terms of minimum throughput.
%
%
%

In~\cite{zhao2025clustering}, a centralized hierarchical UAV architecture is proposed to mitigate onboard resource constraints and inter-UAV resource competition in multi-UAV networks, where a central UAV coordinates multiple access UAV-BSs serving ground UEs. In addition, DRL–based hierarchical and multi-UAV control schemes are applied for trajectory optimization and resource allocation in dynamic environments. Their proposed method generalizes to varying numbers of ground UEs without retraining and demonstrates superior performance compared to baseline methods in terms of both energy efficiency and UE fairness.
However, their strong reliance on GPS information limits their effectiveness in disaster scenarios.
To enhance system capacity, Wang~\textit{et al.}~\cite{wang2022joint} address the joint optimization of UAV placement, resource allocation, and computation offloading. Due to the inherently non-convex nature of this problem, conventional optimization methods are insufficient, motivating the use of DRL. Specifically, Double Deep Q-Network (DDQN) and Deep Deterministic Policy Gradient (DDPG) are employed to optimize UAV positioning and minimize latency in highly dynamic environments.
Their proposed algorithm exhibits fast convergence and lower latency.
He~\textit{et al.}~\cite{he2024deep} investigate 3D UAV trajectory optimization to improve UE access rate, fairness, throughput, and energy efficiency. The problem was formulated as an optimization task and addressed using DRL. By employing DDQN and DDPG, UAV-BSs learn optimal flight altitudes and high UE-density regions to enhance system performance under dynamic conditions. Numerous experimental tests conducted under different UE distributions show that their suggested algorithm successfully determines the ideal altitude for maximizing coverage.
However, the DRL-based approaches in~\cite{tang2020minimum, wu2025radio, zhao2025clustering, wang2022joint, he2024deep} rely on discrete state-action spaces (e.g., DDQN) and are prone to training instability and sensitivity to hyperparameter tuning. Likewise, value-based continuous-control methods such as DDPG often exhibit convergence instability and limited scalability in high-dimensional and noisy control settings.

\par To address these challenges, more stable policy gradient methods like trust region policy optimization (TRPO) are introduced \cite{schulman2015trust}, enforcing a strict KL-divergence constraint to ensure stable policy updates and improving performance in continuous control tasks such as UAV trajectory optimization. For example, motivated by circumstances where terrestrial infrastructure is difficult to deploy or may be damaged (e.g., disaster/emergency situations), Ho~\textit{et al.}\cite{ho2021uav}~present a DRL system for managing a UAV that provides wireless service to \emph{critical demand areas}. An agent learns a control policy to enhance service delivery while taking into account UAV operation limits, particularly energy/efficiency considerations, in the form of a sequential decision-making task. In their evaluation of policy-gradient RL, which includes TRPO, for UAV motion/trajectory control, the authors note that in dynamic situations, TRPO can provide steady learning and enhanced energy efficiency when compared to DDPG. 
\par Considering~\cite{schulman2015trust, ho2021uav}, TRPO provides stable policy updates but still suffers from high computational complexity due to its constrained optimization and second-order approximations. To address these limitations, Proximal Policy Optimization (PPO) is introduced as an efficient alternative \cite{schulman2017proximal}, replacing the trust region constraint with a clipped surrogate objective that preserves training stability while significantly reducing computational overhead.
 Alwarafy~\textit{et al.}~\cite{alwarafy2025deep} address joint trajectory design and resource allocation for secure and energy-efficient UAV networks in the presence of ground eavesdroppers by employing a PPO-based DRL framework. Their proposed approach jointly optimizes UAV 3D positioning, power allocation, and energy harvesting under secrecy rate, energy utilization efficiency, and interference constraints. Their simulation results show that PPO achieves faster convergence, higher utility values, and improved stability compared to DDPG and other benchmark methods in highly dynamic UAV network environments.
In our previous study~\cite{11162280}, a single-agent single-UAV framework based on PPO was proposed to learn efficient UAV trajectories using radio signal sensing. The proposed approach demonstrated strong adaptability and coverage performance across diverse UE mobility patterns, including static, random, linear, and circular motions.
In another previous work~\cite{akhtarshenas2026ppo}, PPO was applied for high altitude platform system (HAPS)-BS positioning in maritime networks. However, the framework was constrained by simplified mobility assumptions and limited environmental dynamics. In this work, these limitations are addressed through a more flexible multi-UAV framework with richer state representations and improved adaptability to dynamic disaster-response scenarios.
Li~\textit{et al.}~\cite{li2023uav} optimize UAV trajectories for effective spectrum cartography in dynamic emitter environments. To handle sparse and delayed feedback, a PPO-based algorithm is proposed alongside a backtracking advantage function. Their approach provides improved spectrum cartography accuracy and reduced energy consumption.
A UAV path planning for dynamic disaster-response missions under kinematic constraints is studied in \cite{mowla2025iot}. A controller based on PPO is suggested to produce realistic, real-time trajectories. Based on simulation results, PPO performs better than RL baselines and traditional planners in terms of mission success and path efficiency, proving its resilience in quickly evolving disaster scenarios. 
\subsection{Motivation and Contributions}
In real-world emergency scenarios, UAV-based communication systems operate under highly dynamic and uncertain conditions that are not well captured by single-UAV, static, or idealized environment models. Much of the existing literature assumes a single-UAV architecture with full a priori knowledge of UE locations and strong reliance on GPS information or predefined spatial grids, which limits applicability in practical deployments where UE mobility is unpredictable and environmental information is incomplete. Moreover, conventional optimization methods and several DRL approaches (e.g., Q-learning and DDQN) rely on discretized state-action representations, while policy-gradient methods such as DDPG and TRPO suffer from scalability limitations, high training variance, and increased complexity. These constraints hinder real-time adaptability and autonomous decision-making in rapidly changing emergency environments. 
To address these challenges, we leverage PPO, which has demonstrated robust and sample-efficient learning with stable policy updates for UAV trajectory control in related literature \cite{alwarafy2025deep, li2023uav, mowla2025iot, 11162280}. 

Building on our previous studies~\cite{11162280, akhtarshenas2026ppo}, we propose a realistic and adaptive PPO-driven framework for UAV trajectory planning.
Unlike value-based methods such as DQN, DDQN, and DDPG, our PPO approach avoids state–action space discretization and exhibits stable convergence in high-dimensional continuous control settings. Moreover, PPO eliminates the computational complexity associated with the KL constraint in TRPO while maintaining robust policy updates. Unlike prior DRL-based UAV positioning approaches that rely heavily on GPS-based localization or simplified spatial representations, the proposed framework leverages real-time radio-sensing measurements derived from practical reference signals, enabling efficient and autonomous multi-UAV coordination under dynamic emergency conditions.
Our key contributions are summarized as follows:
\begin{itemize}
    \item We propose a continuous-control framework that enables smooth and flexible UAV-BS trajectory adaptation under dynamic emergency deployment scenarios.
    
    \item We replace GPS-based positioning with UE reference-signal sensing and angle-of-arrival (AoA) measurements, enabling robust UAV-BS operation without requiring prior knowledge of UE locations.
    
    \item We employ circular statistics for AoA processing within the state space, improving navigation stability and reducing abrupt UAV-BS directional variations during trajectory adaptation.
    
    \item We design an enhanced PPO-based UAV-BS positioning framework that integrates spatial and radio measurements, including UAV-BS locations, downlink received power, SINR, and AoA information, together with sigmoid-based reward shaping to achieve improved learning stability, faster convergence, and higher throughput performance.
    
    \item Simulation results demonstrate that the proposed PPO framework consistently outperforms DQN and DDPG in terms of convergence stability, mean reward, throughput performance, and generalization capability under identical initialization and simulation conditions.
\end{itemize}

\begin{figure*}[t]
    \centering

    \begin{subfigure}[t]{0.495\linewidth}
        \centering
        \includegraphics[
            height=6.8cm,
            width=\linewidth,
            keepaspectratio
        ]{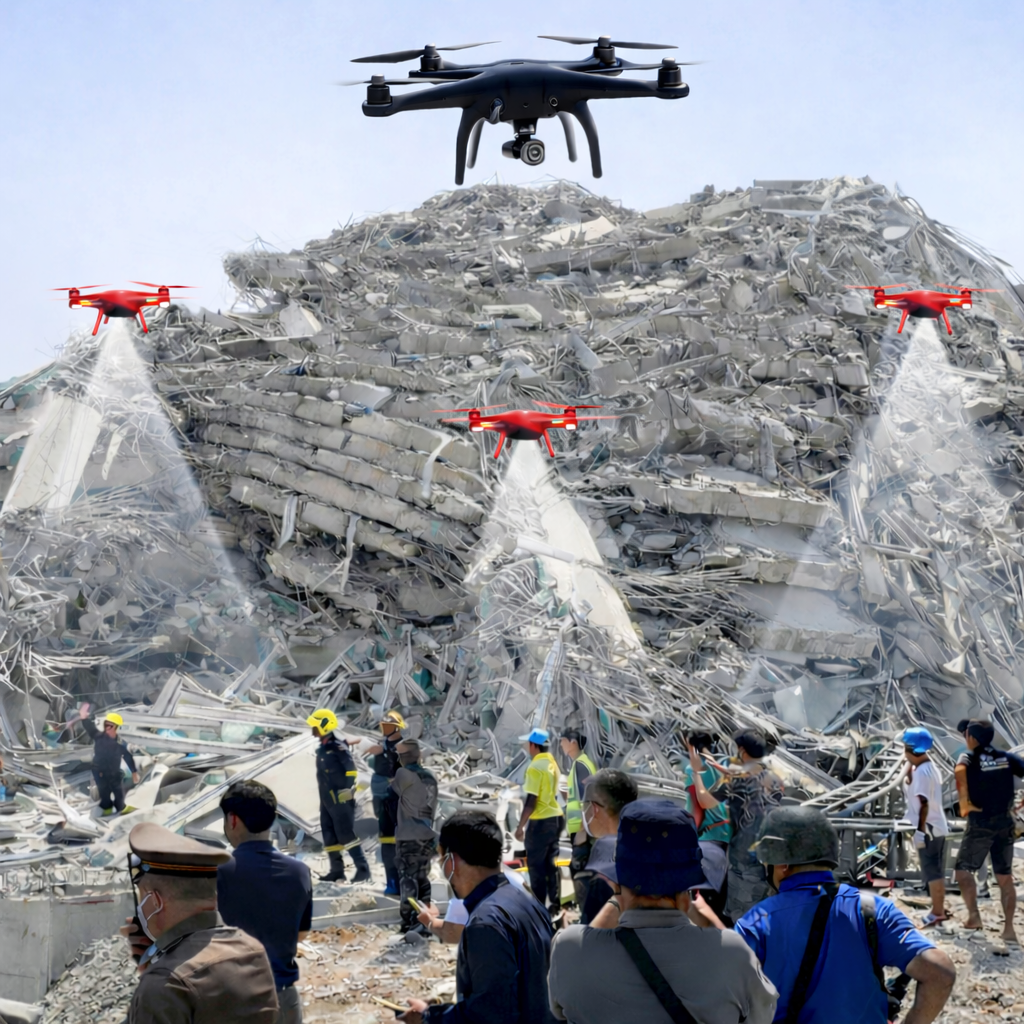}
        \caption{Earthquake-induced disaster scenario~\cite{detroitcatholic_myanmar_earthquake_2025}}
        \label{fig:earthquake}
    \end{subfigure}
    \hfill
    \begin{subfigure}[t]{0.495\linewidth}
        \centering
        \includegraphics[
            height=6.8cm,
            width=\linewidth,
            keepaspectratio
        ]{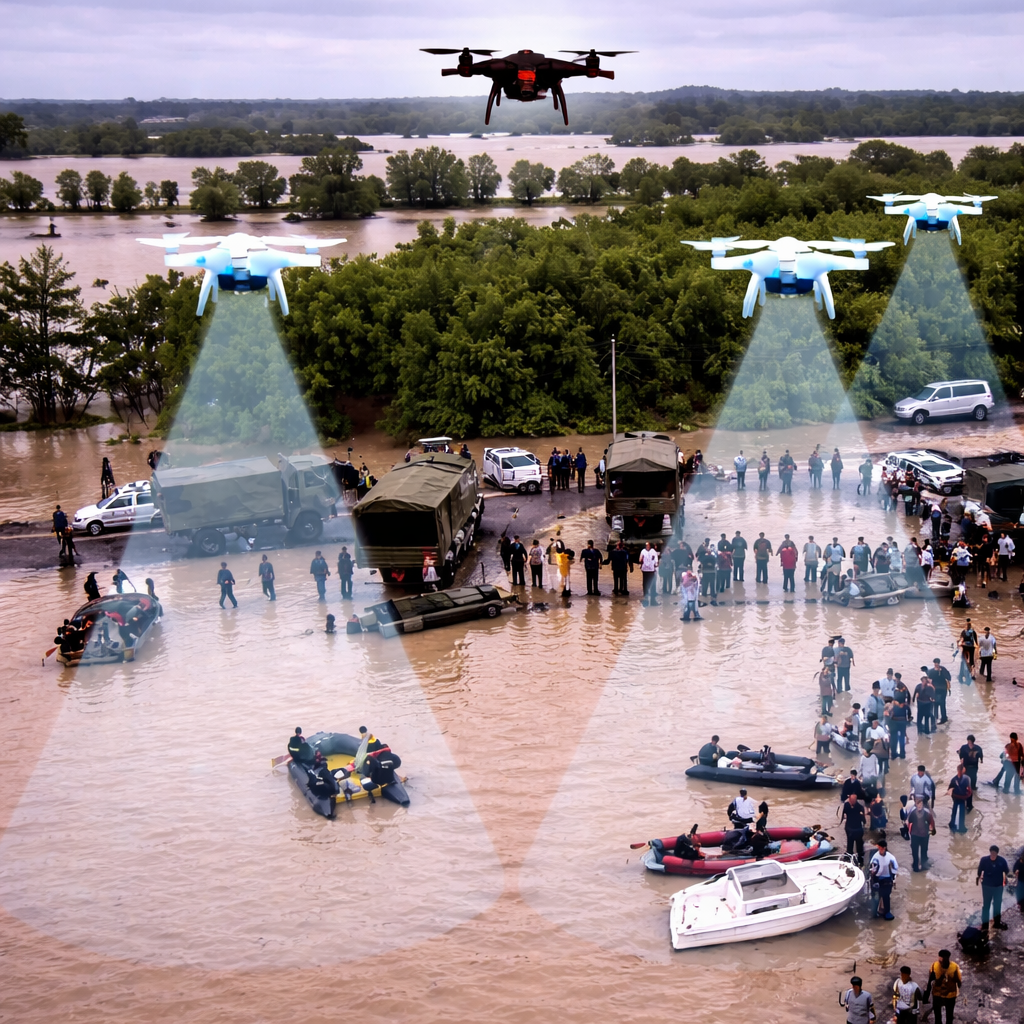}
        \caption{Storm-induced disaster scenario~\cite{istock_flood_rescue_466992392}}
        \label{fig:storm}
    \end{subfigure}

    \caption{UAV-BS-assisted communication in disaster-affected urban environments.}
    \label{fig:storm_}
\end{figure*}


\section{System Model}
\label{sec:system_model}

\subsection{System Description}
\label{System_network}

In this section, we investigate an emergency wireless network consisting of one central UAV functioning as a controller and multiple serving UAV-BSs functioning as aerial BSs, deployed to support the ground UEs.
As depicted in Fig.~\ref{fig:storm_}, a hierarchical control architecture is adopted, where the central black UAV-BS acts just as a control agent and the surrounding red UAV-BSs (Fig~\ref{fig:earthquake}) and white UAV-BSs (Fig~\ref{fig:storm})  provide communication services to ground UEs.
The network operates in both downlink and uplink directions using time division multiple access (TDMA). 

\subsubsection{UE set definition}\label{UE_SET_HS}
Let $\mathcal{U} = \{u_{1}, u_{2}, \ldots, u_{U}\}$ represent the set of first-responder UEs, where each element corresponds to an individual UE device deployed within the disaster area. The total number of UEs is denoted by $U = |\mathcal{U}|$, representing the cardinality of the ground UE set.

The spatial location of the $u^{\text{th}}$ UE is defined by the three-dimensional coordinate vector ${{\boldsymbol{\rho}}}^{\mathrm{U}}_u = [x^{\mathrm{U}}_u, y^{\mathrm{U}}_u, z^{\mathrm{U}}_u]^\top$, where $x^{\mathrm{U}}_u$, $y^{\mathrm{U}}_u$, and $z^{\mathrm{U}}_u$ denote the corresponding Cartesian coordinates. Aggregating the positions of all UEs, the overall UE position matrix is expressed as $\mathbf{X}^{\mathrm{U}} = \left[{{\boldsymbol{\rho}}}^{\mathrm{U}}_1, \ldots, {{\boldsymbol{\rho}}}^{\mathrm{U}}_u, \ldots, {{\boldsymbol{\rho}}}^{\mathrm{U}}_U\right]^{\top}$, where $\mathbf{X}^{\mathrm{U}} \in \mathbb{R}^{U \times 3}$, with each row representing the spatial coordinates $(x,y,z)$ of an individual UE.

The UE distribution follows a hotspot (HS)-based spatial clustering model, where UEs are grouped into localized communication regions. 
Let $\mathcal{H} = \{h_1, \ldots, h_h, \ldots, h_H\}$ denote the set of HSs, where $H = |\mathcal{H}|$ represents the total number of hotspots. The spatial location of the $h^{\text{th}}$ HS is defined by the three-dimensional coordinate vector ${{\boldsymbol{\rho}}}^{\mathrm{H}}_h = [x^{\mathrm{H}}_h, y^{\mathrm{H}}_h, z^{\mathrm{H}}_h]^\top$, where the coordinates correspond to the geometric center of the associated hotspot region. Aggregating the locations of all HSs, the hotspot position matrix is expressed as $\mathbf{X}^{\mathrm{H}} = \left[{{\boldsymbol{\rho}}}^{\mathrm{H}}_1, \ldots, {{\boldsymbol{\rho}}}^{\mathrm{H}}_h, \ldots, {{\boldsymbol{\rho}}}^{\mathrm{H}}_H\right]^{\top}$, where $\mathbf{X}^{\mathrm{H}} \in \mathbb{R}^{H \times 3}$.
Furthermore, for the $h^{\text{th}}$ HS, let $r_h$ denote the hotspot radius, and define the set of hotspot radii as $\mathcal{R} = \{r_1, \ldots, r_h, \ldots, r_H\}$.

In this paper, we assume homogeneous HS sizes, i.e., $r_h = r$ for all $h$. Furthermore, each HS is assumed to contain the same number of UEs. Thus, the number of UEs served by each HS is given by $N = \frac{|\mathcal{U}|}{|\mathcal{H}|}$.

As a default case, we assume that the UEs associated to each HS follow the same linear motion mobility model with speed $v$ (m/s). In addition, static, circular motion, and cosine-pattern motion mobility patterns are considered to further evaluate the system performance as follows:

\begin{itemize}

   \item \textit{Static UEs (No Mobility)}: Multiple HSs, each composed of a group of UEs distributed within a predefined coverage region, remain stationary during the simulation, representing a static UE deployment scenario.
   
   \item \textit{Linear Motion}: Multiple HSs move along independent linear trajectories with constant velocity, representing coordinated group mobility in straight-line motion.
    
   \item \textit{Circular Motion}: Multiple HSs move along independent circular trajectories with predefined radii and constant velocity, modeling coordinated rotational group mobility.

   \item \textit{Cosine-Pattern Motion}: Multiple HSs follow independent cosine-shaped trajectories with constant velocity and periodic vertical displacement, representing oscillatory mobility behavior.

\end{itemize}

\subsubsection{UAV-BS set definition}
Let ${\boldsymbol{\rho}}^{C} = [x^{\mathrm{C}}, y^{\mathrm{C}}, z^{\mathrm{C}}]^{\top}$ represent the three-dimensional position vector of the central controller UAV.
Similarly, let $\mathcal{D} = \{1, \ldots, d, \ldots, D\}$ denote the set of serving UAV-BSs, where $D = |\mathcal{D}|$ is the total number of UAV-BSs. 
The three-dimensional position of the $d^{\text{th}}$ UAV-BS is represented by ${\boldsymbol{\rho}}^{\mathrm{D}}_d = [x^{\mathrm{D}}_d, y^{\mathrm{D}}_d, z^{\mathrm{D}}_d]^{\top}$, where $x$, $y$, and $z$ denote the Cartesian coordinates in three-dimensional space.

It is assumed that the central UAV is located at the geometric center of the considered area.
The collective positions of all UAV-BSs are organized in the matrix
$\mathbf{X}^{\mathrm{D}} = 
\left[
{\boldsymbol{\rho}}^{\mathrm{D}}_1,\ldots,
{\boldsymbol{\rho}}^{\mathrm{D}}_d,
\ldots,
{\boldsymbol{\rho}}^{\mathrm{D}}_D
\right]^{\top},$
where $\mathbf{X}^{\mathrm{D}}_i \in \mathbb{R}^{D \times 3}$, with rows corresponding to the spatial coordinates $(x,y,z)$ and columns corresponding to the $D$ UAV-BSs. Since inter-UAV communication is not the primary focus of this study, perfect links between the central UAV and the UAV-BSs are assumed. Additional notation is summarized in Table~\ref{tab:notation}.
\begin{table}[t]
\centering
\caption{Notation Definitions}
\label{tab:notation}
\renewcommand{\arraystretch}{1.3}
\begin{tabular}{|c|l|}
\cline{1-2}
\multicolumn{2}{|c|}{\textbf{Communication-based Notations}} \\ 
\cline{1-2}
Symbol & Description \\ \hline
${\bm{\rho}}$ & Coordinate (position vector) \\ 
\hline
$\mathrm{X}$& position matrix\\
\hline
$P^{\mathrm{tx}}$ & Transmit power \\ \hline
$P^{\mathrm{rx}}$ & Received power \\ \hline
$n_0$ & BS noise figure \\ \hline
$B$ & Network bandwidth \\ \hline
$G$ & Channel gain \\ \hline
$G^{\rm a}$ & Antenna gain \\ \hline
$G^{\rm p}$ & Path gain \\ \hline
$G^{\rm s}$ & Shadow fading \\ \hline
$G^{\rm ff}$ & Fast fading \\ \hline
P & DL received power \\ \hline
$\gamma$ & Signal-to-Interference-plus-Noise Ratio (SINR) \\ \hline
$\sigma^2$ & Noise power \\ \hline
$T$ & episode length \\ \hline
$R$ & Network throughput \\ \hline
$\alpha$ & Angle of arrival \\ \hline
$\beta$ & Agent movement direction \\ \hline
$r_{\max}$ & Agent's maximum movement distance \\ \hline
\cline{1-2}
\multicolumn{2}{|c|}{\textbf{RL-based Notations}} \\ \cline{1-2}
$\eta$ & Learning rate \\ \hline
$\xi$ & Discount factor \\ \hline
$\lambda$ & GAE parameter \\ \hline
$\epsilon$ & PPO clip parameter \\ \hline
$\theta$ & Policy parameters \\ \hline
$\phi$ & Value function parameters \\ \hline
\cline{1-2}
\multicolumn{2}{|c|}{\textbf{Frequently-used mathematical operators}} \\ 
\cline{1-2}
${x}$ & Scalar variable \\ \hline
${\bm{x}}$ & Vector \\ \hline
$\bar{\bm{x}}$ & Mean (average) of $\bm{x}$ \\ \hline
$[\,\cdot\,]^\top$ & Transpose operator \\ \hline
$\tilde{\bm{x}}$ & Time-varying (instantaneous) vector \\ \hline
$\psi(\cdot)$ & Circular mean operator \\ \hline
$\omega(\cdot)$ & Circular standard deviation operator \\ \hline
\end{tabular}
\end{table}
\subsection{Channel Model}
We consider a network operating over a bandwidth $B$ at a carrier frequency $f$. Each radio link within the network is subject to both slow and fast fading.
Users are multiplexed over orthogonal frequency resources (e.g., physical resource blocks, PRBs), indexed by k.
The overall channel gain between the $u^{\text{th}}$ UE belonging to any HS and the base station of the $d^{\text{th}}$ UAV on the $k^{\text{th}}$ frequency resource is denoted by $G_{u,d,k}$.
This composite gain is modeled as the product of several contributing factors: antenna gain ($G^{\rm a}$), path gain ($G^{\rm p}$), shadow fading ($G^{\rm s}$), and fast fading ($G^{\rm ff}$), as given by
\begin{equation}
    G_{u,d,k} \left(\boldsymbol{\rho}^\mathrm{U}_u,\,
\boldsymbol{\rho}^\mathrm{D}_d\right) = G^{\rm a}_{u,d} \cdot G^{\rm p}_{u,d} \cdot G^{\rm s}_{u,d} \cdot \left|G^{\rm ff}_{u,d,k}\right|^2.
\end{equation}

In this study, we adopt the Urban Macro (UMa) channel models specified by the Third Generation Partnership Project (3GPP) in TR 36.814~\cite{3gpptr36814} 
to compute each component of the channel gain described above, with the following modifications: the BS antennas are assumed to be omnidirectional, and the multi-path fading is modeled using a Rician distribution.

\subsection{UE Received Power}

The power received by the $u^{\text{th}}$ UE from the $d^{\text{th}}$ UAV-BS on the $k^{\text{th}}$ frequency resource is given by
\begin{equation}
    P^{\rm rx}_{{u,d,k}}(\boldsymbol{\rho}^{\mathrm{U}}_u, \boldsymbol{{\rho}}^{\mathrm{D}}_d)
    = P^{\rm tx}_{d,k} \cdot G_{u,d,k}(\boldsymbol{{\rho}}^{\mathrm{U}}_u, \boldsymbol{{\rho}}^{\mathrm{D}}_d).
    \label{eq_intro:Received_power}
\end{equation}
where $P^{\rm tx}_{d,k}$ denotes the transmit power of the BS onboard the $d^{\text{th}}$ UAV-BS on the $k^{\text{th}}$ frequency resource.

\subsection{UE Signal Quality}

This subsection defines the signal quality experienced by a UE
when receiving signals from UAV-BSs. The signal quality experienced by the $u^{\text{th}}$ UE 
from the $d^{\text{th}}$ UAV-BS on the $k^{\text{th}}$ frequency resource is quantified by the signal-to-interference-plus-noise ratio (SINR), denoted by $\gamma_{u,d,k}$. It is computed as~\cite{goldsmith2005wireless}
\begin{equation}
    \gamma_{u,d,k}({\boldsymbol{\rho}}^{\mathrm{U}}_u, \mathbf{X}^{\mathrm{D}})
    =
    \frac{P^{\mathrm{rx}}_{{u,d,k}}({\boldsymbol{\rho}}^{\mathrm{U}}_u, {\boldsymbol{\rho}}^{\mathrm{D}}_d)}
    {\sum_{\substack{d' = 1 \\ d' \neq d}}^{D}
    P^{\mathrm{rx}}_{u,d',k}({\boldsymbol{\rho}}^{\mathrm{U}}_u, {\boldsymbol{\rho}}^{\mathrm{D}}_{d'}) + \sigma^2_k}.
    \label{eq_intro:sinr}
\end{equation}
where $\sigma^2_k$ denotes the noise power in the $k^{\text{th}}$ frequency resource, and $\mathbf{X}^{\mathrm{D}}$ represents the set of positions of all UAV-BSs.

\subsection{UE throughput}
The achievable throughput
for a UE served by a UAV-BS is defined based on the Shannon-Hartley theorem. Accordingly, the achievable throughput for the $u^{\text{th}}$ UE 
connected to the $d^{\text{th}}$ UAV-BS on the $k^{\text{th}}$ frequency resource can be expressed as~\cite{goldsmith2005wireless, shannon1948mathematical}
\begin{equation}
    R_{u,d,k}(\boldsymbol{\rho}^{\mathrm{U}}_u, \mathbf{X}^{\mathrm{D}})
    =
    B_k \log_2 \left(1 + \gamma_{u,d,k}(\boldsymbol{\rho}^{\mathrm{U}}_u, \mathbf{X}^{\mathrm{D}})\right).
    \label{eq_intro:rate}
\end{equation}
where $B_k$ denotes the bandwidth allocated to the $k^{\text{th}}$ frequency resource.

If a scheduler is employed to ensure fair distribution of the available resources among the UEs within the cell—such as a round-robin scheme—the achievable throughput can be expressed as~\cite{shannon1948mathematical, tse2005fundamentals}
\begin{equation}
    R_{u,d}(\boldsymbol{\rho}^{\mathrm{U}}_u, \mathbf{X}^{\mathrm{D}})
    =
    \frac{B}{U} \log_2 \left(1 + \bar{\gamma}_{u,d}(\boldsymbol{\rho}^{\mathrm{U}}_u, \mathbf{X}^{\mathrm{D}})\right).
    \label{eq_intro:mean_rate}
\end{equation}
where $\frac{B}{U}$ denotes the average bandwidth allocated per UE, and $\bar{\gamma}_{u,d}$ represents the effective SINR averaged over the allocated frequency resources allocated to the UE.


\subsection{Angle of Arrival (AoA)}
Here, we introduce the angle-of-arrival (AoA) information used in the proposed scheme. It is important to highlight that, unlike many conventional approaches, the proposed scheme does not rely on GPS data from the UEs for optimization. This design choice is motivated by the potential unavailability or unreliability of GPS signals in emergency scenarios. Instead, our method assumes the use of reference signals, along with AoA estimations derived from them, as surrogate information for localization.
In more detail, we assume that the BS onboard the $d^{\text{th}}$ UAV is capable of estimating the AoA, denoted by $\alpha_{u,d}$, of the reference signals transmitted by the $u^{\text{th}}$ UE. 
This estimation is performed using an antenna array and appropriate signal processing techniques. 


{\color{red}
}

\section{UAV Positioning Problem Statement}

This section presents the DRL framework that is utilized to tackle the optimization problem that is the subject of this work.
In particular, we aim to determine 
where the UAV-BSs should be placed to maximize UEs' network performance. 
An objective function that captures fairness in the attainable throughput is used to first characterize the problem. Then, to effectively train UAV-BS positioning methods in this dynamic and complicated environment, a DRL-based approach utilizing PPO is presented.

\subsection{Objective Function}

We analyze our proposed system over a finite UE 
movement and UAV-BS flight period, $T$, indexed by discrete time steps $t \in \{0,1,\ldots,T\}$. For example, at time step $t$, the location of the $u^{\text{th}}$ UE is denoted by $\boldsymbol{\rho}^{\mathrm{U}}_{u,t} = [x^{\mathrm{U}}_{u,t},\, y^{\mathrm{U}}_{u,t},\, z^{\mathrm{U}}_{u,t}]^{\top}$, while the location of the $d^{\text{th}}$ UAV-BS is given by $\boldsymbol{\rho}^{\mathrm{D}}_{d,t} = [x^{\mathrm{D}}_{d,t},\, y^{\mathrm{D}}_{d,t},\, z^{\mathrm{D}}_{d,t}]^{\top}$.

The objective of this work is to determine, in real-time, the optimal UAV-BS positions $\mathbf{X}^\mathrm{D}_{t}$ that maximize the total fair throughput, denoted by \( R_{\text{fair}} \), for the UEs. For notational simplicity, the explicit time index \(t\) is omitted in the remainder of this paper, and the analysis is conducted at a representative fixed time step. 
Unless otherwise stated, any time-indexed metric or parameter refers to its values over sequential time steps $[t, t-1, \ldots]$.

The total fair throughput is defined as the sum of
the logarithms of the throughput of all UEs, i.e.,

\begin{equation}
R_{\mathrm{fair}}(\mathbf{X}^{\mathrm{U}}, \mathbf{X}^{\mathrm{D}})
=
\sum_{d=1}^{D}
\sum_{u \in \mathcal{U}_d}
\log_{10}
\left(
R_{u,d}
\left(
\boldsymbol{\rho}^{\mathrm{U}}_u,
\mathbf{X}^{\mathrm{D}}
\right)
\right),
\label{eq_intro:fair_rate}
\end{equation}
where $\mathcal{U}_d$ denotes the set of UEs associated with the $d^{\text{th}}$ UAV-BS, with $|\mathcal{U}_d| = N$ in the considered setup. Moreover, $R_{u,d}(\cdot)$ denotes the achievable throughput of the $u^{\text{th}}$ UE served by the $d^{\text{th}}$ UAV-BS.
%
%
This formulation ensures a balanced approach in which improvements in transmission rates for UEs with lower rates are prioritized over those with already higher rates, thereby promoting fairness across the network. 
With this objective in mind, the optimization problem can be formally stated as:
\begin{subequations}
\begin{align}
        \max_{{{\mathbf{X}}}^\mathrm{D}} \quad 
    & R_{\text{fair}}(\mathbf{X}^\mathrm{U}, \mathbf{X}^\mathrm{D}) \label{eq:opt_main} \\
    \text{s.t.} \quad 
    & -X_{\max} \leq x^{\mathrm{U}}_{u,x} \leq X_{\max}, 
    && \forall u \in \mathcal{U}, \label{eq:8a} \\
    & -Y_{\max} \leq y^{\mathrm{U}}_{u,y} \leq Y_{\max}, 
    && \forall u \in \mathcal{U}, \label{eq:8b} \\
    & R_{u}(\boldsymbol{\rho}^\mathrm{U}_u, \mathbf{X}^\mathrm{D}) 
    \geq R_{u}^{\min}, 
    && \forall u \in \mathcal{U}, \label{eq:8e} \\
    & \mathrm{HS}_i \overset{\text{connected}}{\leftrightarrow} 
    \mathrm{UAV\mbox{-}BS}_d \;\Rightarrow\; d = i, 
    && \forall i \in \mathcal{D}. \label{eq:8f}
\end{align}
\end{subequations}
where $\mathbf{X}^\mathrm{D}$ denotes the UAV-BS positions to be optimized, while $\boldsymbol{\rho}_u^\mathrm{U}$ and $\mathbf{X}^\mathrm{U}$ represent the (unknown) position of the $u^{\text{th}}$ UE and the set of positions of all UEs, respectively, constrained within the deployment region.
Each HS operates within the specified horizontal deployment bounds along the $x$ and $y$ axes thanks to Constraint eq.~\eqref{eq:8a} and Constraint eq.~\eqref{eq:8b}. 
%
%
In terms of quality-of-service, Constraint eq.~\eqref{eq:8e} guarantees that every user meets a minimal throughput requirement.
Constraint eq.~\eqref{eq:8f} maintains a one-to-one association by requiring that UEs across each HS $\mathrm{HS}_i$ are only connected to its matching UAV-BS$_i$.
Optimizing the UAV-BS position at each time step is a trajectory optimization problem. On the other hand, real-time UAV-BS trajectory optimization is very difficult due to the problem's large dimensionality, non-convexity, stochastic nature, and nonlinearity. The constant movement of the UEs, along with dynamic fluctuations in the radio environment, necessitates adaptive and rapid decision-making capabilities that conventional optimization methods frequently fail to deliver adequately.

To address these challenges, we use DRL, specifically PPO, for UAV-BS trajectory optimization. DRL enables adaptive and real-time decision-making in stochastic, high-dimensional, and non-convex environments without requiring explicit system modeling. Compared to other RL approaches, value-based methods such as DQN and DDQN rely on action-space discretization, which limits their applicability in continuous control problems. Actor-critic methods such as DDPG and TRPO can handle continuous action spaces, but they are often prone to training instability, hyperparameter sensitivity, and higher computational complexity. In contrast, PPO provides a favorable balance between stable policy updates, sample efficiency, and implementation simplicity, making it particularly suitable for UAV-BS trajectory optimization in dynamic environments.



\subsection{DRL-based PPO}
In RL, an agent interacts with an environment and learns to make decisions by receiving feedback in the form of rewards. This interaction is typically formalized through four fundamental components:
\begin{itemize}
    \item \textbf{State spaces ($S$):} The set of all possible situations or states the agent can perceive.
    \item \textbf{Action Space ($A$):} The set of decisions or moves the agent can take in a given state.
    \item \textbf{Reward ($R$):} Scalar feedback signals received after taking an action, used to guide learning.
    \item \textbf{Policy ($\pi$):} A mapping from states to actions that defines the agent’s behavior.
\end{itemize}

The proposed UAV-BS trajectory optimization problem is formulated as a Markov decision process (MDP). At each time step $t$, the agent interacts with the environment through states $s_t \in \mathcal{S}$, actions $a_t \in \mathcal{A}$, and rewards $r_t$, with transitions represented as $(s_t, a_t, r_t, s_{t+1})$, where $s_{t+1}$ denotes the subsequent state~\cite{sutton2018reinforcement}. By employing PPO, the agent aims to learn a policy $\pi$ that maximizes the total fair throughput defined in eq.~\eqref{eq_intro:fair_rate}, treated as the expected long-term return (see Algorithm~\ref{alg:ppo_uav_single}).

PPO, similar to its predecessor TRPO, is an on-policy, model-free algorithm belonging to the actor-critic family~\cite{schulman2015trust}. It extends the REINFORCE algorithm~\cite{williams1992simple} by incorporating a value function estimator, which improves training stability and sample efficiency. In particular, PPO employs an advantage function to estimate the relative benefit of an action compared to the expected value of the current state, thereby enabling more stable and efficient policy updates. The advantage function is defined as:
\begin{equation}
     \hat{A}_t = r_t - V(s_t).
\end{equation}
where \( s_t \) and \( r_t \) denote the state and reward at time step \( t \), respectively, and \( V(s_t) \) represents the estimated value of state \( s_t \) under the current policy.
This advantage-based approach, coupled with the use of a value function, characterizes PPO as an actor-critic method, enabling the simultaneous optimization of both the policy \( \pi \) and the value function \( V(s) \).

    
    

\subsection{PPO Key Details}

PPO employs a clipped objective function to balance efficiency and stability, while remaining relatively simple to implement compared to other policy optimization methods. The algorithm optimizes the policy by maximizing a surrogate objective function~\cite{schulman2017proximal}, which is defined as:
\begin{equation}
    \begin{aligned}
    L^{\text{CLIP}}(\theta) = \hat{\mathbb{E}}_t \left[
        \min\left(
            r_t(\theta)\hat{A}_t,\ 
            \text{clip}(r_t(\theta), 1 - \epsilon, 1 + \epsilon)\hat{A}_t
        \right)
    \right].
    \end{aligned}
    \label{eq:clipped_loss}
\end{equation}
where \( \theta \) denotes the parameters of the neural networks used in the policy, \( \hat{A}_t \) is the estimated advantage at time step \( t \), \( \epsilon \) is a clipping hyperparameter that controls the extent of policy updates, and \( r_t(\theta) \) is the probability ratio between the new and old policies, measuring how much the policy has changed between updates.
Eq.~\eqref{eq:clipped_loss} constrains policy updates by limiting the probability ratio, thereby preventing excessively large updates that could destabilize training.

Specifically, \( r_t(\theta) \) denotes the ratio between the probability of taking action \( a_t \) in state \( s_t \) under the current policy \( \pi_\theta \) and the probability of taking the same action under the previous policy \( \pi_{\theta_{\text{old}}} \) ~\cite{schulman2017proximal}:
\begin{equation}
    r_t(\theta) = \frac{\pi_\theta(a_t \mid s_t)}{\pi_{\theta_{\text{old}}}(a_t \mid s_t)}.
\end{equation}
PPO differs from TRPO by using a clipping mechanism instead of enforcing a KL-divergence constraint. This clipping strategy prevents excessively large policy updates, thereby improving training stability by limiting the objective function when policy changes become too aggressive.
This discourages excessive policy updates by taking the minimum of the clipped and unclipped objectives. Additionally, PPO employs generalized advantage estimation (GAE)~\cite{schulman2015high} to improve the accuracy of the advantage function. 
In GAE,
an additional parameter $\lambda$ balances the bias-variance trade-off in advantage estimation, while the discount factor $\xi$ accounts for future benefits. This results in more reliable and efficient policy updates, making PPO well-suited for complex operation such as UAV-BS positioning. 

\section{Proposed PPO Implementation}

Without loss of generality, and considering a centralized learning framework with a single-agent, multi-UAV-BS setup, the following experiments analyze how PPO optimizes the positions of the UAV-BSs to efficiently serve the UEs across multiple corresponding HSs.


\begin{algorithm}[htb]
\caption{PPO Algorithm (Actor-Critic)}
\label{alg:ppo_uav_single}
\begin{algorithmic}[1]
\setlength{\baselineskip}{1.1\baselineskip} 

\STATE \textbf{Input:} rollout horizon $T$, epochs $K$, minibatch size $M\le T$, discount $\xi$, GAE parameter $\lambda$, clip $\epsilon$
\STATE \textbf{Initialize:} policy parameters $\theta$ \textit{(Actor)}, value parameters $\phi$ \textit{(Critic)}
\STATE $\theta_{\mathrm{old}} \leftarrow \theta$ \textit{(Actor)}

\FOR{iteration $=1,2,\dots$}
    \STATE Reset environment; initialize UAV-BS states and moving-HS states
    \FOR{$t=1,2,\dots,T$}
        \STATE Obs $\boldsymbol{s}_t \leftarrow \big[\boldsymbol{\rho}^{D}_{d,t},\ \{{{\boldsymbol{p}}_{{h,t}}},\bm{\gamma}_{h,t}, \psi_{{\bm\alpha}_{h,t}}, \omega_{\bm{\alpha}_{h,t}} \}_{h=1}^{|\mathcal{H}|}\big]$
        \STATE Sample action $a_t \sim \pi_{\theta_{\mathrm{old}}}(\cdot \mid s_t)$ \textit{(Actor)}
        \STATE Execute $a_t$; 
        \STATE Receive reward $r_t \leftarrow \sum_{u=1}^{U} R_{u,t}$
        \STATE Store $(s_t,a_t,r_t,\log\pi_{\theta_{\mathrm{old}}}(a_t\mid s_t))$ 
        \STATE Evaluate value $V_{\phi}(s_t)$ \textit{(Critic)}
    \ENDFOR

    \STATE Compute advantages $\{\hat{A}_t\}_{t=1}^{T}$ using GAE \textit{(Critic)}:
    \STATE \hspace{1em}$\delta_t \leftarrow r_t + \xi V_{\phi}(s_{t+1}) - V_{\phi}(s_t)$
    \STATE \hspace{1em}$\hat{A}_t \leftarrow \sum_{l=0}^{T-t}(\xi\lambda)^l\,\delta_{t+l}$
    \STATE Compute returns $\hat{R}_t \leftarrow \hat{A}_t + V_{\phi}(s_t)$ \textit{(Critic)}

    \FOR{epoch $=1,2,\dots,K$}
        \STATE Shuffle data, split into minibatches of size $M$
        \FOR{each minibatch}
            \STATE Compute probability ratio \textit{(Actor)}:
            \STATE \hspace{1em}$\rho_t(\theta)=\exp\big(\log\pi_{\theta}(a_t\mid s_t)-\log\pi_{\theta_{\mathrm{old}}}(a_t\mid s_t)\big)$
            \STATE Compute $L^{\mathrm{clip}}(\theta)$ \textit{(Actor)}
            \STATE Compute value loss $L^{V}(\phi)$ \textit{(Critic)}
            \STATE Update policy parameters $\theta$ using Adam \textit{(Actor)}
            \STATE Update value parameters $\phi$ using Adam \textit{(Critic)}
        \ENDFOR
    \ENDFOR

    \STATE $\theta_{\mathrm{old}} \leftarrow \theta$ \textit{(Actor)}
\ENDFOR
\end{algorithmic}
\end{algorithm}

\subsection{State-Space Definition}
In this subsection, we define the state-space representation employed by the DRL agent for the proposed UAV-BS trajectory optimization problem. 
Each UAV-BS operates over episodes of length $T$, where at each time step it interacts with the environment. Accordingly, we formulate a time-indexed state representation for the proposed single-agent, multi-UAV-BS system, capturing the dynamics associated with the $d^{\text{th}}$ UAV-BS and the $u^{\text{th}}$ UE, as follows:

\begin{equation}
{\mathbf{X}}^{\mathrm{D}}_d(t) =
\left[
{\bm{\rho}}_{d,t-T}^{\mathrm{D}}, \dots, \bm{\rho}_{d,t-i}^{\mathrm{D}}, \dots, \bm{\rho}_{d,t}^{\mathrm{D}}
\right],
\end{equation}

\begin{equation}
{\bm{p}}_{u,h,d}(t)
=
\left[
p_{u,h,d,t-T}, \dots, p_{u,h,d,t-i}, \dots, p_{u,h,d,t}
\right],
\end{equation}


\begin{equation}
    {\bm{\gamma}}_{u,h,d}(t) = [\gamma_{u,h,d,t-T},\dots,\gamma_{u,h,d,t-i}, \dots, \gamma_{u,h,d,t}],
\end{equation}

\begin{equation}
   {\bm{\alpha}}_{u,h,d}(t) = [\alpha_{u,h,d,t-T}, \dots,\alpha_{u,h,d,t-i}, \dots, \alpha_{u,h,d,t}].
\end{equation}
where ${\mathbf{X}}^{\mathrm{D}}_d(t)$ denotes the time-dependent UAV-BS position matrix. 
The time-varying vectors ${\bm{p}}_{u,h,d}(t)$, ${\bm{\gamma}}_{u,h,d}(t)$, and ${\bm{\alpha}}_{u,h,d}(t)$ represent the received power, SINR, and angle of arrival (AoA), respectively, for the $u^{\text{th}}$ UE in the $h^{\text{th}}$ HS served by the $d^{\text{th}}$ UAV-BS across an episode of length $T$.
The indices $\{u,h,d,t\}$ denote the UE, HS, UAV-BS, and time-step indices, respectively. The AoA is estimated from UE reference signals received at the UAV-BS.
The overall state, including spatial and communication statistics of the $u^{\text{th}}$ UE in the $h^{\text{th}}$ HS served by the $d^{\text{th}}$ UAV-BS at time step $i$, is represented as
\begin{equation}\label{eq:ue_stats}
\boldsymbol{{s}}_{t=i} =
\bigl\{\boldsymbol{{\rho}}^{\mathrm{D}}_{d,i},\,
{p_{u,h,d,i}} ,\,
\mathrm{\gamma}_{u,h,d,i},\,
\mathrm{\alpha}_{u,h,d,i}\bigr\}.
\end{equation}
where $\alpha_{u,h,d,i} \in (-\pi,\pi]$ represents the AoA of the $u^{\text{th}}$ UE, for $u \in \{1,\ldots,N\}$ and $i \in \{0,\ldots,T-1\}$.
We now introduce the full state space of the proposed UAV-BS-assisted wireless network under the full observability. 
For the network described in Section~\ref{System_network}, consisting of $D$ UAV-BSs and $H$ HSs, we assume a one-to-one association between UAV-BSs and HSs, such that the $h^{\text{th}}$ HS is served by the $d^{\text{th}}$ UAV-BS with $d = h$ and $D=H$. 
Accordingly, the UAV-BS position matrix, UE downlink received power matrix, SINR matrix,
and AoA matrix at time step $i$ are 
redefined
to jointly represent all UAV-BS-HS pairs as follows:

\begin{equation}
\mathbf{X}^{\mathrm{D}}_{i}
=
\begin{bmatrix}
x^{\mathrm{D}}_{1,i}, & y^{\mathrm{D}}_{1,i}, & z^{\mathrm{D}}_{1,i} \\
\vdots  & \vdots & \vdots \\
x^{\mathrm{D}}_{d,i}, & y^{\mathrm{D}}_{d,i}, & z^{\mathrm{D}}_{d,i} \\
\vdots & \vdots & \vdots \\
x^{\mathrm{D}}_{D,i} , & y^{\mathrm{D}}_{D,i} , & z^{\mathrm{D}}_{D,i}
\end{bmatrix},
\end{equation}

\begin{equation}
\label{power_downlink}
\mathbf{P}_{i}
=
\begin{bmatrix}
p_{1,1,1,i} & \cdots & p_{u,1,1,i} & \cdots & p_{N,1,1,i} \\
\vdots & \ddots & \vdots & \ddots & \vdots \\
p_{1,h,d,i} & \cdots & p_{u,h,d,i} & \cdots & p_{N,h,d,i} \\
\vdots & \ddots & \vdots & \ddots & \vdots \\
p_{1,H,D,i} & \cdots & p_{u,H,D,i} & \cdots & p_{N,H,D,i}
\end{bmatrix},
\end{equation}

\begin{equation}
\label{SINR_matrix}
\boldsymbol{\Gamma}_{i}
=
\begin{bmatrix}
\gamma_{1,1,1,i} & \cdots & \gamma_{u,1,1,i} & \cdots & \gamma_{N,1,1,i} \\
\vdots & \ddots & \vdots & \ddots & \vdots \\
\gamma_{1,h,d,i} & \cdots & \gamma_{u,h,d,i} & \cdots & \gamma_{N,h,d,i} \\
\vdots & \ddots & \vdots & \ddots & \vdots \\
\gamma_{1,H,D,i} & \cdots & \gamma_{u,H,D,i} & \cdots & \gamma_{N,H,D,i}
\end{bmatrix},
\end{equation}

\begin{equation}
\label{AOA_matrix}
\boldsymbol{\mathrm{A}}_{i}
=
\begin{bmatrix}
\alpha_{1,1,1,i} & \cdots & \alpha_{u,1,1,i} & \cdots & \alpha_{N,1,1,i} \\
\vdots & \ddots & \vdots & \ddots & \vdots \\
\alpha_{1,h,d,i} & \cdots & \alpha_{u,h,d,i} & \cdots & \alpha_{N,h,d,i} \\
\vdots & \ddots & \vdots & \ddots & \vdots \\
\alpha_{1,H,D,i} & \cdots & \alpha_{u,H,D,i} & \cdots & \alpha_{N,H,D,i}
\end{bmatrix}.
\end{equation}
Here, the UE index $u$ is repeated across the columns to indicate UEs with the same local index in different HSs. In contrast, the HS index $h$ and UAV-BS index $d$ remain fixed within each row, representing UEs associated with the same HS and served by the same UAV-BS.
Moreover, $\mathbf{X}^{\mathrm{D}}_{i} \in \mathbb{R}^{D \times 3}$, while $\mathbf{P}_{i}$, $\boldsymbol{\Gamma}_{i}$, and $\mathbf{A}_{i} \in \mathbb{R}^{H \times N}$.
Each UAV-BS treats the UEs within a given HS as a single aggregated entity. Therefore, communication statistics are averaged over the $N$ UEs in each HS. Specifically, the downlink received power and SINR are computed using the arithmetic mean. In contrast, for AoA measurements, we employ circular statistics, namely the circular mean and circular standard deviation, which are more appropriate for angular data and provide improved robustness and accuracy for UAV-BS navigation. A formal proof of this choice is provided in the Appendix.
Thus, the circular mean of ${\alpha}_{h,d,i}$ is computed as follows \cite{mardia2009directional}:

\begin{equation}
\begin{aligned}
\psi_{\bm{\alpha}_{h,d,i}}
&=
\arg\!\left(
\frac{1}{N}\sum_{u=1}^{N} e^{\mathrm{j}\alpha_{u,h,d,i}}
\right)
\\
&=
\arg\!\left(
\sum_{u=1}^{N}\cos\alpha_{u,h,d,i}
+
\mathrm{j}\sum_{u=1}^{N}\sin\alpha_{u,h,d,i}
\right).
\end{aligned}
\label{eq:circ_mean}
\end{equation}

Equivalently, this circular mean can be computed using the two-argument arctangent of the summed sine and cosine components.

To define the circular standard deviation, we first introduce the mean resultant length, given by
\begin{equation}\label{eq:resultant_length}
R_{h,d,i}
=
\left\lVert \frac{1}{N}\sum_{u=1}^{N} e^{\mathrm{j}\alpha_{u,h,d,i}} \right\rVert
\in[0,1].
\end{equation}

The circular standard deviation is then defined as follows \cite{mardia2009directional}:
\begin{equation}\label{eq:circ_std}
\omega_{\boldsymbol{\alpha}_{h,d,i}}
=
\sqrt{-2\ln R_{h,d,i}}.
\end{equation}

Now, by applying the arithmetic mean to the downlink received power eq.~\eqref{power_downlink} and SINR eq.~\eqref{SINR_matrix}, and the circular mean and circular standard deviation to the AoAeq.~\eqref{AOA_matrix} across the user dimension (i.e., over $N$ users) for each HS, we obtain the following vectors:

\begin{equation}
{{\boldsymbol{p}}}_{i} =
\left[
p_{1,1,i}, \dots, p_{h,d,i}, \dots, p_{H,D,i}
\right]^{\top},
\end{equation}

\begin{equation}
{\boldsymbol{\gamma}}_{i} =
\left[
\gamma_{1,1,i}, \dots, \gamma_{h,d,i}, \dots, \gamma_{H,D,i}
\right]^{\top},
\end{equation}

\begin{equation}
\boldsymbol{\psi}_{\boldsymbol{\mathrm{A}}_{i}} =
\left[
\psi_{\bm{\alpha}_{1,1,i}}, \dots, \psi_{\bm{\alpha}_{h,d,i}}, \dots, \psi_{\bm{\alpha}_{H,D,i}}
\right]^{\top}.
\end{equation}

\begin{equation}
\boldsymbol{\omega}_{\boldsymbol{\mathrm{A}}_{i}} =
\left[
\omega_{\bm{\alpha}_{1,1,i}}, \dots, \omega_{\bm{\alpha}_{h,d,i}}, \dots, \omega_{\bm{\alpha}_{H,D,i}}
\right]^{\top}.
\end{equation}

\begin{figure*}[!t]
\centering
\begin{equation}\label{eq:state_matrix}
\scalebox{0.9}{$
\begin{aligned}
\mathbf{S}_{i}
&=
\left[
\mathbf{X}^{\mathrm{D}}_{i} \;
\mathbf{p}_{i} \;
\boldsymbol{\gamma}_{i} \;
\boldsymbol{\psi}_{\boldsymbol{\mathrm{A}}_{i}} \;
\boldsymbol{\omega}_{\boldsymbol{\mathrm{A}}_{i}}
\right]
\in \mathbb{R}^{D \times 7}
=
\left[
\begin{array}{
c
c
c@{\hspace{2.2em}}
c@{\hspace{2.2em}}
c@{\hspace{2.2em}}
c@{\hspace{2.2em}}
c}
x^{\mathrm{D}}_{1,i} & y^{\mathrm{D}}_{1,i} & z^{\mathrm{D}}_{1,i} 
& p_{1,1,i}
& \gamma_{1,1,i}
& \psi_{\boldsymbol{\alpha}_{1,1,i}}
& \omega_{\boldsymbol{\alpha}_{1,1,i}}
\\
\vdots & \vdots & \vdots
& \vdots
& \vdots
& \vdots
& \vdots
\\
x^{\mathrm{D}}_{d,i} & y^{\mathrm{D}}_{d,i} & z^{\mathrm{D}}_{d,i}
& p_{h,d,i}
& \gamma_{h,d,i}
& \psi_{\boldsymbol{\alpha}_{h,d,i}}
& \omega_{\boldsymbol{\alpha}_{h,d,i}}
\\
\vdots & \vdots & \vdots
& \vdots
& \vdots
& \vdots
& \vdots
\\
x^{\mathrm{D}}_{D,i} & y^{\mathrm{D}}_{D,i} & z^{\mathrm{D}}_{D,i}
& p_{D,i}
& \gamma_{H,D,i}
& \psi_{\boldsymbol{\alpha}_{H,D,i}}
& \omega_{\boldsymbol{\alpha}_{H,D,i}}
\end{array}
\right].
\end{aligned}
$}
\end{equation}
\end{figure*}  
where $p_{1,1,i} = \frac{1}{N} \sum_{u=1}^{N} p_{u,1,1,i}$ denotes the average downlink received power over all $N$ UEs associated with the $1$-st HS served by the $1$-st UAV-BS at time step $i$. By aggregating the previously defined matrices and vectors, the state $\boldsymbol{S}_i$ (i.e., the state space at time step $i$) is formulated as an extended state matrix, as given in eq.~ \eqref{eq:state_matrix} 
where $d \in \{1, 2, \ldots, D\}$ and $h \in \{1, 2, \ldots, H\}$.
Here, $\mathbf{X}^{\mathrm{D}}_i \in \mathbb{R}^{D \times 3}$, where each row corresponds to the spatial coordinates $(x,y,z)$ of a UAV-BS, and $\boldsymbol{p}_{i}, \boldsymbol{\gamma}_i, \boldsymbol{\psi}_{\boldsymbol{\alpha}_i}, \boldsymbol{\omega}_{\boldsymbol{\alpha}_i} \in \mathbb{R}^{H \times 1}$.
Consequently, 
Each HS is assigned to a unique UAV-BS, meaning that no UAV-BS serves more than one HS. Therefore, when the numbers of UAV-BSs and HSs are equal (i.e., $D = H$), the state matrix satisfies $\mathbf{S}_i \in \mathbb{R}^{D \times 7}$.
The right-hand side
in eq.~\eqref{eq:state_matrix} 
is important because it provides a suitable structure as an input to the neural network. Each row, e.g., 
\begin{equation}
\boldsymbol{s}_{d,i} =
\begin{bmatrix}
x^{\mathrm{D}}_{d,i} &
y^{\mathrm{D}}_{d,i} &
z^{\mathrm{D}}_{d,i} &
p_{h,d,i} &
\gamma_{h,d,i} &
\psi_{\boldsymbol{\alpha}_{h,d,i}} &
\omega_{\boldsymbol{\alpha}_{h,d,i}}
\end{bmatrix}.
\end{equation}
is treated as a single input feature vector.

\subsection{Action Space}
The action specifies the movement decision of each UAV-BS at every step. 
Given the continuous action space, the agent selects two control variables: the movement direction and displacement magnitude. 
The direction is represented by the angle $\beta_d$, and the displacement by the distance $r_{d}$. 
Accordingly, the action space is defined as $\mathcal{A} = \{(\beta_d, r_d)\}$, where $\beta_d \in [-180^\circ, 180^\circ)$ and $r_{d} \in [0, r_{\max}]$. 
Here, $\beta_d$ is measured with respect to the east direction, and $r_{{\max}}$ denotes the maximum displacement per step, determined by the UAV-BS speed and the size of the operating environment.

The proposed framework leverages the ability of PPO to operate directly in continuous action spaces. 
By avoiding action discretization, the agent explores a continuous movement-action space, enabling flexible UAV-BS repositioning,
and fine-grained spatial adaptation. 
This is particularly important in the considered scenario, where even small position adjustments can lead to significant variations in network performance.
\subsection{Reward Definition}
\label{reward_definition}
Within the proposed framework, the reward function is designed to maximize the total fairness-aware throughput $R_{\mathrm{fair}}$ defined in eq.~\eqref{eq_intro:fair_rate}.
To ensure numerical stability and consistent scaling during training, a min–max normalization is applied to the reward signal. 
This normalization mitigates large variations in reward magnitude and facilitates stable and efficient convergence of the learning process.
In particular, we adopt and compare two parametric reward functions based on widely used nonlinear activation functions, namely the sigmoid and hyperbolic tangent functions~\cite{dubey2022activation}.
The sigmoid reward function $z_\text{sig}$ is defined as follows,
\begin{equation}\label{eq:sigmoid}
    z_\text{sig}(R_{\text{fair}})
    =
    \left(
    1+\exp\!\left(
    -c_s\left( R_{\text{fair}}-c_m \right)
    \right)
    \right)^{-1},
\end{equation}
while the hyperbolic tangent reward function $z_\text{tanh}$ is defined as
\begin{equation}\label{eq:tanh}
    z_\text{tanh}(R_{\text{fair}})
    =
    \tanh\!\left(
    c_s\left(R_{\text{fair}}-c_m\right)
    \right).
\end{equation}
Here, $c_s > 0$ controls the slope with respect to $R_{\text{fair}}$, whereas $c_m$ sets the location of the transition inflection point.

To tune these parameters, we consider two representative scenarios. 
In the first scenario, UAV-BSs are positioned directly above their corresponding hotspots. 
In the second scenario, UAV-BS positions evolve during an episode. 
For both cases, the average throughput over the entire episode is evaluated across multiple realizations, and the parameters $c_s$ and $c_m$ are tuned to achieve appropriate reward scaling and sensitivity to input variations, thereby guiding the learning process.

In Section~\ref{Reward_sim_sigmoid}, the results demonstrate that properly scaled sigmoid-based reward functions provide more reliable convergence, improved stability, and superior throughput performance during both the training and evaluation phases. These findings highlight the importance of smooth and well-conditioned reward shaping for stable PPO training dynamics.

\begin{table}[!t]
\centering
\caption{Simulation Environment and PPO Configuration}
\label{tab:simulation_ppo_parameters}
\footnotesize
\setlength{\tabcolsep}{3pt}
\renewcommand{\arraystretch}{1.2}
\begin{tabular}{|c|c|}
\hline
\textbf{Parameter} & \textbf{Value} \\
\hline

\multicolumn{2}{|c|}{\textbf{Simulation Environment}} \\
\hline
Evaluation Area ($L \times W$) & $400\,\mathrm{m} \times 1200\,\mathrm{m}$ \\
$L$ & $[-200,\,200]\,\mathrm{m}$ \\
$W$ & $[-600,\,600]\,\mathrm{m}$ \\
UE Altitude within HS & $1.5\,\mathrm{m}$ \\
Number of HSs ($H$) & 3 \\
HS Radius & $0.1\,\mathrm{m}$ \\
UE Density per HS & 10 \\
UE Speed within HS & $8\,\mathrm{m/s}$ \\
Number of UAV-BSs ($D$) & 3 \\
UAV-BS Altitude & $50\,\mathrm{m}$ \\
UAV-BS Initial Speed & $30\,\mathrm{m/s}$ \\
Maximum Displacement ($r_{\max}$) & $20\,\mathrm{m}$ \\
Carrier Frequency ($f$) & $2\,\mathrm{GHz}$ \\
System Bandwidth ($B$) & $5\,\mathrm{MHz}$ \\
Transmit Power ($P^{\mathrm{tx}}$) & $43\,\mathrm{dBm}$ \\
Noise Figure & $5\,\mathrm{dB}$ \\
Propagation Model & 3GPP UMa (TR~36.814) \\
Antenna Type & Omnidirectional \\
Channel Model & Rician \\
\hline

\multicolumn{2}{|c|}{\textbf{PPO Training Hyperparameters}} \\
\hline
Num. episodes & 22{,}524 \\
Max steps / episode & 128 \\
Discount factor ($\gamma$) & 0.99 \\
Learning rate ($\eta$) & $3 \times 10^{-5}$ \\
Batch size & 128 \\
Epochs per episode & 15 \\
GAE parameter ($\lambda$) & 0.95 \\
PPO clip parameter ($\epsilon$) & 0.2 \\
Entropy coefficient & 0.1 \\
State memory size & 2 \\
Evaluation policy & Stochastic policy sampling \\
\hline

\multicolumn{2}{|c|}{\textbf{PPO Neural Network Architecture}} \\
\hline
Hidden Layers & 3 \\
Neurons per Hidden Layer & 128 \\
Maximum Gradient Norm & 1.0 \\
\hline
\end{tabular}
\end{table}


\begin{table}[h]
\centering
\caption{Spatial Configuration of HSs and UAV-BSs.}
\label{tab:combined_setup}

\begin{subtable}{\linewidth}
\centering
\caption{Hotspot initial positions for both training and evaluation.}
\label{tab:hs_positions}
\renewcommand{\arraystretch}{1.2}+
\begin{tabular}{c c}
\hline
Entities & $\mathbf{X}^{\mathrm{H}}_0$ \\
\hline
\vspace{0.8mm}
$\begin{bmatrix}
\boldsymbol{\rho}^{\mathrm{H}}_1 \\
\boldsymbol{\rho}^{\mathrm{H}}_2 \\
\boldsymbol{\rho}^{\mathrm{H}}_3
\end{bmatrix}$
\vspace{0.8mm}
&
$\begin{bmatrix}
-170 & -470 & 1.5 \\
170  & 0    & 1.5 \\
-170 & 470  & 1.5
\end{bmatrix}$ \\[-0.5mm]
\hline
\end{tabular}
\renewcommand{\arraystretch}{1}
\end{subtable}

\vspace{0.4cm}

\begin{subtable}{\linewidth}
\centering
\caption{UAV-BS deployment scenarios for the evaluation process.}
\label{tab:uav_scenarios}
\renewcommand{\arraystretch}{1.3}
\begin{tabular}{c c c}
\hline
Scenario & Entities & $\mathbf{X}^{\mathrm{D}}$ \\
\hline
a &
$\begin{bmatrix}
\boldsymbol{\rho}^{\mathrm{D}}_1 \\
\boldsymbol{\rho}^{\mathrm{D}}_2 \\
\boldsymbol{\rho}^{\mathrm{D}}_3 \\
\end{bmatrix}$
&
$\begin{bmatrix}
0 & 180 & 50 \\
0 & 0   & 50 \\
0 & -180 & 50 \\

\end{bmatrix}$ \\
\hline
b &
$\begin{bmatrix}
\boldsymbol{\rho}^{\mathrm{D}}_1 \\
\boldsymbol{\rho}^{\mathrm{D}}_2 \\
\boldsymbol{\rho}^{\mathrm{D}}_3 \\
\end{bmatrix}$
&
$\begin{bmatrix}
-150 & -350 & 50 \\
150  & 0    & 50 \\
-150 & 350  & 50 \\
\end{bmatrix}$ \\
\hline
c &
$\begin{bmatrix}
\boldsymbol{\rho}^{\mathrm{D}}_1 \\
\boldsymbol{\rho}^{\mathrm{D}}_2 \\
\boldsymbol{\rho}^{\mathrm{D}}_3 \\
\end{bmatrix}$
&
$\begin{bmatrix}
-180 & 0 & 50 \\
0    & 0 & 50 \\
180  & 0 & 50 \\
\end{bmatrix}$ \\
\hline
d &
$\begin{bmatrix}
\boldsymbol{\rho}^{\mathrm{D}}_1 \\
\boldsymbol{\rho}^{\mathrm{D}}_2 \\
\boldsymbol{\rho}^{\mathrm{D}}_3 \\
\end{bmatrix}$
&
$\begin{bmatrix}
-180 & -380 & 50 \\
0    & 0    & 50 \\
180  & 380  & 50 \\
\end{bmatrix}$ \\
\hline
\end{tabular}
\renewcommand{\arraystretch}{1}
\end{subtable}

\end{table}


 \begin{table}[t]
\centering
\caption{Performance level for throughput and convergence}
\label{tab:performance_levels}
\footnotesize
\setlength{\tabcolsep}{14 pt} 
\renewcommand{\arraystretch}{1.2}
\begin{tabular}{|l|c|}
\hline
\textbf{Metric Level} & \textbf{Numerical Definition} \\
\hline
\multicolumn{2}{|c|}{\textbf{Throughput Performance}} \\
\hline
\textbf{Best} & $\geq 30$~Mbps \\
Strong & $[25, 30)$~Mbps \\
Good & $[20, 25)$~Mbps \\
Moderate & $[15, 20)$~Mbps \\
Weak & $< 15$~Mbps \\
\hline
\multicolumn{2}{|c|}{\textbf{Convergence Speed}} \\
\hline
\textbf{Fast} & $\leq 12$k episodes \\
Medium & $12$k-$15$k episodes \\
Slow & $> 15$k episodes \\
\hline
\end{tabular}
\end{table}

\section{UAV-BS Positioning Training and Evaluation}

In this section, we investigate the UAV-BS positioning and trajectory optimization problem. 
The first subsection details the simulation environment, PPO training hyperparameters, mobility models, and performance evaluation metrics adopted in this work. The second subsection presents the proposed approach improvement
and simulation setup,
along with the corresponding numerical results and their discussion.

\subsection{Experimental Setup }
To verify our claims and evaluate the effectiveness of the proposed methodology, we conduct a comprehensive study under diverse UE mobility patterns. The performance is assessed from two perspectives: communication-level metrics, represented by network throughput, and AI-level metrics, represented by the learning reward.
To clearly illustrate the performance of the proposed model, we provide training and evaluation plots, statistically aggregated results reported as mean $\pm$ standard deviation over multiple random seeds, and qualitative performance comparison tables. These results offer both quantitative performance evaluation and interpretable insights into the system behavior.

To this end, we consider a rectangular deployment area of dimensions $400\,\mathrm{m} \times 1200\,\mathrm{m}$, within which three UAV-BSs operate with a system bandwidth of $5\,\mathrm{MHz}$ at a carrier frequency of $2\,\mathrm{GHz}$ and an altitude of $50\,\mathrm{m}$. 
The evaluation setup follows the system model described in Section~\ref{sec:system_model}, adopting the 3GPP UMa propagation model (see Table~\ref{tab:simulation_ppo_parameters}).


\subsubsection{PPO Algorithm Configuration}
Key parameters and architectural configurations for the proposed PPO-based algorithm are as follows. 
Training is conducted over the 22{,}524 episodes, with each episode consisting of a fixed number of frames (128). 
A constant learning rate ($\eta=3 \times 10^{-5}$)
is assumed, and both the actor and critic networks adopt a multi-layer architecture consisting of three hidden layers, each with 128 neurons.
The PPO algorithm leverages a discount factor ($\xi=0.99$) and a GAE parameter ($\lambda=0.95$) to balance immediate and future rewards. 
To guarantee stable learning, gradient clipping is used with a bounded norm, and a clipping parameter ($\epsilon=0.2$) is applied in the surrogate objective. 
Moreover, we consider a finite state memory size ($M=2$) for our model, which enables the agent to capture UE mobility patterns without directly observing UE positions. Hyperparameters were selected empirically based on convergence stability and preliminary tuning experiments.

\begin{figure*}[!htb]
\centering
\textbf{\small Reward Shaping Impact on PPO Training Dynamics}\\[6pt]
\begin{subfigure}[!htb]{0.48\textwidth}
\centering
\includegraphics[width=\linewidth,height=5cm]{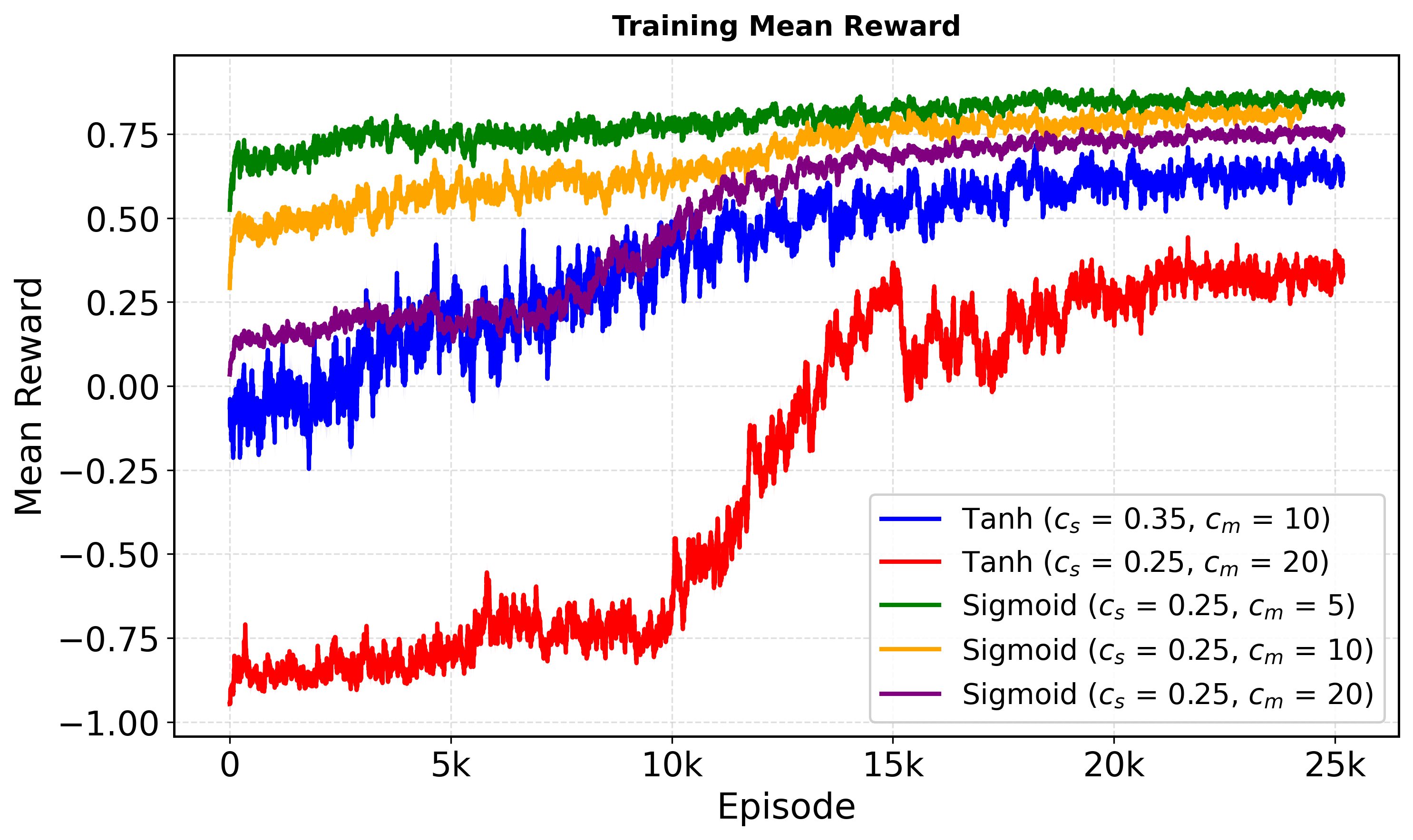}
\caption{Reward evolution during training}
\label{fig:rand_mean_r}
\end{subfigure}
\hfill
\begin{subfigure}[!htb]{0.48\textwidth}
\centering
\includegraphics[width=\linewidth,height=5cm]{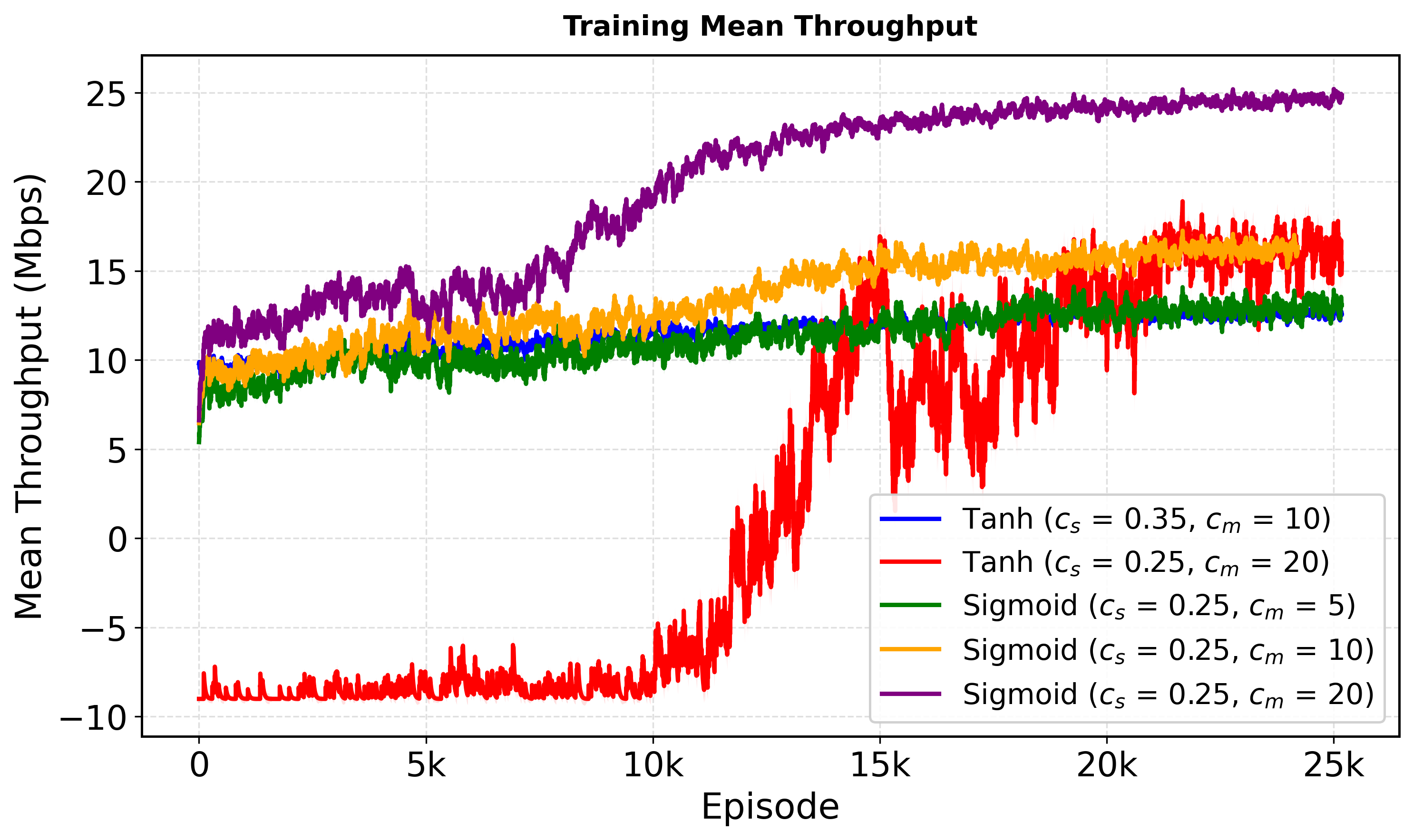}
\caption{Throughput performance during training}
\label{fig:rand_mean_th_f}
\end{subfigure}

\vspace{8pt}


\caption{Comparison of reward shaping strategies during training. Sigmoid-based functions show more stable convergence and better performance than tanh-based designs.}
\label{fig:reward_comparison_train}
\end{figure*}

\subsubsection{UE Mobility Scenarios}
We evaluate the proposed algorithm under several UE mobility patterns with increasing trajectory complexity, while keeping the initial positions fixed (See Table~\ref{tab:hs_positions}). The considered mobility models, introduced in Section~\ref{System_network}, Subsection~\ref{UE_SET_HS}, are simulated using the following configurations:

\begin{itemize}

   \item \textit{Static UEs (No Mobility)}: Three HSs, each consisting of 10 UEs arranged within a circular region of radius \(0.1\).
   
   \item \textit{Linear Motion}: Three HSs move along independent straight-line trajectories at a speed of \(8\,\mathrm{m/s}\).
    
   \item \textit{Circular Motion}: Three HSs move along independent circular trajectories with a radius of \(50\,\mathrm{m}\) at a constant speed of \(8\,\mathrm{m/s}\).

   \item \textit{Cosine-Pattern Motion}: Three HSs follow independent cosine-shaped trajectories at \(8\,\mathrm{m/s}\), with vertical displacement varying between \(-15\,\mathrm{m}\) and \(+15\,\mathrm{m}\).

\end{itemize}

\subsection{Training and evaluation phases}
Using the numerical results, we  
represent the performance of the proposed approach in both the training and evaluation phases. 
During training, the initial locations of the UAV-BSs are selected randomly within the predefined environment. 
For evaluation, predefined and fixed initial UAV-BS positions are used to assess the agent’s performance under consistent conditions. 
Specifically, four distinct evaluation positions are considered, each chosen from different regions of the playground to provide a comprehensive and fair evaluation of the agent’s behavior through the entire area (see Table~\ref{tab:simulation_ppo_parameters}).
Accordingly, we consider two types of points used in the experiments:
\begin{itemize}
    \item Random-point training: A random point is generated inside the playground during the training phase.
    \item Evaluation point: A predefined and fixed point in Table~\ref{tab:uav_scenarios} used to evaluate the UAV-BSs after specific iterations (e.g., every 500 iterations in our simulation).
\end{itemize}
\subsubsection{Performance evaluation level}
We assume that the maximum achievable total network throughput is obtained when each UAV-BS is positioned at the top of its corresponding HS. Based on this upper bound,
the evaluation throughput is categorized into five qualitative performance levels: \emph{Best}, \emph{Strong}, \emph{Good}, \emph{Moderate}, and \emph{Weak}. These levels are defined based on the maximum and average overall network throughput obtained from the considered system setup, which are 35 Mbps and 18 Mbps, respectively. These reference values correspond to an ideal scenario without DRL control, where the UAV-BSs perfectly follow the HSs by matching their direction and speed while continuously maintaining optimal service coverage.
In addition, to enable a systematic comparison of learning dynamics across scenarios, the convergence speed is categorized into three qualitative levels: \emph{Fast}, \emph{Medium}, and \emph{Slow}. It is worth noting that the proposed algorithm was evaluated under various reward-function and state-space configurations, with the fastest convergence observed at approximately 12k training episodes. Therefore, this value is adopted as the reference convergence speed in our evaluations.
The corresponding numerical thresholds for both throughput and convergence speed are summarized in Table~\ref{tab:performance_levels}.


\subsubsection{Ablation Study}
\begin{figure*}[!t]
\centering
\textbf{\small Training Performance Under Different State Space Designs}\\[6pt]

\begin{subfigure}[t]{0.48\textwidth}
\centering
\includegraphics[width=\linewidth,height=5cm]{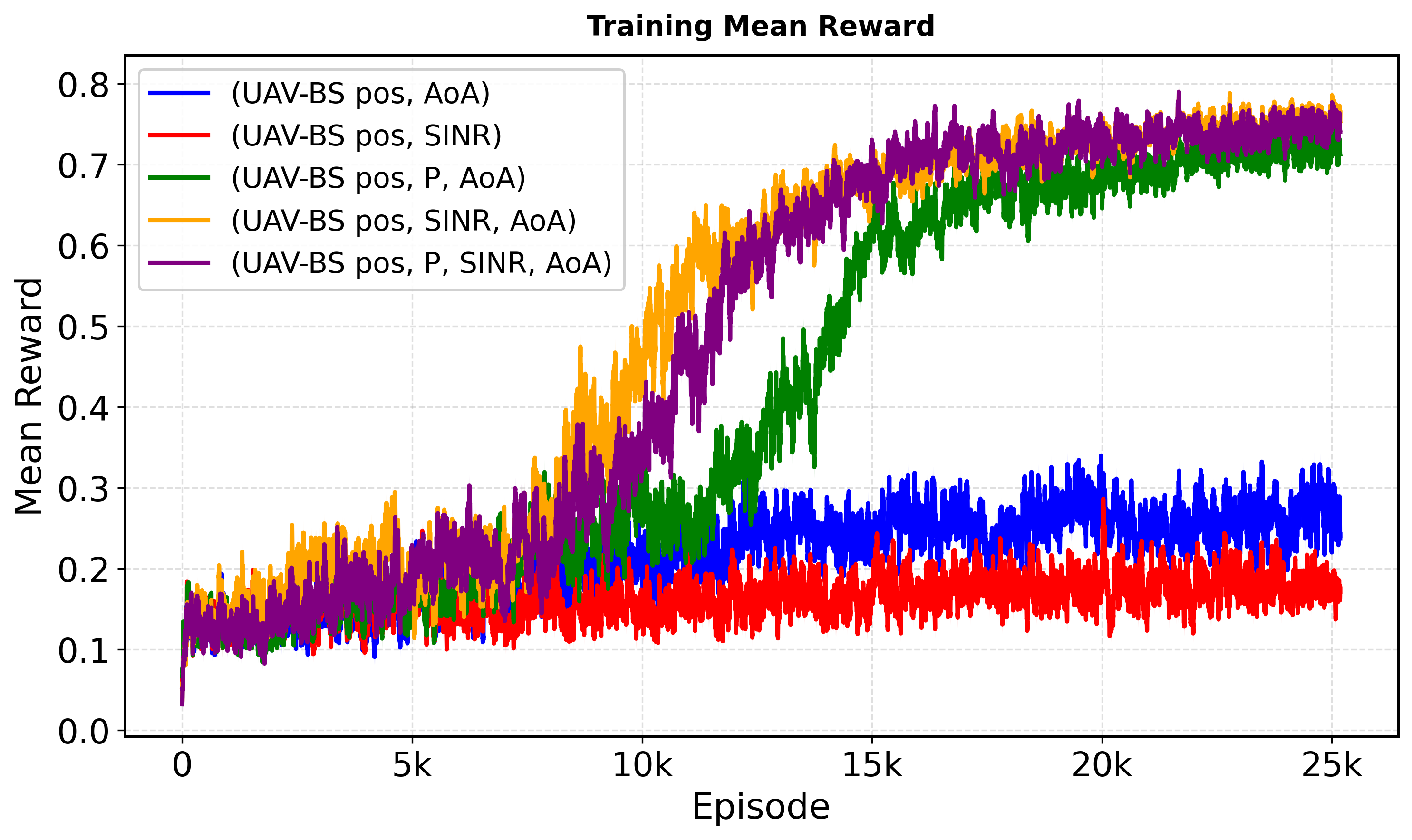}
\caption{Reward convergence behavior during training}
\label{fig1:obs_mean_reward}
\end{subfigure}
\hfill
\begin{subfigure}[t]{0.48\textwidth}
\centering
\includegraphics[width=\linewidth,height=5cm]{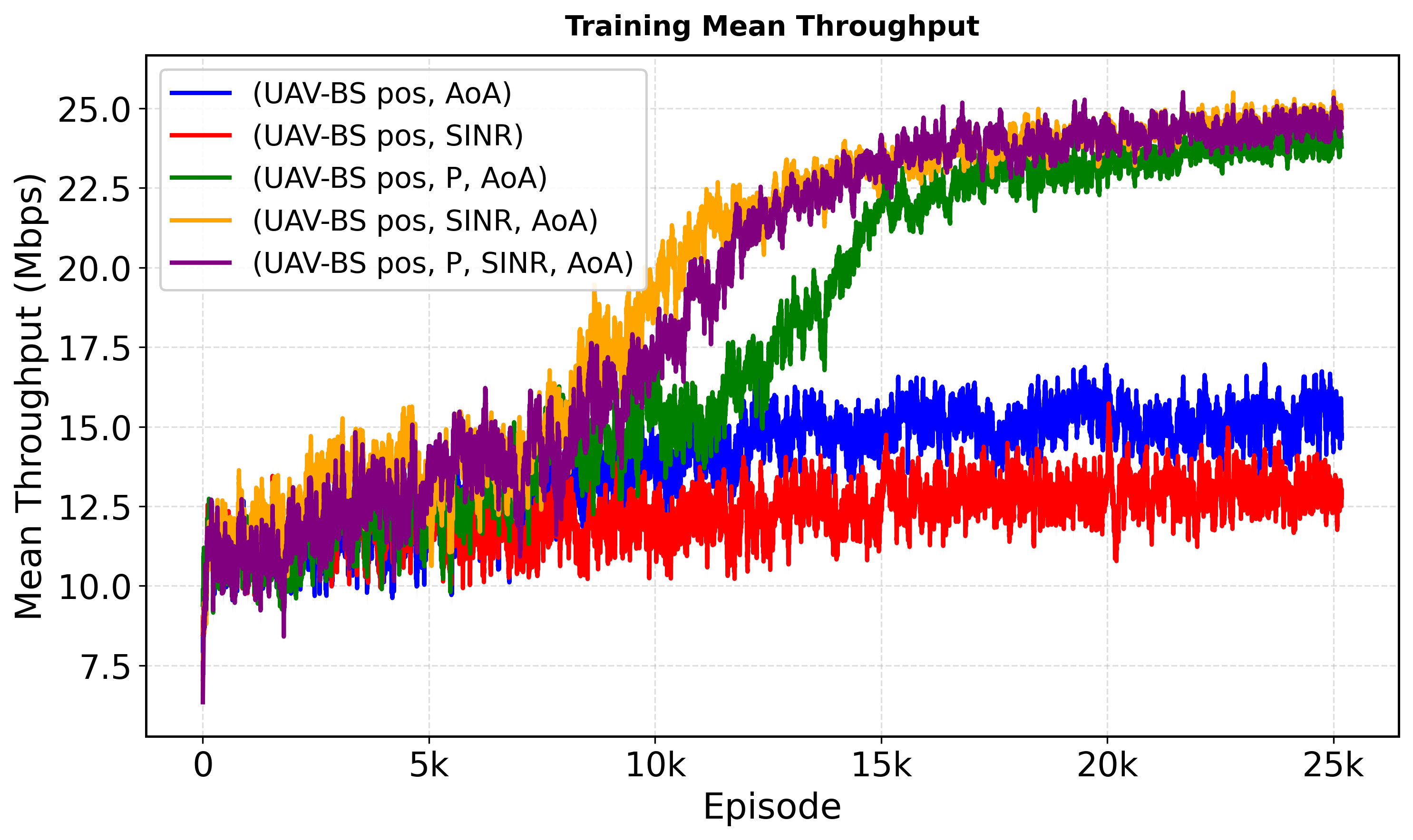}
\caption{Throughput performance during training}
\label{fig1:obs_mean_throughput}
\end{subfigure}

\caption{Training performance comparison for different state space designs under random initialization.}
\label{fig:obs_train}

\vspace{3mm}

\captionof{table}{Steady-state performance comparison of different state space designs.}
\label{tab:obs_state_space}

\footnotesize
\setlength{\tabcolsep}{6pt}
\renewcommand{\arraystretch}{1.2}

\begin{tabular}{|l|c|c|c|}
\hline
\multicolumn{1}{|c|}{\textbf{State Space}} &
\multicolumn{1}{c|}{\textbf{Mean Reward}} &
\multicolumn{1}{c|}{\textbf{Mean Throughput (Mbps)}} &
\multicolumn{1}{c|}{\textbf{Assessment}} \\
\hline
UAV-BS pos + AoA & $\approx 0.25\text{-}0.35$ & $\approx 14\text{-}17$ & Poor \\
\hline
UAV-BS pos + SINR & $\approx 0.10\text{-}0.25$ & $\approx 10\text{-}15$ & Worst \\
\hline
UAV-BS pos + P + AoA & $\approx 0.65\text{-}0.75$ & $\approx 22\text{-}24$ & Moderate \\
\hline
\textbf{UAV-BS pos + SINR + AoA} &
\boldsymbol{$\approx 0.75\text{-}0.80$} &
\boldsymbol{$\approx 24\text{-}26$} &
\textbf{Strong} \\
\hline
\textbf{UAV-BS pos + P + SINR + AoA} &
\boldsymbol{$\approx 0.80\text{-}0.85$} &
\boldsymbol{$\approx 25\text{-}27$} &
\textbf{Strong} \\
\hline
\end{tabular}

\end{figure*}

Ablation studies are conducted in the simulation section, particularly in the reward-function and state-space refinement subsections, to evaluate the impact of key design choices on the overall system performance.
Specifically, individual reward-function components and state-space features are modified, removed, or added to analyze their influence on the learning stability and network throughput.
The resulting performance variations are then compared to identify the contribution and significance of each design element.
The following simulation subsections provide a detailed discussion of the reward-design parameters and state-space configurations that most strongly affect the proposed system performance (See table.~\ref{tab:obs_state_space}).

\begin{figure*}[!htb]
\centering
\textbf{\small Throughput Performance of Reward Functions Under Various Initial UAV-BS Positions}\\[6pt]

\begin{subfigure}[t]{0.48\textwidth}
\centering
\includegraphics[width=\linewidth,height=5cm]{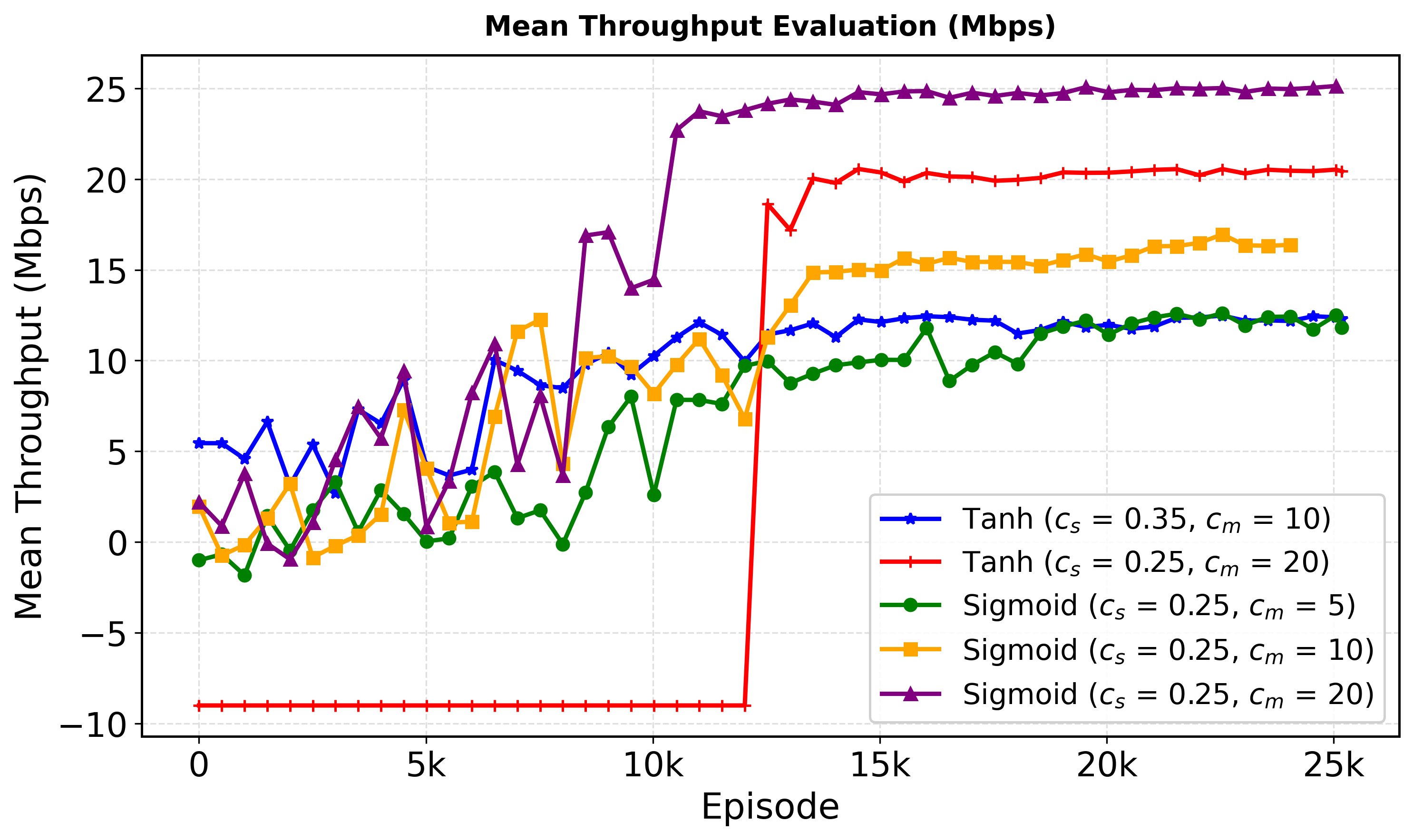}
\caption{Evaluation for UAV-BS Initial Deployment (Scenario~a)}
\label{RF_EVAL1}
\end{subfigure}%
\hfill%
\begin{subfigure}[t]{0.48\textwidth}
\centering
\includegraphics[width=\linewidth,height=5cm]{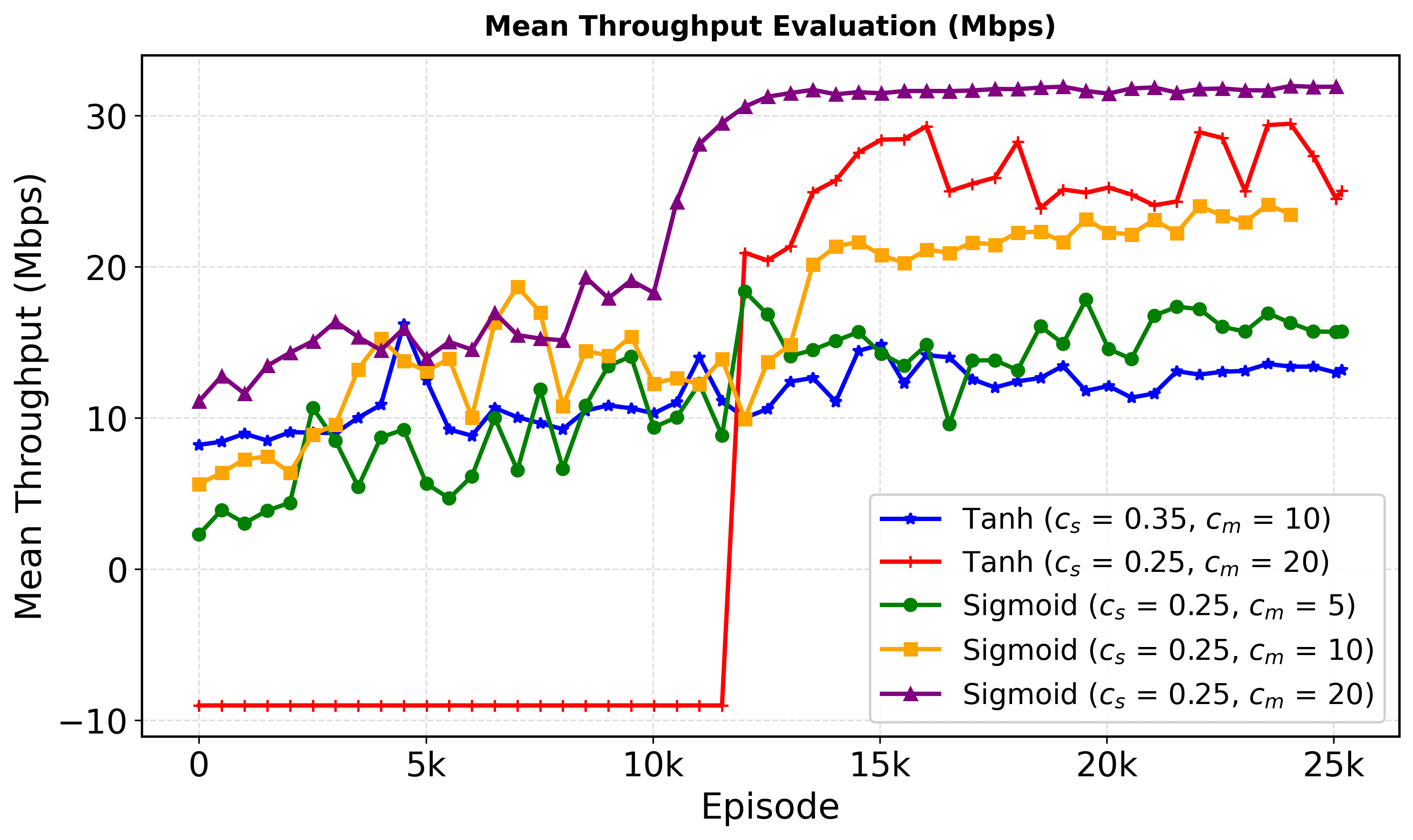}
\caption{Evaluation for UAV-BS Initial Deployment (Scenario~b)}
\label{RF_EVAL2}
\end{subfigure}

\vspace{8pt}

\begin{subfigure}[t]{0.48\textwidth}
\centering
\includegraphics[width=\linewidth,height=5cm]{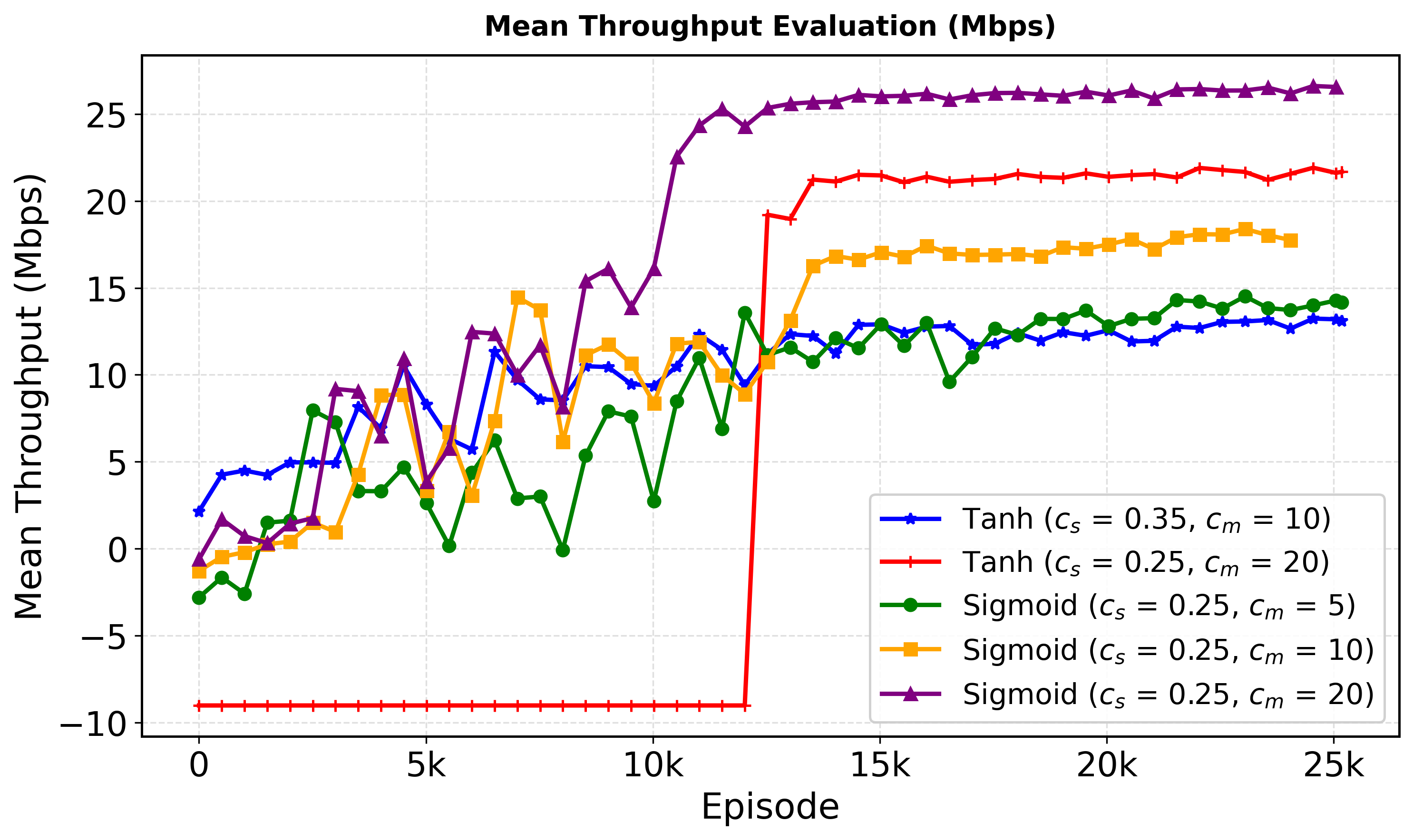}
\caption{Evaluation for UAV-BS Initial Deployment (Scenario~c)}
\label{RF_EVAL3}
\end{subfigure}%
\hfill%
\begin{subfigure}[t]{0.48\textwidth}
\centering
\includegraphics[width=\linewidth,height=5cm]{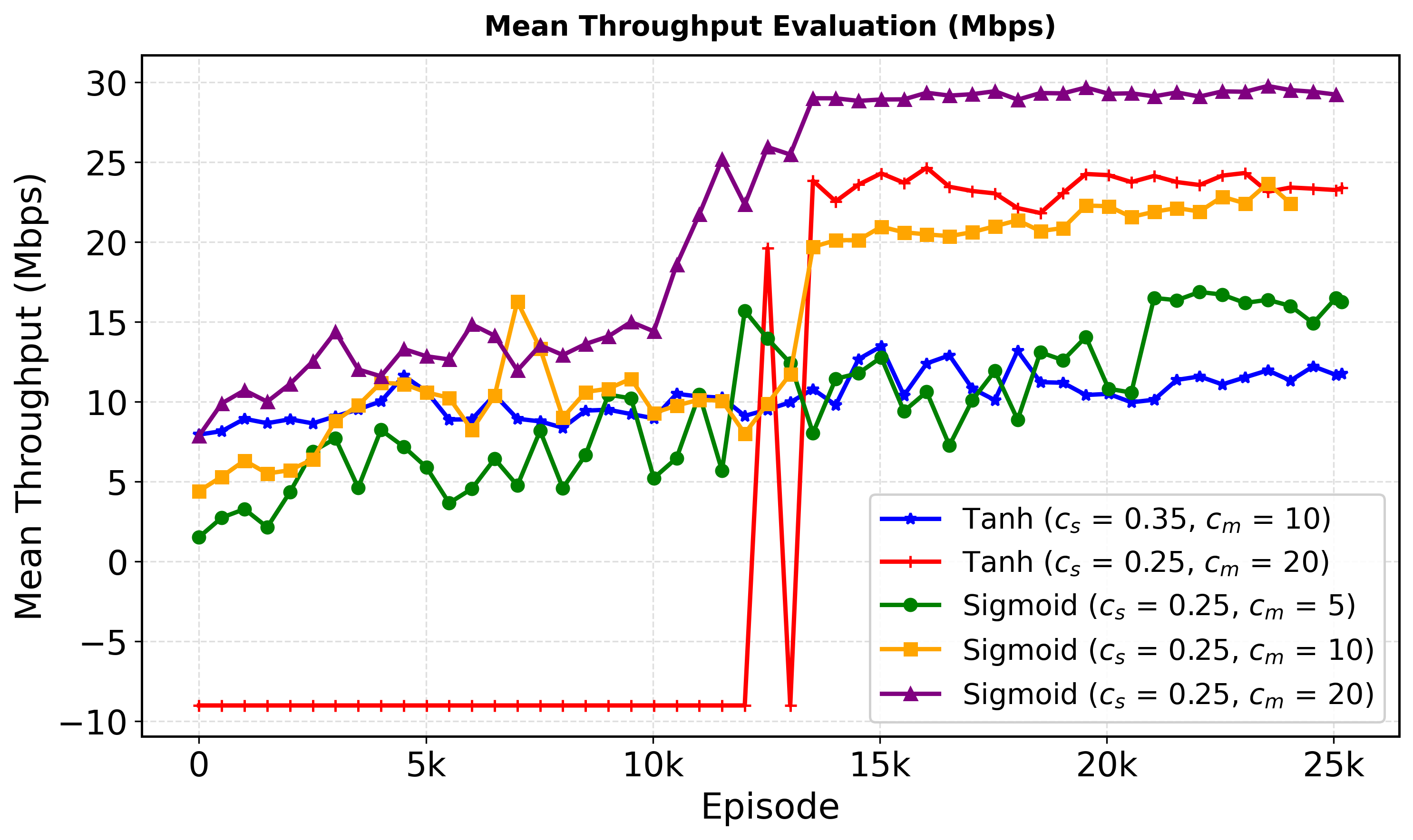}
\caption{Evaluation for UAV-BS Initial Deployment (Scenario~d)}
\label{RF_EVAL4}
\end{subfigure}

\caption{Performance comparison of reward functions across four deployment scenarios. The sigmoid reward with $(c_s = 0.25, c_m = 20)$ achieves the fastest convergence and highest throughput, while tanh-based rewards remain slower and less stable.}
\label{fig:reward_comparison_eval}
\end{figure*}

\section{Simulation Numerical Results and Discussion}
This section describes the proposed approach enhancement and simulation setup, as well as the numerical findings and discussions.
The default scenario in the simulations assumes a linear motion for the UE mobility model. Static, circular, and cosine-pattern motions are also used to evaluate the PPO algorithm's generalization ability (see Table~\ref{tab:simulation_ppo_parameters}).

\subsection{Reward Design Configuration}\label{Reward_sim_sigmoid}
As we explained in subsection~\ref{reward_definition}, the reward is derived from the fairness-aware sum throughput $R_{\mathrm{fair}}$ defined in eq.~ \eqref{eq_intro:fair_rate}. In our setup, $D=3$ UAV-BSs serve $H=3$ HSs under a one-to-one association, assuming full observability.
To ensure numerical stability, the reward signal is first normalized using min-max normalization. 
The normalized value is then mapped to the final reward through nonlinear activation functions defined in eq.~\eqref{eq:sigmoid} and eq.~\eqref{eq:tanh}. 
The sigmoid function produces reward values in the range $(0,1)$, whereas the hyperbolic tangent maps the reward to the interval $(-1,1)$.
In selecting values of $\{c_s,c_m\}$, we consider the maximum and average overall network throughput, obtained from our model setup, which are 35 Mbps and 18 Mbps, respectively. To obtain an efficient reward function, we consider a range of values for the reward-shaping parameters, with $c_s \in [0.15, 0.4]$ and $c_m \in [5, 30]$. 

\subsubsection{Simulation Results}

From the parameter ranges $c_s \in [0.15, 0.4]$ and $c_m \in [5, 30]$, the representative subset $c_s \in \{0.25, 0.35\}$ and $c_m \in \{5, 10, 20\}$ is selected for evaluation under the linear-motion setting. Training and evaluation results are presented in Fig.~\ref{fig:reward_comparison_train} and Fig.~\ref{fig:reward_comparison_eval}, respectively.

Fig.~\ref{fig:reward_comparison_train} shows that sigmoid-based rewards consistently achieve higher and more stable rewards than tanh-based designs. This behavior results from smoother gradient transitions that reduce saturation effects and improve policy-update stability. In particular, the sigmoid configuration $(c_s = 0.25, c_m = 20)$ achieves the most stable convergence, reaching final rewards around $0.85$-$0.90$ with low variance. In contrast, tanh-based rewards, especially $(c_s = 0.25, c_m = 20)$, exhibit delayed convergence, oscillatory behavior, and prolonged negative-reward phases.
A similar trend is observed in throughput performance. As shown in Fig.~\ref{fig:reward_comparison_train}(b), the sigmoid reward with $(c_s = 0.25, c_m = 20)$ achieves the highest and most stable throughput, converging near $24$-$25$~Mbps. The sigmoid configuration $(c_s = 0.25, c_m = 10)$ provides moderate performance around $17$~Mbps, whereas tanh-based rewards generally stabilize at lower throughput levels between approximately $12$ and $20$~Mbps while exhibiting slower and less stable convergence.

The evaluation results in Fig.~\ref{fig:reward_comparison_eval} further confirm these observations across all UAV-BS deployment scenarios. The sigmoid reward with $(c_s = 0.25, c_m = 20)$ consistently achieves the fastest convergence and highest final throughput, ranging from approximately $25$ to $32$~Mbps depending on the scenario. The sigmoid configuration $(c_s = 0.25, c_m = 10)$ and the tanh configuration $(c_s = 0.25, c_m = 20)$ provide intermediate performance, while tanh $(0.35,10)$ and sigmoid $(0.25,5)$ exhibit the slowest convergence and lowest throughput, generally remaining below approximately $15$-$17$~Mbps.
Although the absolute throughput varies across deployment scenarios, the relative ranking of the reward functions remains consistent. Increasing the margin parameter $c_m$ improves both convergence speed and throughput for sigmoid-based rewards, whereas tanh-based rewards remain more sensitive to parameter scaling and less stable during training.
Overall, the results demonstrate that properly scaled sigmoid reward functions provide faster convergence, improved stability, and consistently higher throughput during both training and evaluation, highlighting the importance of smooth reward shaping for stable PPO-based UAV-BS trajectory optimization.

\begin{figure*}[!htb]
\centering
\textbf{\small Throughput Performance Under Different State Representations for Various Initial UAV-BS Positions}\\[6pt]

\begin{subfigure}[t]{0.48\textwidth}
\centering
\includegraphics[width=\linewidth,height=5cm]{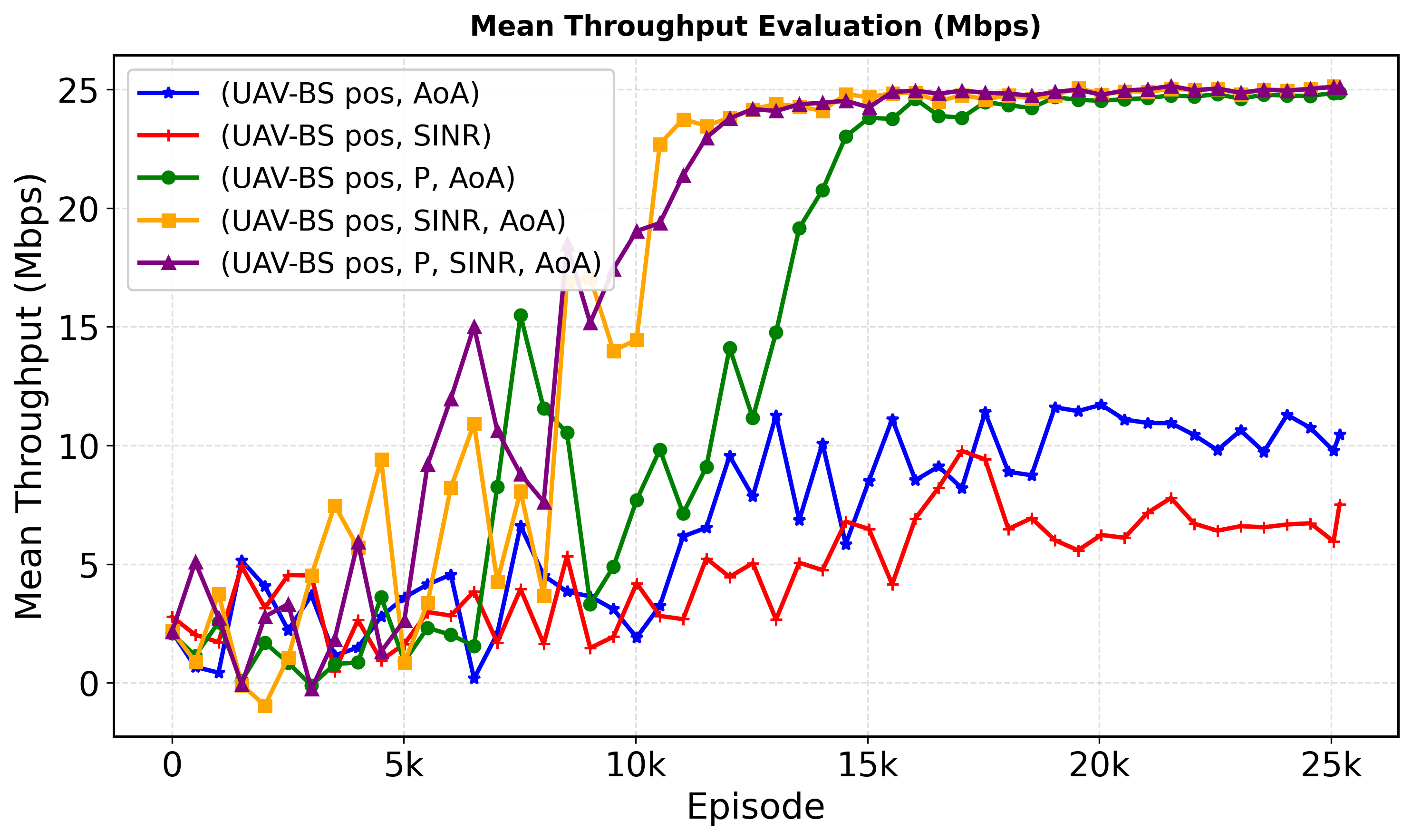}
\caption{Evaluation for UAV-BS Initial Deployment (Scenario~a)}
\label{obs_eval1}
\end{subfigure}%
\hfill%
\begin{subfigure}[t]{0.48\textwidth}
\centering
\includegraphics[width=\linewidth,height=5cm]{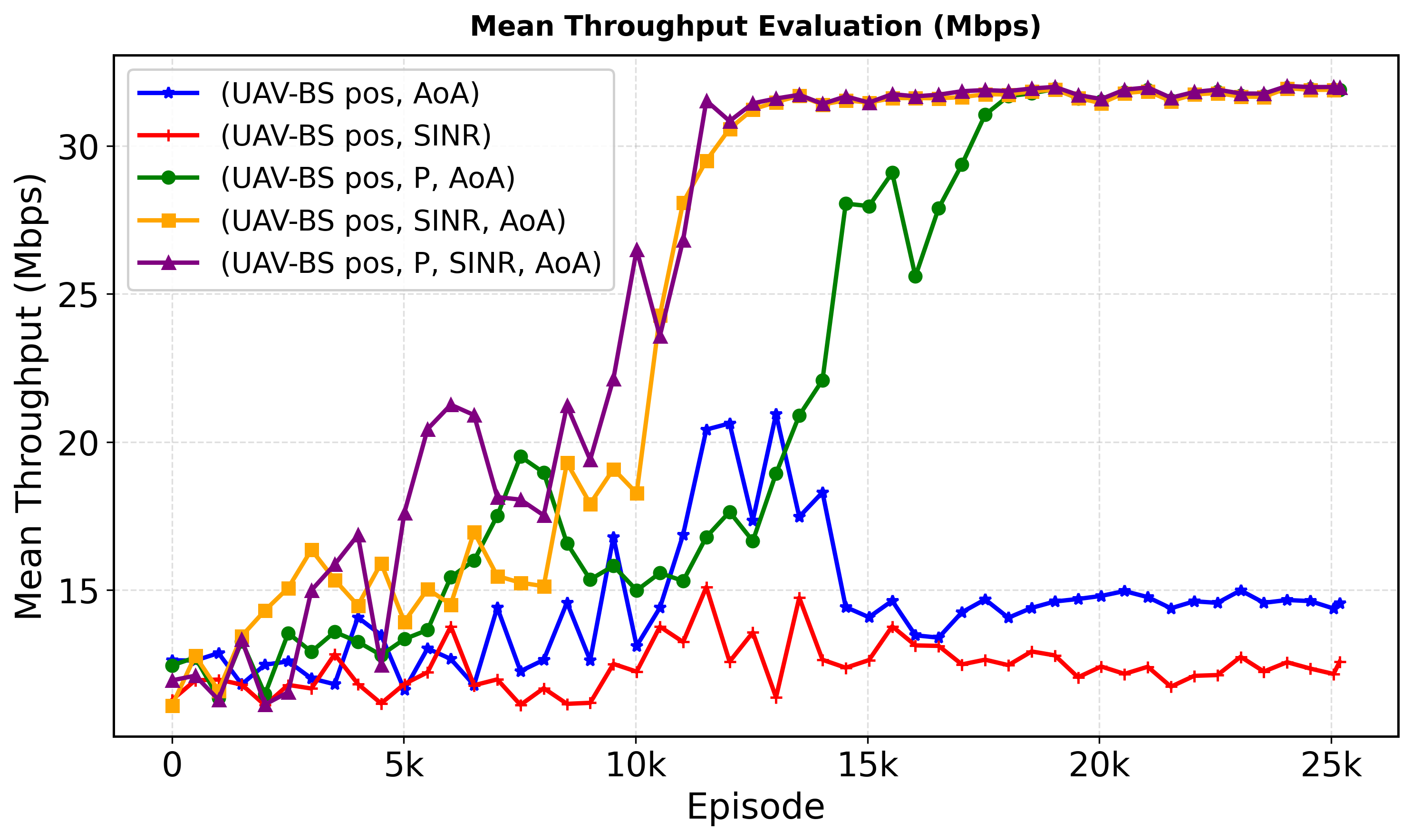}
\caption{Evaluation for UAV-BS Initial Deployment (Scenario~b)}
\label{obs_eval2}
\end{subfigure}

\vspace{8pt}

\begin{subfigure}[t]{0.48\textwidth}
\centering
\includegraphics[width=\linewidth,height=5cm]{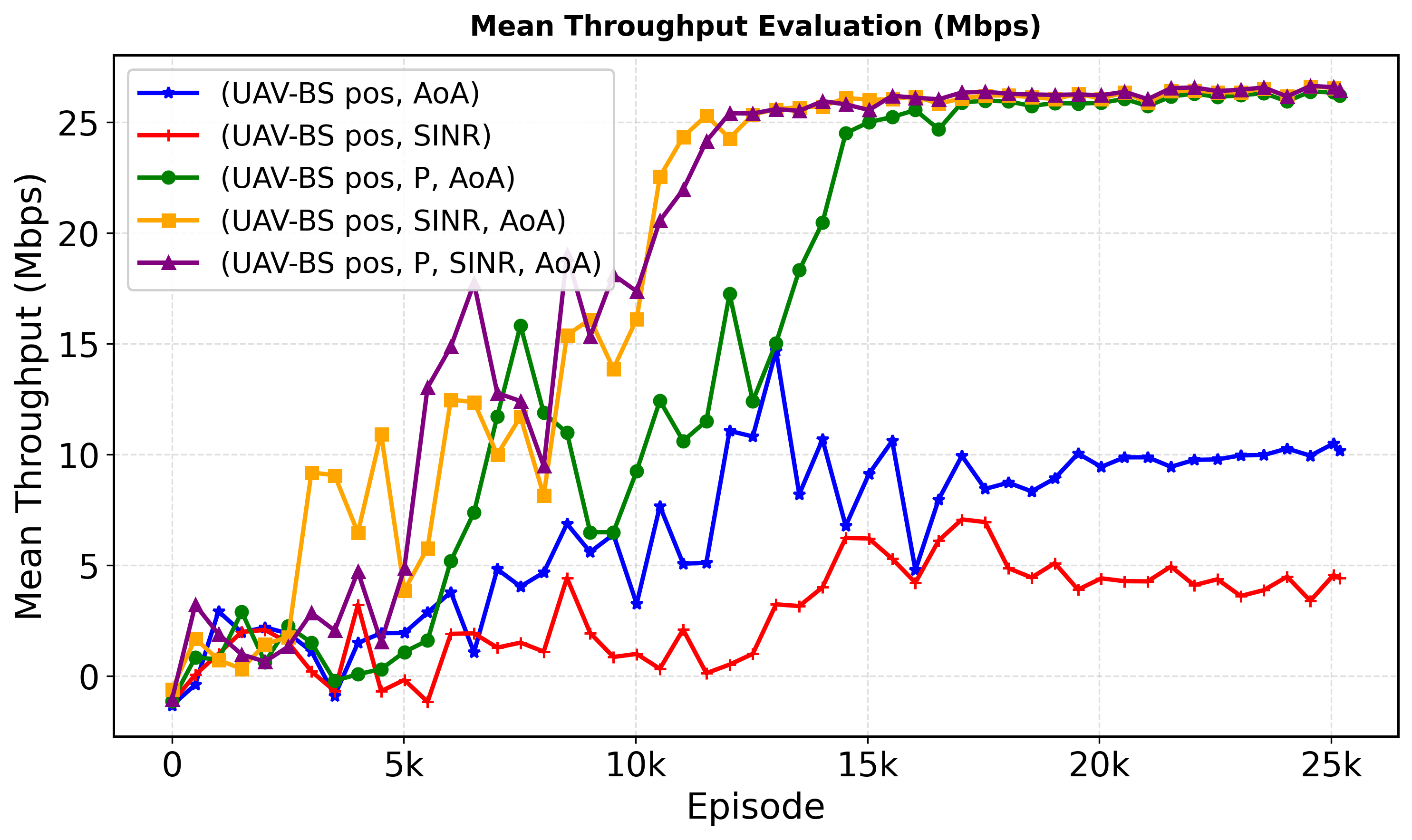}
\caption{Evaluation for UAV-BS Initial Deployment (Scenario~c)}
\label{obs_eval3}
\end{subfigure}%
\hfill%
\begin{subfigure}[t]{0.48\textwidth}
\centering
\includegraphics[width=\linewidth,height=5cm]{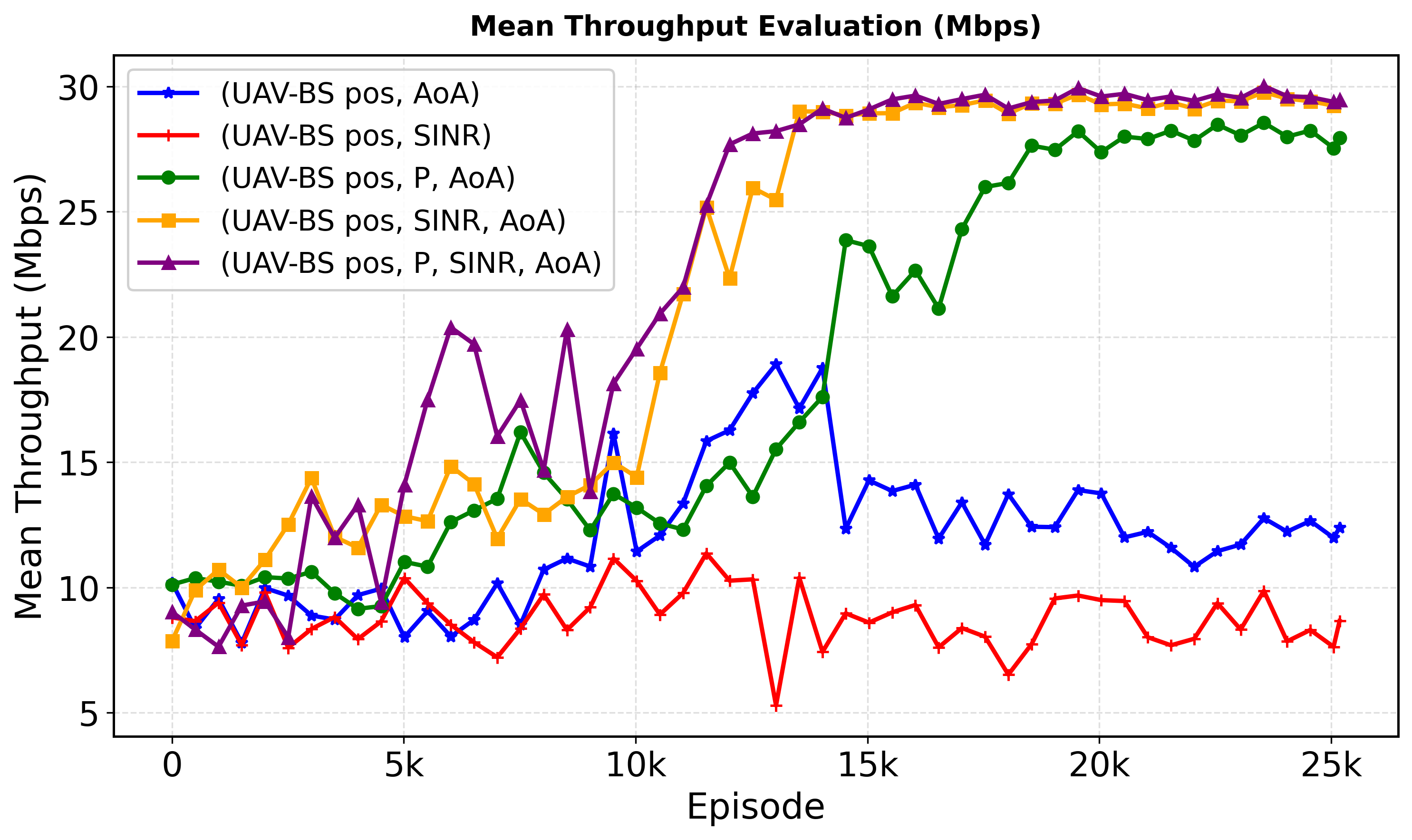}
\caption{Evaluation for UAV-BS Initial Deployment (Scenario~d)}
\label{obs_eval4}
\end{subfigure}

\caption{Performance comparison of state representations across four UAV-BS deployment scenarios. State spaces combining spatial, radio-level, and AoA information achieve faster convergence and higher throughput than reduced representations.}
\label{fig:obs_eval}
\end{figure*}

\subsection{State Space Configuration}
\label{sec:state_space}

Following the reward-function refinement and the selection of sigmoid parameters ($c_s = 0.25$, $c_m = 20$), this subsection defines and analyzes the state space used as input to the proposed RL neural network. 
As described in the previous subsection, $D=3$ UAV-BSs serve $H=3$ HSs under a one-to-one association, assuming full observability.
To evaluate the contribution of different features, we consider both the full state space and reduced variants obtained by excluding specific components. This enables a systematic assessment of the roles of spatial information, link-quality indicators, and AoA statistics in achieving stable learning and high-throughput performance.

\subsubsection{Simulation Results}
\label{observation_depth}

Following the reward-function refinement and the selection of sigmoid parameters ($c_s = 0.25$, $c_m = 20$), this subsection evaluates the impact of different state space designs on multi-UAV-BS coordination under the linear-motion model. Training and evaluation results are presented in Fig.~\ref{fig:obs_train}, Table~\ref{tab:obs_state_space}, and Fig.~\ref{fig:obs_eval}, respectively.

Fig.~\ref{fig:obs_train} shows that the choice of state representation strongly influences learning behavior from the early training stages. State spaces combining UAV-BS position, link-quality indicators, and AoA statistics achieve faster convergence and smoother learning dynamics, whereas reduced representations exhibit slower and noisier progression.
As training progresses, the performance gap becomes more pronounced. Richer state spaces incorporating both directional and radio-level information converge toward stable high-throughput regimes, typically reaching approximately $24$-$27$~Mbps. In contrast, reduced representations stabilize at substantially lower throughput levels, often below approximately $10$-$17$~Mbps. In particular, the configuration using only UAV-BS position and SINR consistently underperforms compared to the configuration combining UAV-BS position with AoA statistics, indicating the stronger contribution of directional information to UAV-BS trajectory adaptation.
The steady-state results summarized in Table~\ref{tab:obs_state_space} further confirm these observations. State spaces combining SINR and AoA statistics achieve the highest rewards and throughput, demonstrating the strongest overall learning effectiveness. Reduced state spaces relying only on AoA or SINR information exhibit weaker convergence and lower throughput performance. Although including received power $P$ improves performance when combined with AoA statistics, the additional gain obtained by combining both $P$ and SINR remains relatively limited.

The evaluation results in Fig.~\ref{fig:obs_eval} show consistent behavior across all deployment scenarios. State spaces combining UAV-BS position with SINR and AoA statistics (orange and purple curves) achieve the fastest convergence and highest throughput, reaching approximately $25$-$33$~Mbps depending on the scenario. Similarly, the configuration including UAV-BS position, received power, and AoA statistics (green curve) also provides strong performance, particularly in Scenarios~b-d.
In contrast, reduced state spaces containing only UAV-BS position with AoA or SINR information (blue and red curves) exhibit slower convergence, larger oscillations, and lower throughput, generally remaining below approximately $10$-$15$~Mbps.

Overall, the results demonstrate that enriching the state space with directional and radio-level information significantly improves PPO convergence speed, throughput stability, and generalization capability. While the complete state space provides the most reliable performance, the reduced configuration combining UAV-BS position, SINR, and AoA statistics offers an effective tradeoff between state complexity and learning performance.

\begin{table*}[t]
\centering
\caption{Performance metrics for linear and circular training. Results are reported as mean $\pm$ standard deviation over 4 runs.}
\label{tab:combined_results}

\begin{subtable}{0.48\textwidth}
\centering
\caption{Linear motion}
\label{tab:combined_results_linear}
\begin{tabular}{c c c c}
\hline
Episode & Mean Reward & Mean Throughput (Mbps) & Reward Std. \\
\hline
1000  & 0.12 $\pm$ 0.02 & 10.5 $\pm$ 1.1 & 0.10 $\pm$ 0.01 \\
5000  & 0.23 $\pm$ 0.03 & 13.5 $\pm$ 1.2 & 0.15 $\pm$ 0.02 \\
10000 & 0.61 $\pm$ 0.04 & 22.0 $\pm$ 1.0 & 0.32 $\pm$ 0.02 \\
15000 & 0.73 $\pm$ 0.03 & 24.5 $\pm$ 0.8 & 0.33 $\pm$ 0.01 \\
20000 & 0.76 $\pm$ 0.02 & 25.0 $\pm$ 0.6 & 0.31 $\pm$ 0.01 \\
\hline
\end{tabular}
\end{subtable}
\hfill
\begin{subtable}{0.48\textwidth}
\centering
\caption{Circular motion}
\label{tab:combined_results_circular}
\begin{tabular}{c c c c}
\hline
Episode & Mean Reward & Mean Throughput (Mbps) & Reward Std. \\
\hline
1000  & 0.11 $\pm$ 0.02 & 10.2 $\pm$ 1.2 & 0.08 $\pm$ 0.01 \\
5000  & 0.25 $\pm$ 0.04 & 14.0 $\pm$ 1.3 & 0.16 $\pm$ 0.02 \\
10000 & 0.65 $\pm$ 0.05 & 22.5 $\pm$ 1.2 & 0.33 $\pm$ 0.02 \\
15000 & 0.75 $\pm$ 0.03 & 24.7 $\pm$ 0.8 & 0.32 $\pm$ 0.01 \\
20000 & 0.78 $\pm$ 0.02 & 25.0 $\pm$ 0.5 & 0.31 $\pm$ 0.01 \\
\hline
\end{tabular}
\end{subtable}

\end{table*}







\begin{figure*}[!t]
\centering
\textbf{\small PPO Training Performance and Stability Across Multiple Runs (Linear motion Scenario)}\\[6pt]

\begin{subfigure}{0.48\textwidth}
\includegraphics[width=\linewidth,height=5cm]{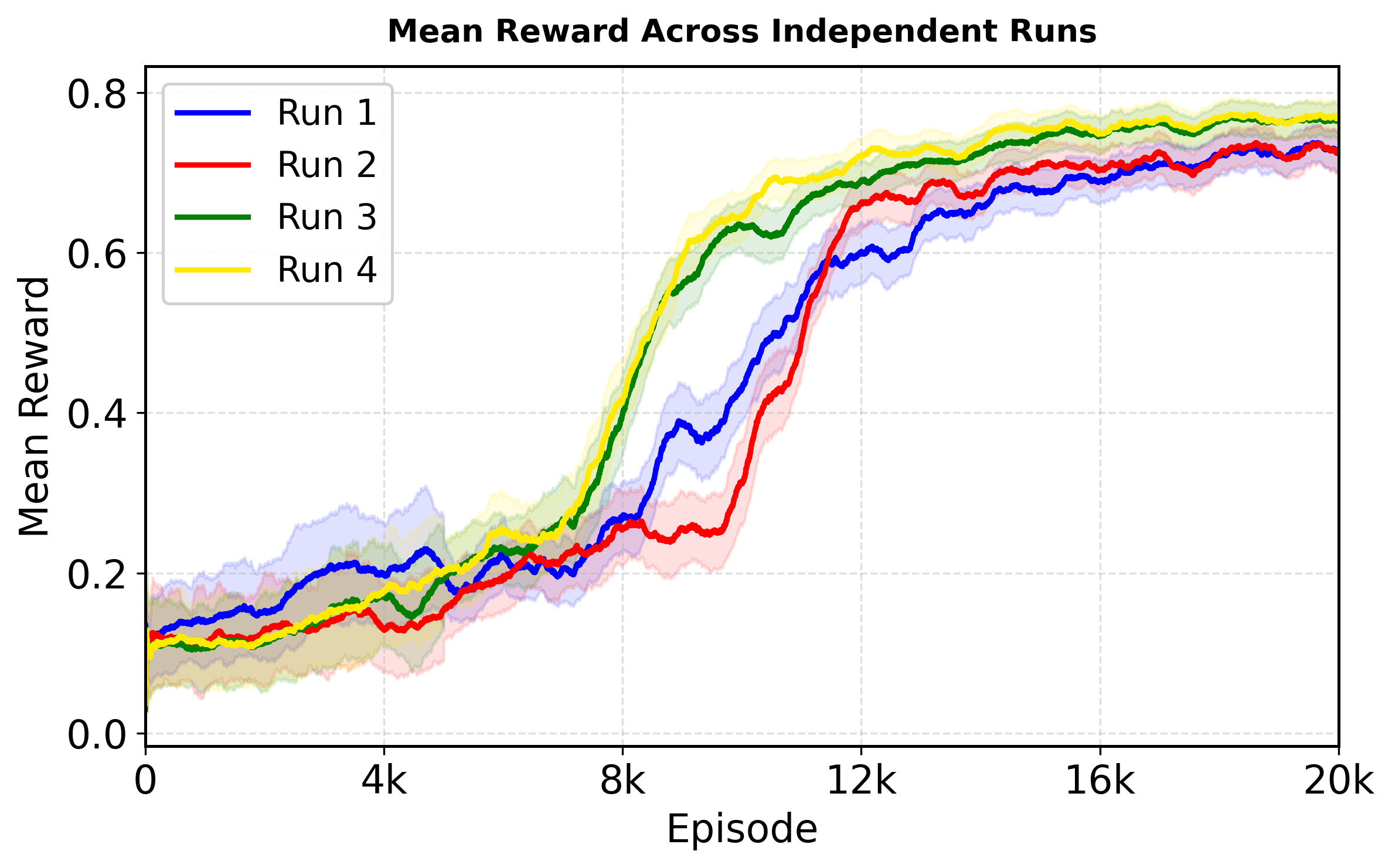}
\caption{Mean reward per episode (Linear motion, individual seeds)}
\end{subfigure}
\hfill
\begin{subfigure}{0.48\textwidth}
\includegraphics[width=\linewidth,height=5cm]{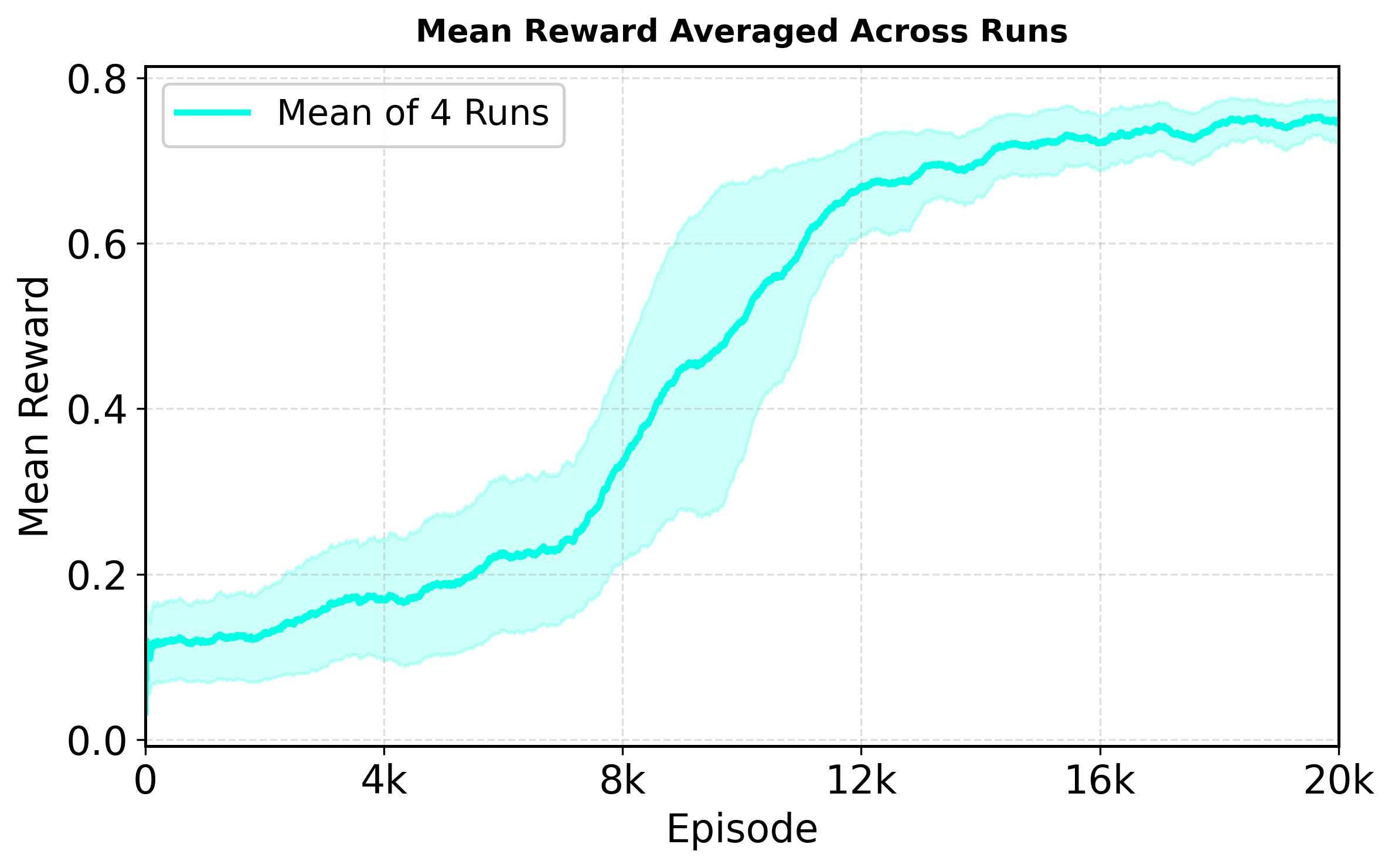}
\caption{Mean reward per episode (Linear motion, averaged across seeds)}
\end{subfigure}

\vspace{0.3cm}

\begin{subfigure}{0.48\textwidth}
\includegraphics[width=\linewidth,height=5cm]{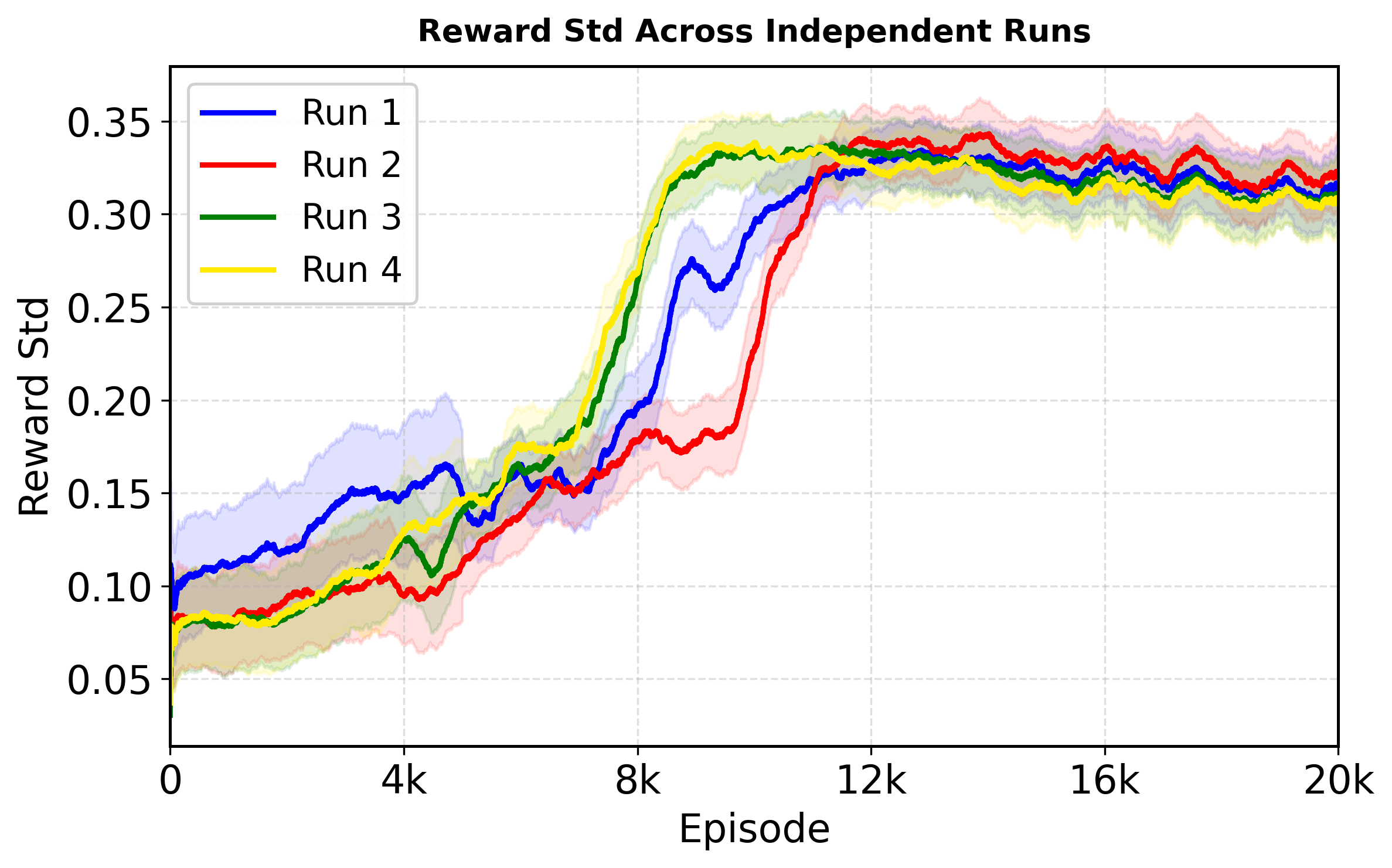}
\caption{Reward std per episode (Linear motion, individual seeds)}
\end{subfigure}
\hfill
\begin{subfigure}{0.48\textwidth}
\includegraphics[width=\linewidth,height=5cm]{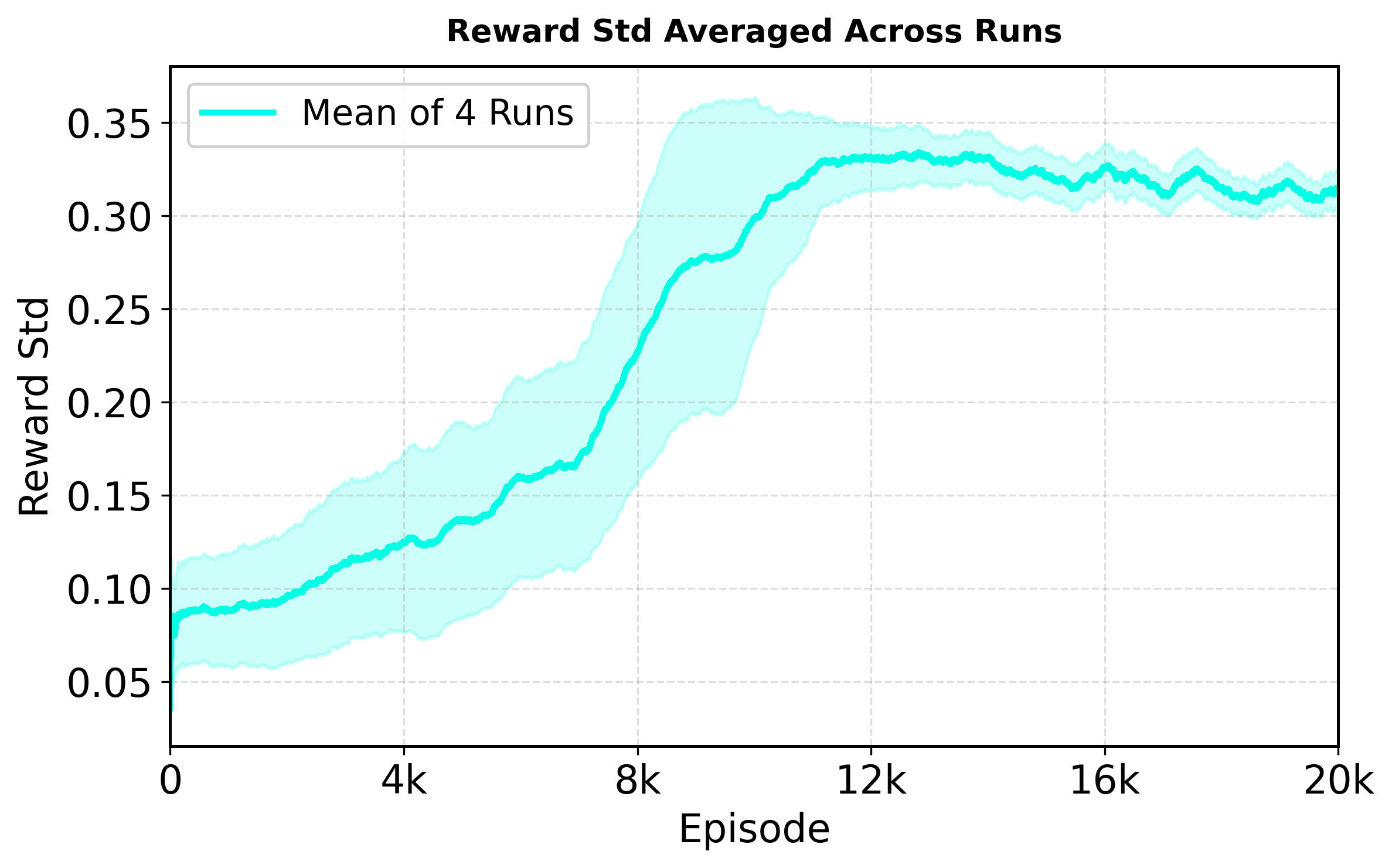}
\caption{Reward std per episode (Linear motion, averaged across seeds)}
\end{subfigure}

\vspace{0.3cm}

\begin{subfigure}{0.48\textwidth}
\includegraphics[width=\linewidth,height=5cm]{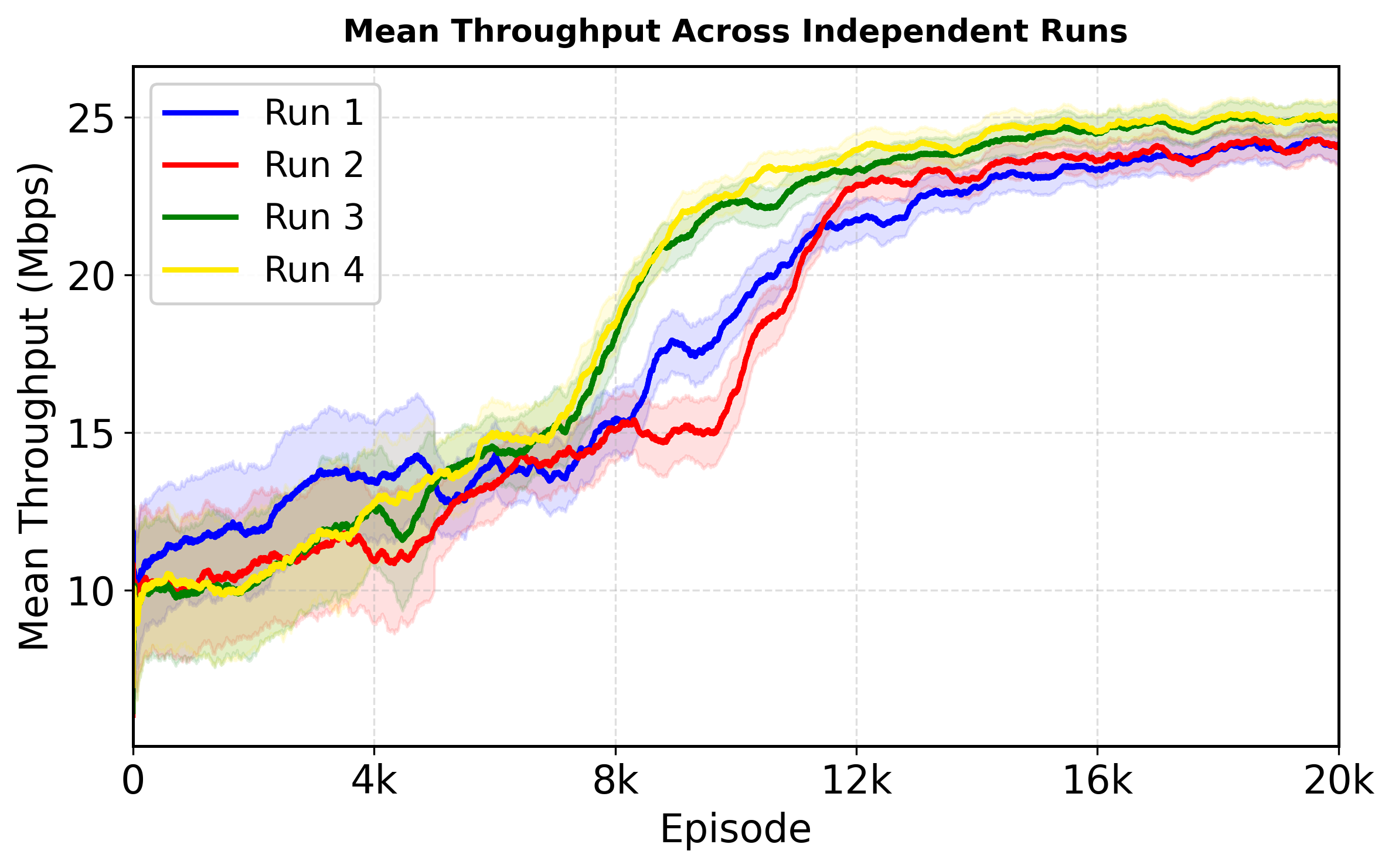}
\caption{Mean throughput per episode (Linear motion, individual seeds)}
\end{subfigure}
\hfill
\begin{subfigure}{0.48\textwidth}
\includegraphics[width=\linewidth,height=5cm]{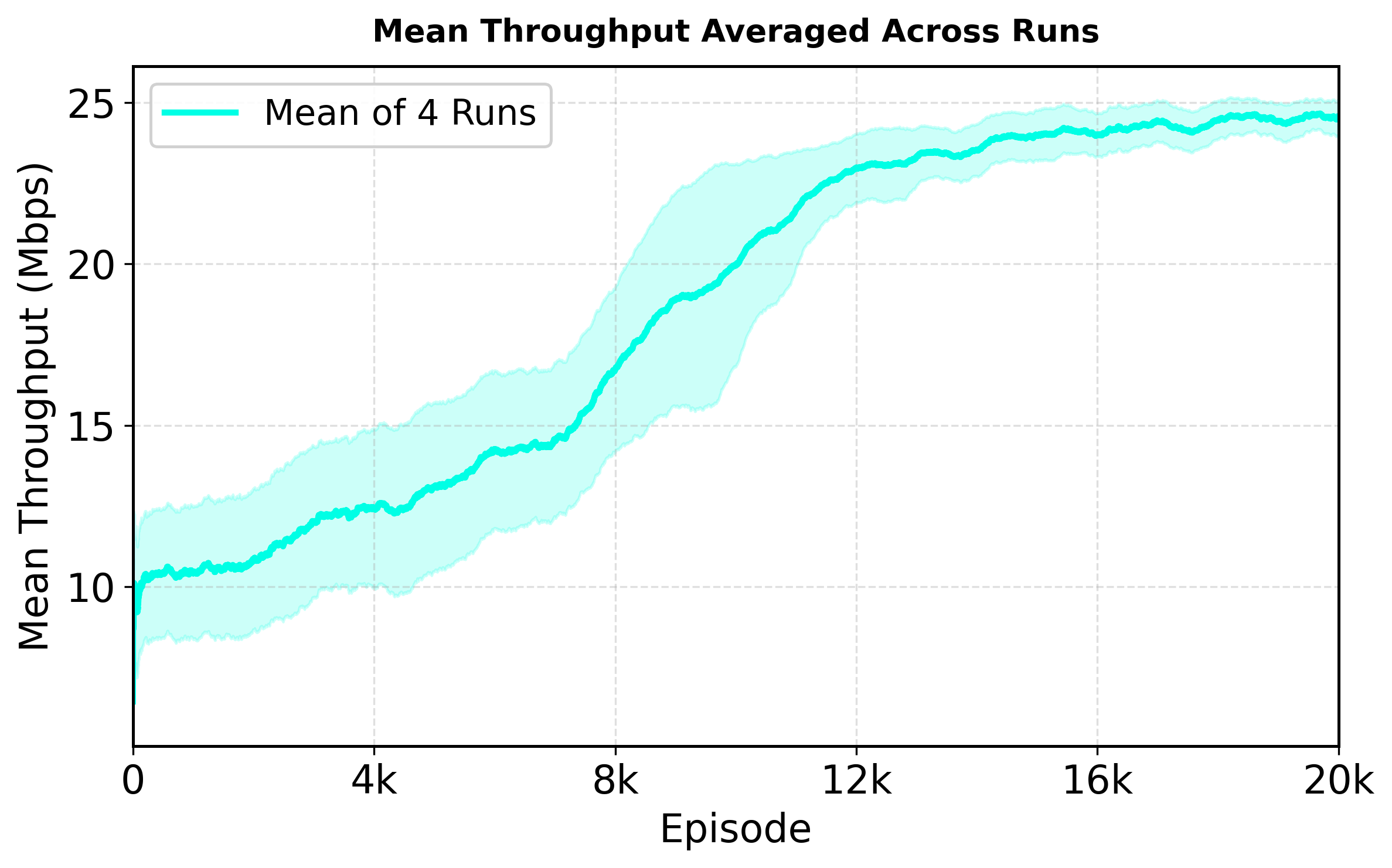}
\caption{Mean throughput (Linear motion, averaged across seeds)}
\end{subfigure}

\caption{Performance in the linear motion scenario. The left column shows results from independent runs with different random seeds, while the right column presents averages across seeds. Solid lines denote mean values per episode, and shaded regions indicate ±1 standard deviation (std). Averaging across seeds reduces stochastic variation and highlights overall learning trends.}
\label{fig:combined_results}
\end{figure*}

\begin{figure*}[!t]
\centering
\textbf{\small PPO Training Performance and Stability Across Multiple Runs (Circular motion Scenario)}\\[6pt]

\begin{subfigure}{0.48\textwidth}
\includegraphics[width=\linewidth,height=5cm]{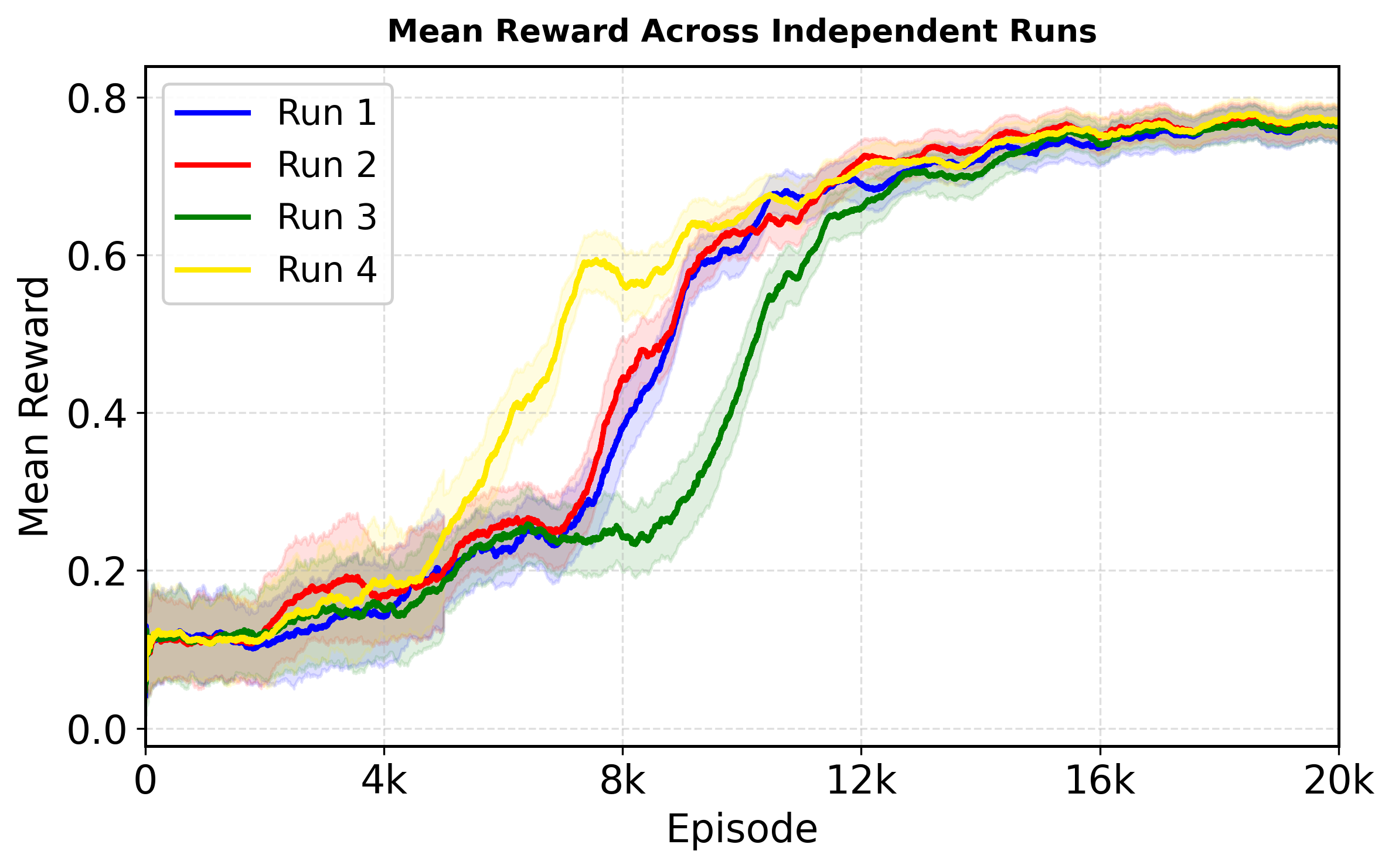}
\caption{Mean reward per episode (Circular motion, individual seeds)}
\end{subfigure}
\hfill
\begin{subfigure}{0.48\textwidth}
\includegraphics[width=\linewidth,height=5cm]{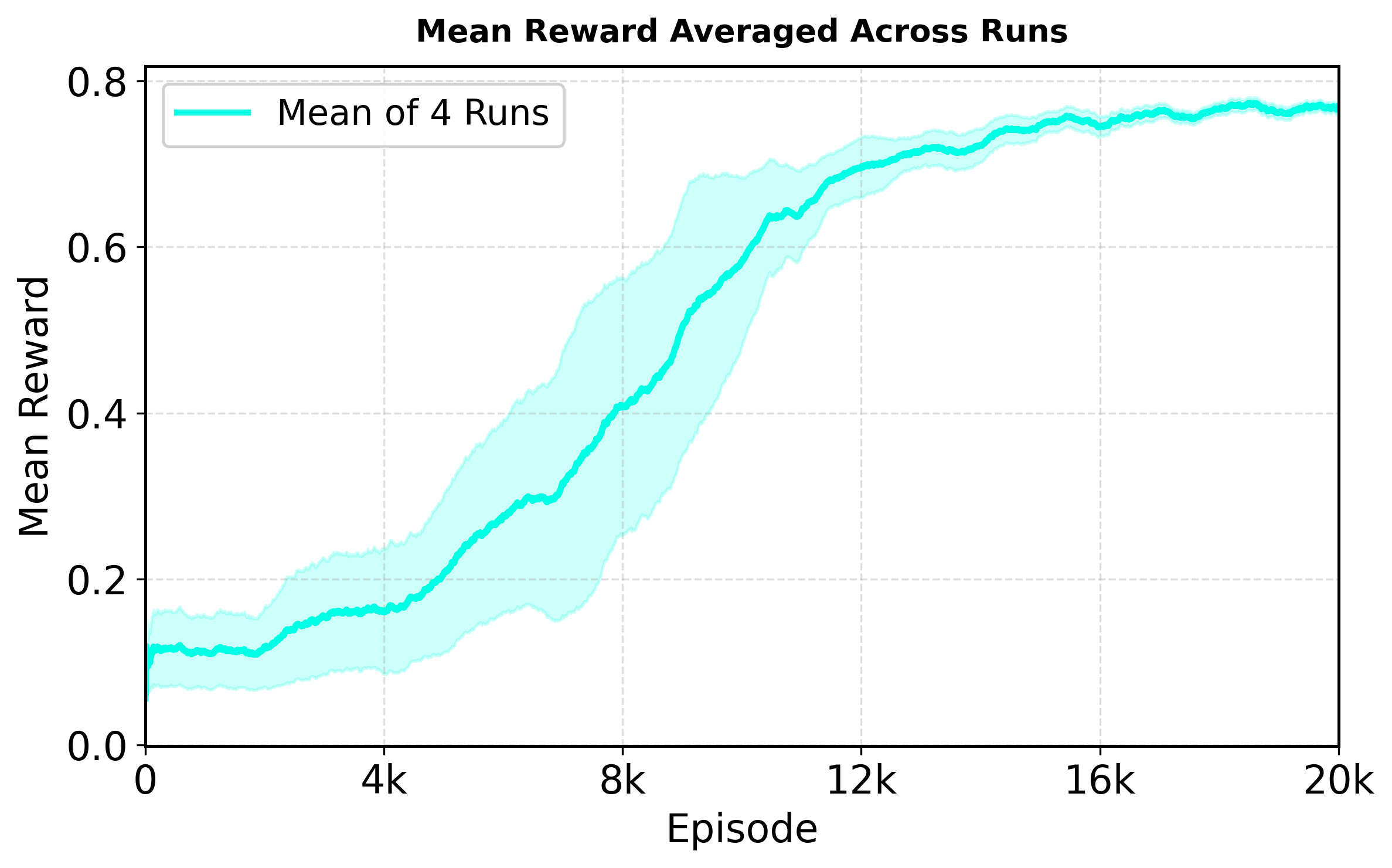}
\caption{Mean reward per episode (Circular motion, averaged across seeds)}
\end{subfigure}

\vspace{0.3cm}

\begin{subfigure}{0.48\textwidth}
\includegraphics[width=\linewidth,height=5cm]{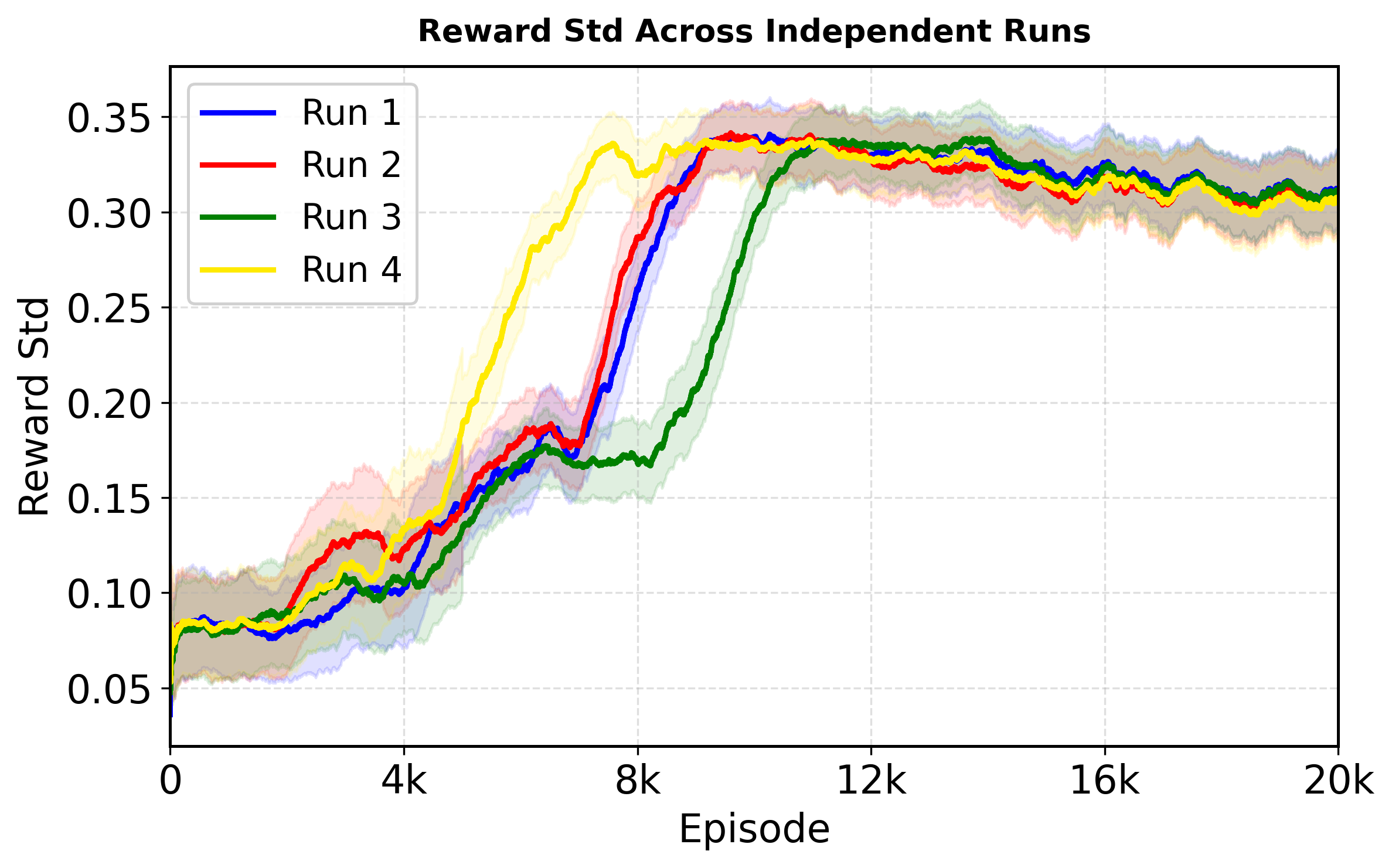}
\caption{Reward std per episode (Circular motion, individual seeds)}
\end{subfigure}
\hfill
\begin{subfigure}{0.48\textwidth}
\includegraphics[width=\linewidth,height=5cm]{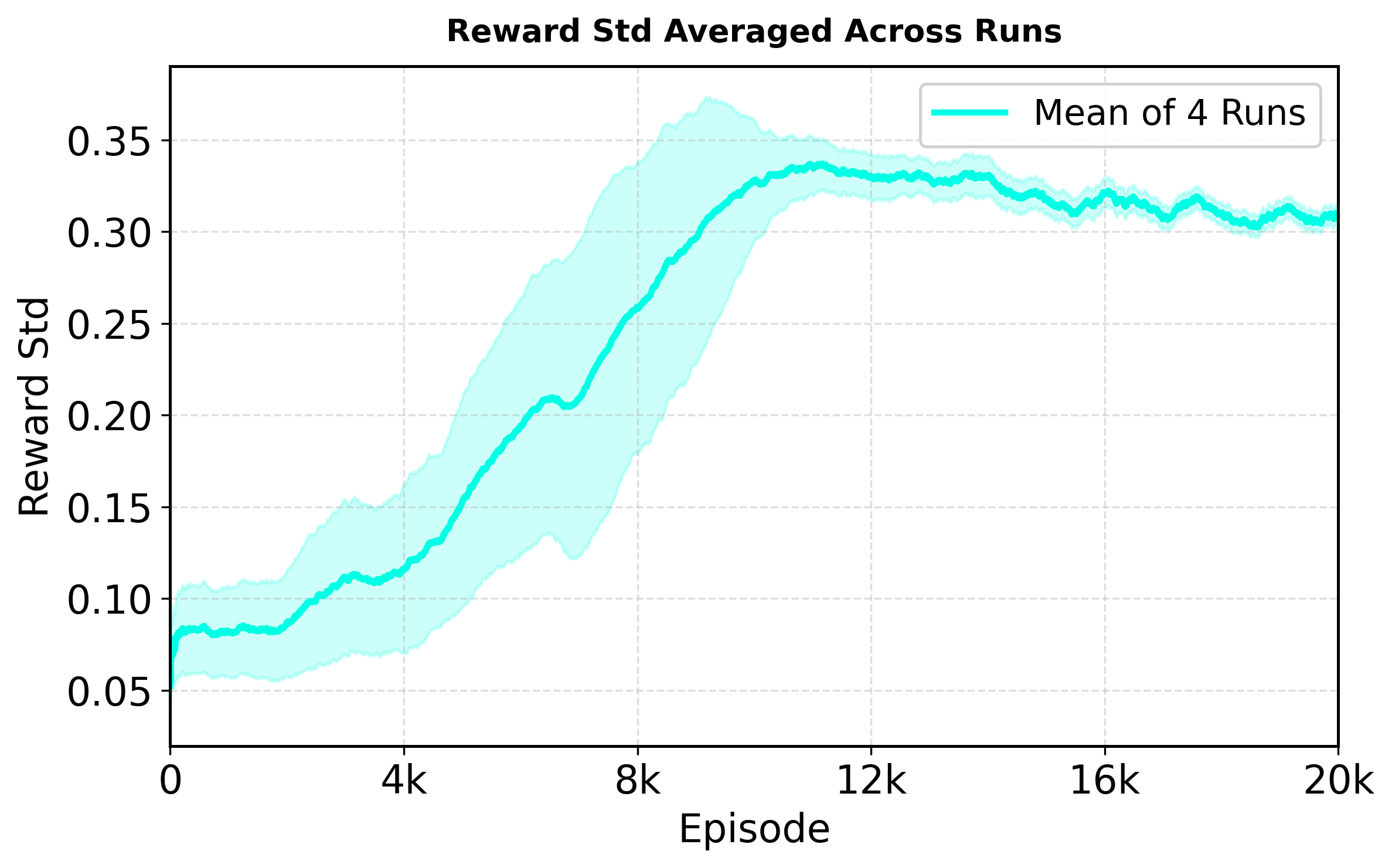}
\caption{Reward std per episode (Circular motion, averaged across seeds)}
\end{subfigure}

\vspace{0.3cm}

\begin{subfigure}{0.48\textwidth}
\includegraphics[width=\linewidth,height=5cm]{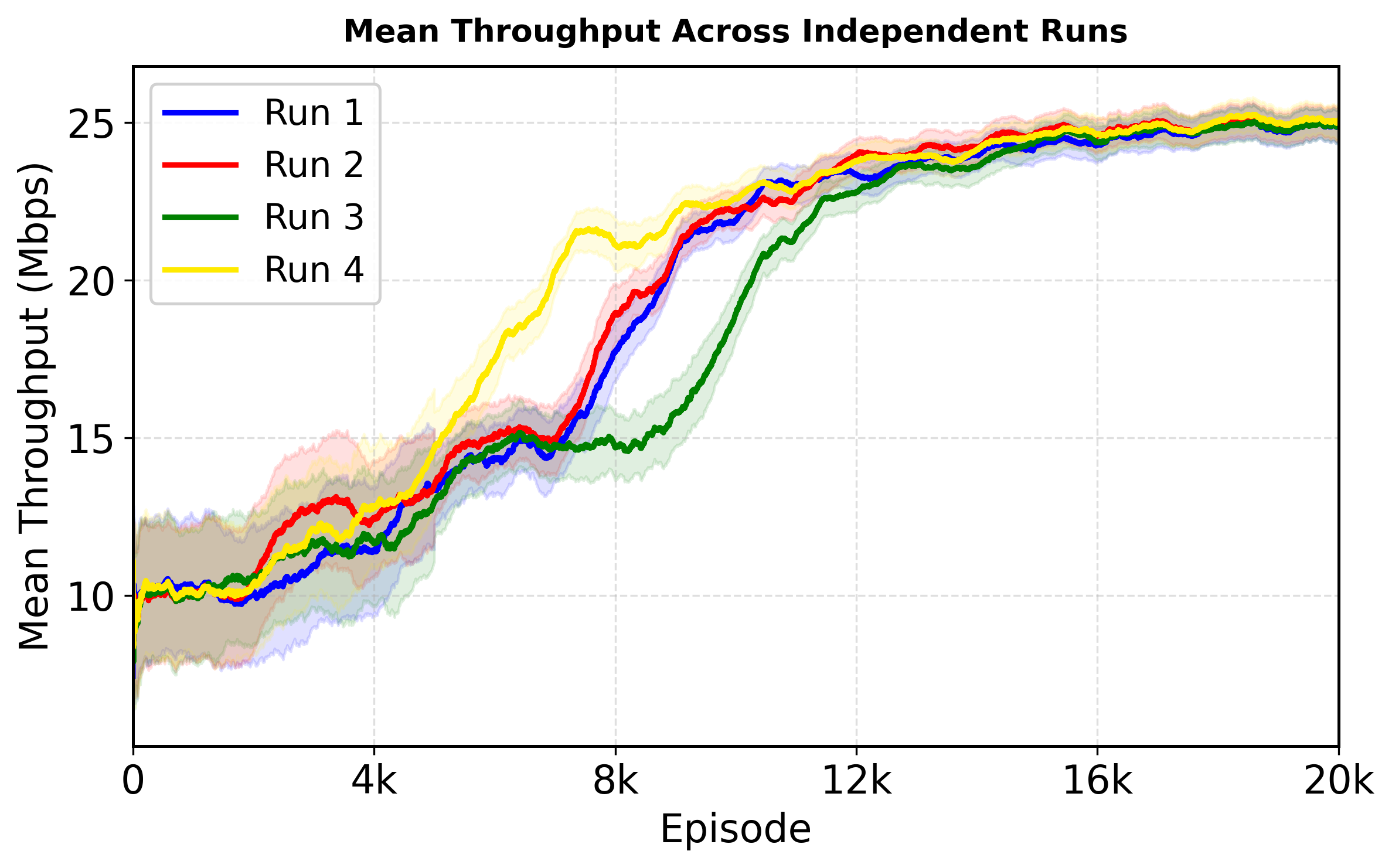}
\caption{Mean throughput per episode (Circular motion, individual seeds)}
\end{subfigure}
\hfill
\begin{subfigure}{0.48\textwidth}
\includegraphics[width=\linewidth,height=5cm]{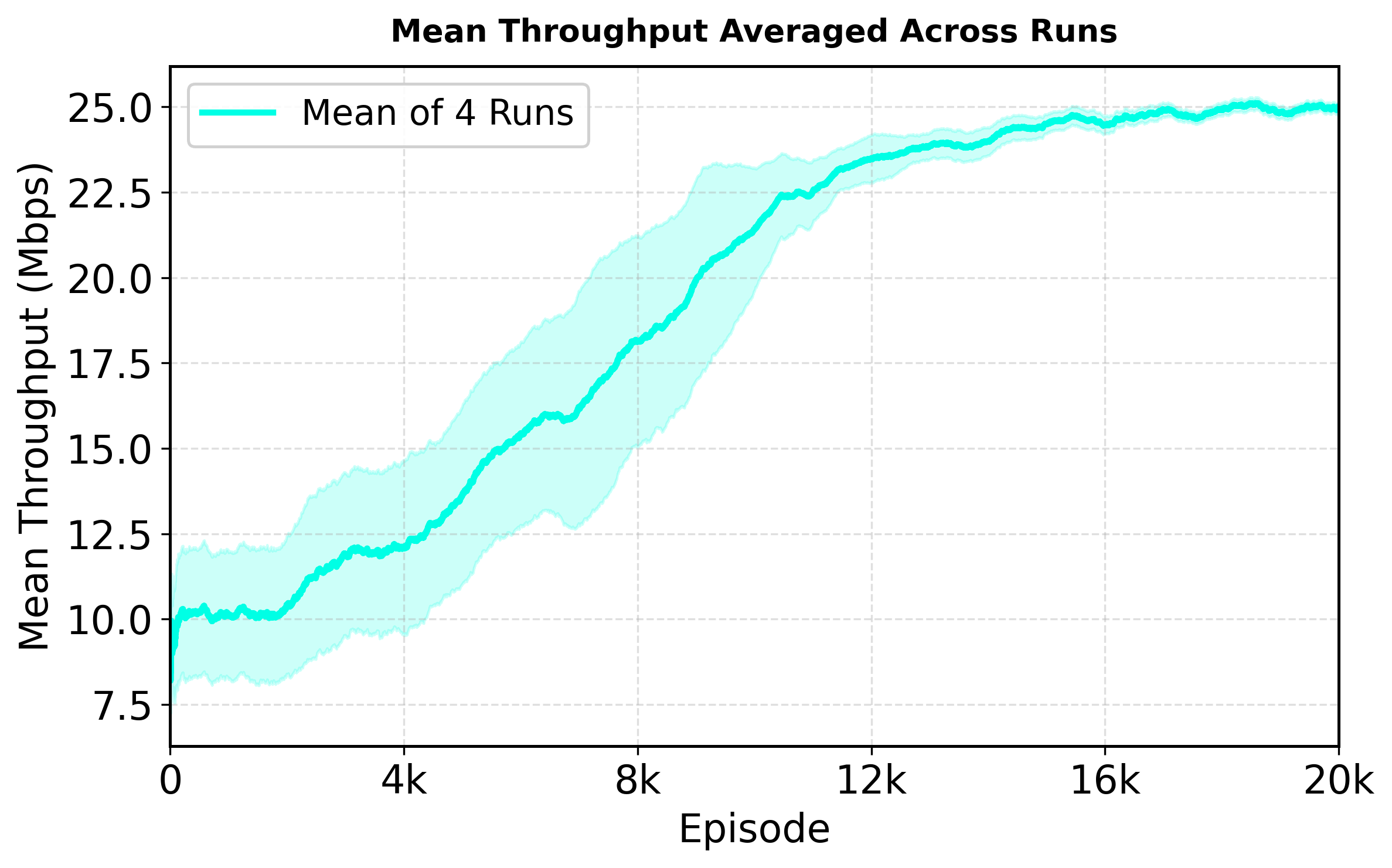}
\caption{Mean throughput (Circular motion, averaged across seeds)}
\end{subfigure}

\caption{Performance in the Circular motion scenario. The left column shows results from independent runs with different random seeds, while the right column presents averages across seeds. Solid lines denote mean values per episode, and shaded regions indicate ±1 standard deviation (std). Averaging across seeds reduces stochastic variation and highlights overall learning trends.}
\label{fig:combined_results_circular}
\end{figure*}
\subsection{Statistical Analysis of Results}

To provide a more rigorous evaluation, all experiments are conducted using four independent runs with different random seeds. Performance is reported as mean $\pm$ standard deviation, while confidence intervals and local temporal variability bands are used to illustrate training stability and variability.
The reported results are derived from metrics monitored throughout training and reflect the learning dynamics of the proposed PPO framework. Two mobility patterns are considered: \textit{Linear motion} and \textit{Circular motion}, representing both structured and dynamic UAV-BS movement conditions.
\subsubsection{Linear Motion}
Table~\ref{tab:combined_results_linear} summarizes the training performance under the linear-motion scenario. Both mean reward and throughput increase steadily throughout training, indicating stable policy improvement and consistent convergence behavior. Although reward variability increases during intermediate stages due to exploration, it gradually stabilizes as training progresses.
Fig.~\ref{fig:combined_results} illustrates the corresponding training dynamics. The individual-run plots show consistent upward trends in both throughput and reward across all runs, indicating robust learning behavior with limited sensitivity to random initialization. The averaged curves further confirm smooth convergence, while the narrowing confidence intervals in later stages indicate reduced variability after policy stabilization. The standard deviation curves show higher variability during the exploration phase, followed by progressive stabilization as convergence is achieved.
\subsubsection{Circular Motion}
Table~\ref{tab:combined_results_circular} summarizes the training performance under the circular-motion scenario. Similar to the linear-motion case, both reward and throughput improve consistently throughout training, demonstrating stable learning and effective policy optimization. Reward variability temporarily increases during the transition phase before stabilizing near convergence.
Fig.~\ref{fig:combined_results_circular} presents the corresponding training dynamics. All runs exhibit stable upward trends in reward and throughput, although the individual trajectories show moderately higher fluctuations during mid-training due to the increased complexity of circular mobility. The averaged plots nevertheless demonstrate stable convergence, with smooth mean trajectories and progressively narrower confidence intervals. The standard deviation curves further highlight this behavior, showing temporary variability growth during exploration followed by stabilization in later stages.
Overall, both mobility scenarios demonstrate stable and reproducible PPO training behavior across multiple random seeds. Mean reward and throughput consistently improve during training, while variability gradually decreases as the policies converge. Although circular motion introduces slightly higher fluctuations and slower stabilization due to its increased trajectory complexity, all runs converge to comparable final performance levels. The narrow confidence intervals observed in later training stages further confirm the robustness and reproducibility of the proposed approach. 
\subsection{Different RL approaches}



\begin{figure*}[t]
    \centering
\textbf{\small PPO vs. DQN and DDPG: Training Performance Comparison}\\[6pt]
    \begin{subfigure}[t]{0.48\linewidth}
        \centering
        \includegraphics[width=\linewidth, height=5.5cm]{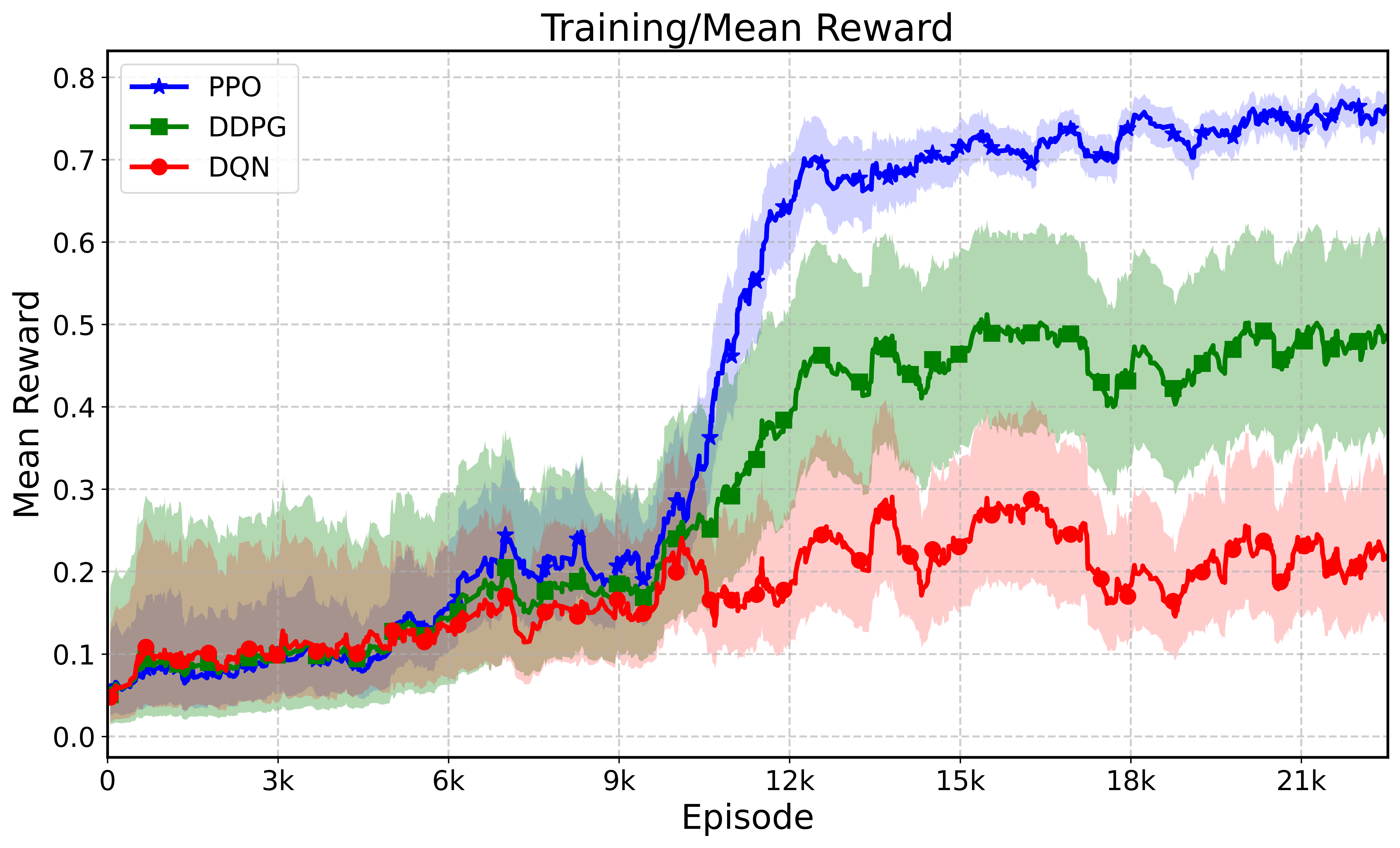}
        \caption{Reward comparison during training}
        \label{fig:DDPG_mean_reward}
    \end{subfigure}
    \hfill
    \begin{subfigure}[t]{0.48\linewidth}
        \centering
        \includegraphics[width=\linewidth, height=5.5cm]{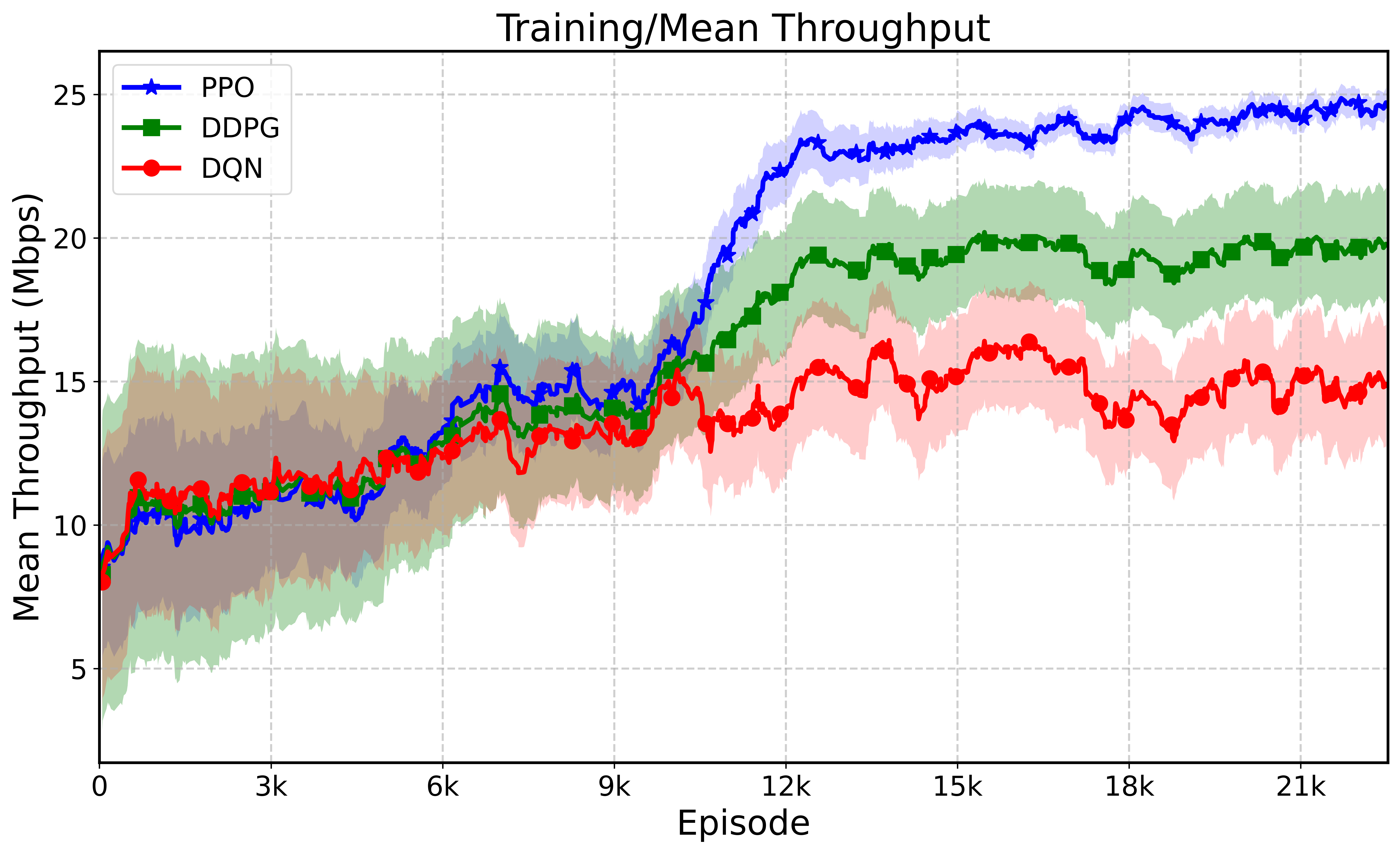}
        \caption{Throughput comparison during training}
        \label{fig:DDPG_mean_throughput}
    \end{subfigure}

    \caption{Training performance comparison of PPO, DQN, and DDPG. PPO shows faster convergence and superior performance in both reward (a) and throughput (b).}
    \label{fig:PPO_DQN_DDPG_training}
\end{figure*}

\begin{table*}[!htb]
\centering
\caption{Training configuration parameters for DQN and DDPG models}
\label{tab:rl_configs}

\begin{subtable}{0.48\textwidth}
\centering
\caption{DQN training configuration}
\label{tab:dqn_config}
\small
\setlength{\tabcolsep}{6pt}
\renewcommand{\arraystretch}{1.1}

\begin{tabular}{lc}
\hline
\textbf{Parameter} & \textbf{Value} \\
\hline
Num. episodes & 22{,}524 \\
Max steps / episode & 128 \\
Discount factor ($\xi$) & 0.99 \\
Learning rate & $3 \times 10^{-4}$ \\
Batch size & 256 \\
Replay buffer size & $1 \times 10^{6}$ \\
Warm-up steps & $1 \times 10^{4}$ \\
Target update interval & 2000 steps \\
Hidden layers & 3 \\
Hidden units per layer & 256 \\
Activation function & ReLU \\
Gradient clipping & 10.0 (L2 norm) \\
Exploration strategy & $\epsilon$-greedy \\
Initial $\epsilon$ & 1.0 \\
Final $\epsilon$ & 0.05 \\
$\epsilon$ decay schedule & Linear \\
$\epsilon$ decay steps & $3 \times 10^{5}$ \\
Evaluation policy & Deterministic (greedy) \\
\hline
\end{tabular}
\end{subtable}
\hfill
\begin{subtable}{0.48\textwidth}
\centering
\caption{DDPG training configuration}
\label{tab:ddpg_config}
\small
\setlength{\tabcolsep}{6pt}
\renewcommand{\arraystretch}{1.1}

\begin{tabular}{lc}
\hline
\textbf{Parameter} & \textbf{Value} \\
\hline
Num. episodes & 22{,}524 \\
Max steps / episode & 128 \\
Discount factor ($\xi$) & 0.99 \\
Actor learning rate & $3 \times 10^{-4}$ \\
Critic learning rate & $1 \times 10^{-4}$ \\
Batch size & 256 \\
Replay buffer size & $1 \times 10^{5}$ \\
Warm-up steps & 250 \\
Target update rate ($\tau$) & 0.005 \\
Hidden layers & 2 \\
Hidden units per layer & 256 \\
Activation function & ReLU \\
Gradient clipping & 1.0 (L2 norm) \\
Exploration strategy & Gaussian noise \\
Exploration noise std & 0.1 \\
Exploration noise clip & 0.5 \\
Reward scaling & 1.0 \\
Evaluation policy & Deterministic \\
\hline
\end{tabular}
\end{subtable}

\end{table*}

We consider two relevant RL baselines, Deep Q-Network (DQN) and Deep Deterministic Policy Gradient (DDPG), for benchmarking and comparison with our proposed algorithm. DQN is a value-based approach that was first put forth for discrete control problems~\cite{mnih2015human}, whereas DDPG is a model-free actor-critic algorithm created for continuous action spaces~\cite{lillicrap2015continuous}. All three algorithms are tested under the same experimental setup to guarantee a fair comparison. Table~\ref{tab:dqn_config} and Table~\ref{tab:ddpg_config} provide summaries of the  DQN and DDPG setups, respectively.


\begin{figure*}[!htb]
\centering
\textbf{\small PPO Generalization Performance Under Static Training Across Mobility Patterns}\\[6pt]

\begin{subfigure}[t]{0.48\textwidth}
\centering
\includegraphics[width=\linewidth,height=5.5cm]{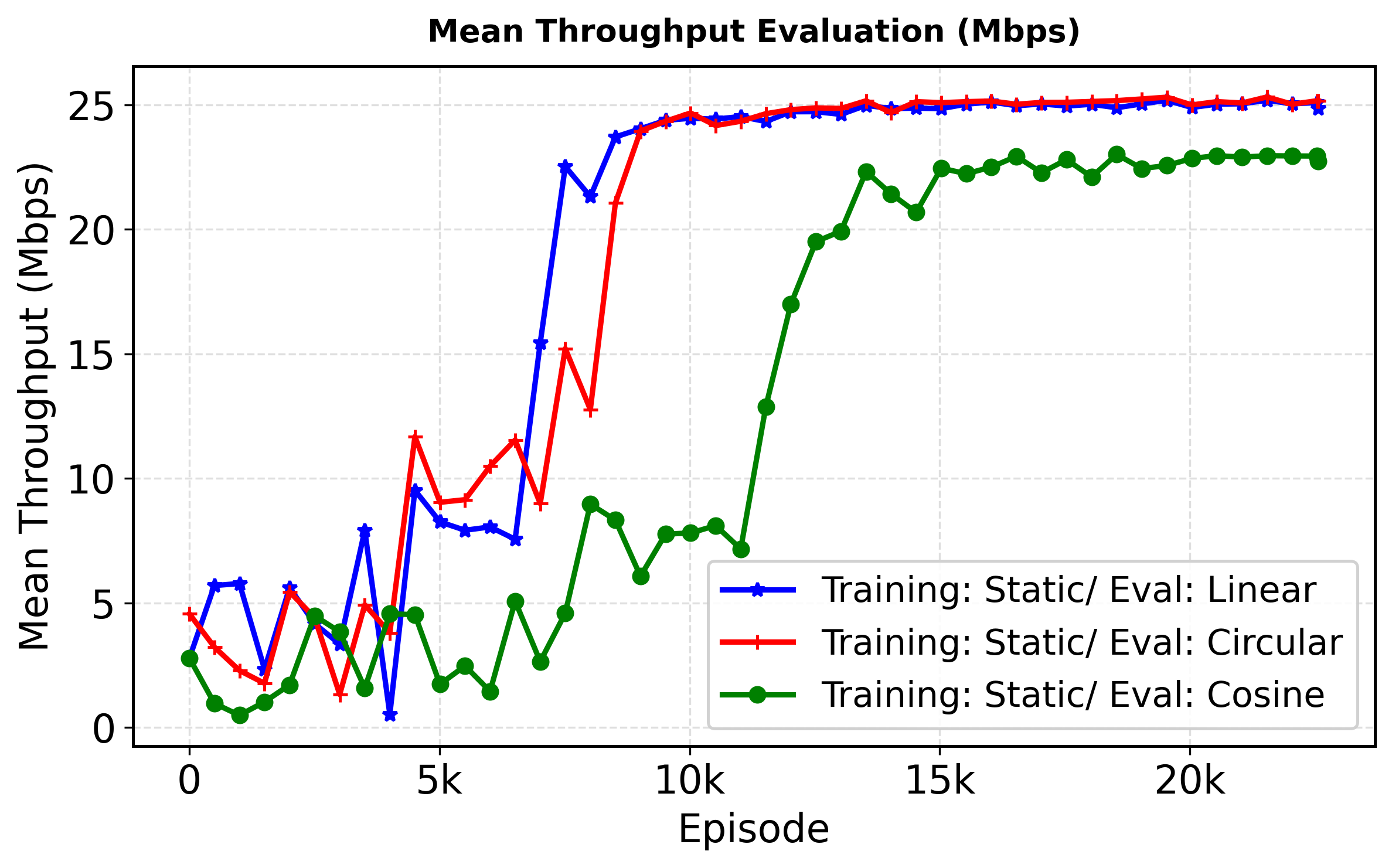}
\caption{Evaluation for UAV-BS Initial Deployment (Scenario~a)}
\label{static_eval1}
\end{subfigure}%
\hfill%
\begin{subfigure}[t]{0.48\textwidth}
\centering
\includegraphics[width=\linewidth,height=5.5cm]{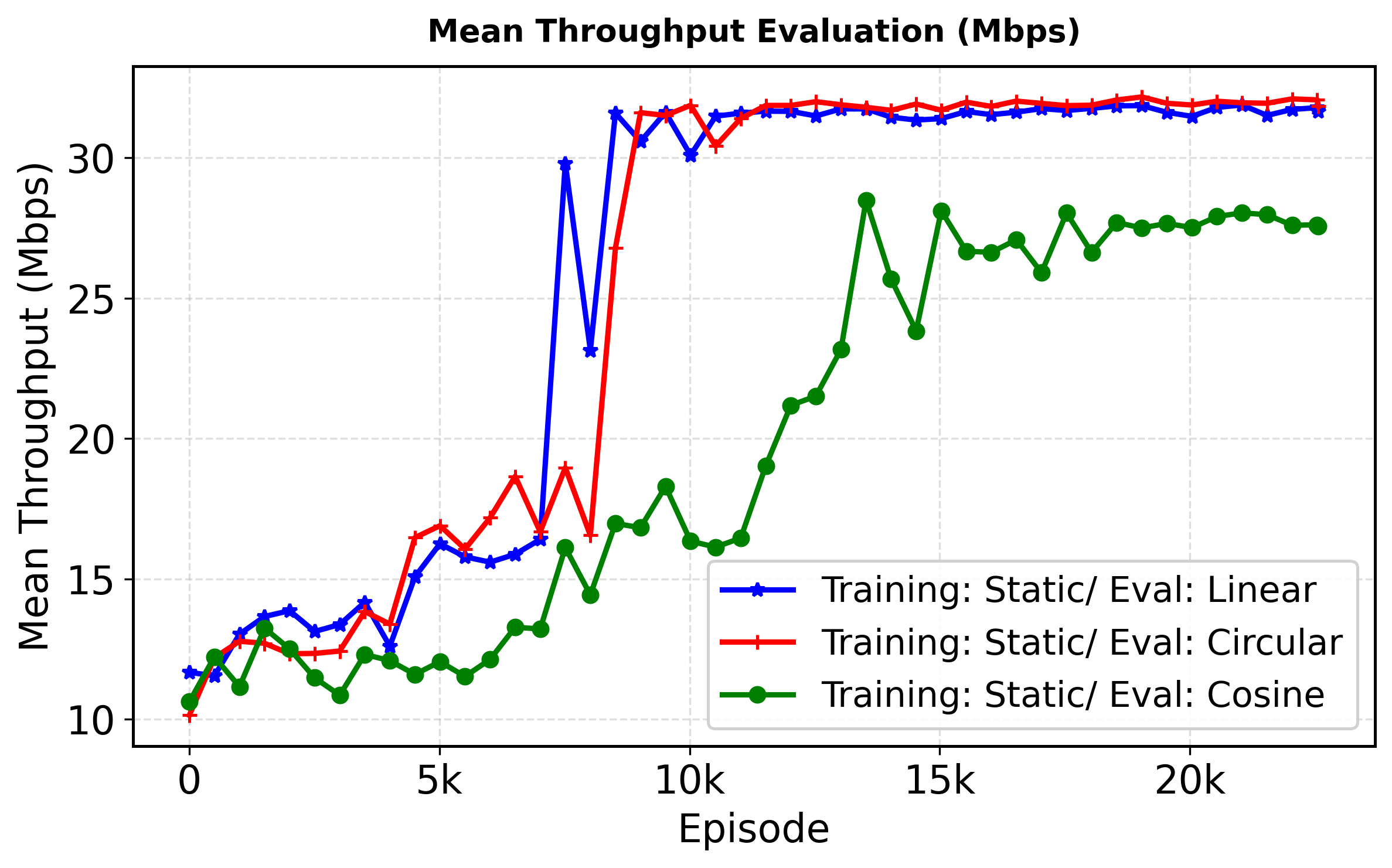}
\caption{Evaluation for UAV-BS Initial Deployment (Scenario~b)}
\label{static_eval2}
\end{subfigure}

\vspace{8pt}

\begin{subfigure}[t]{0.48\textwidth}
\centering
\includegraphics[width=\linewidth,height=5.5cm]{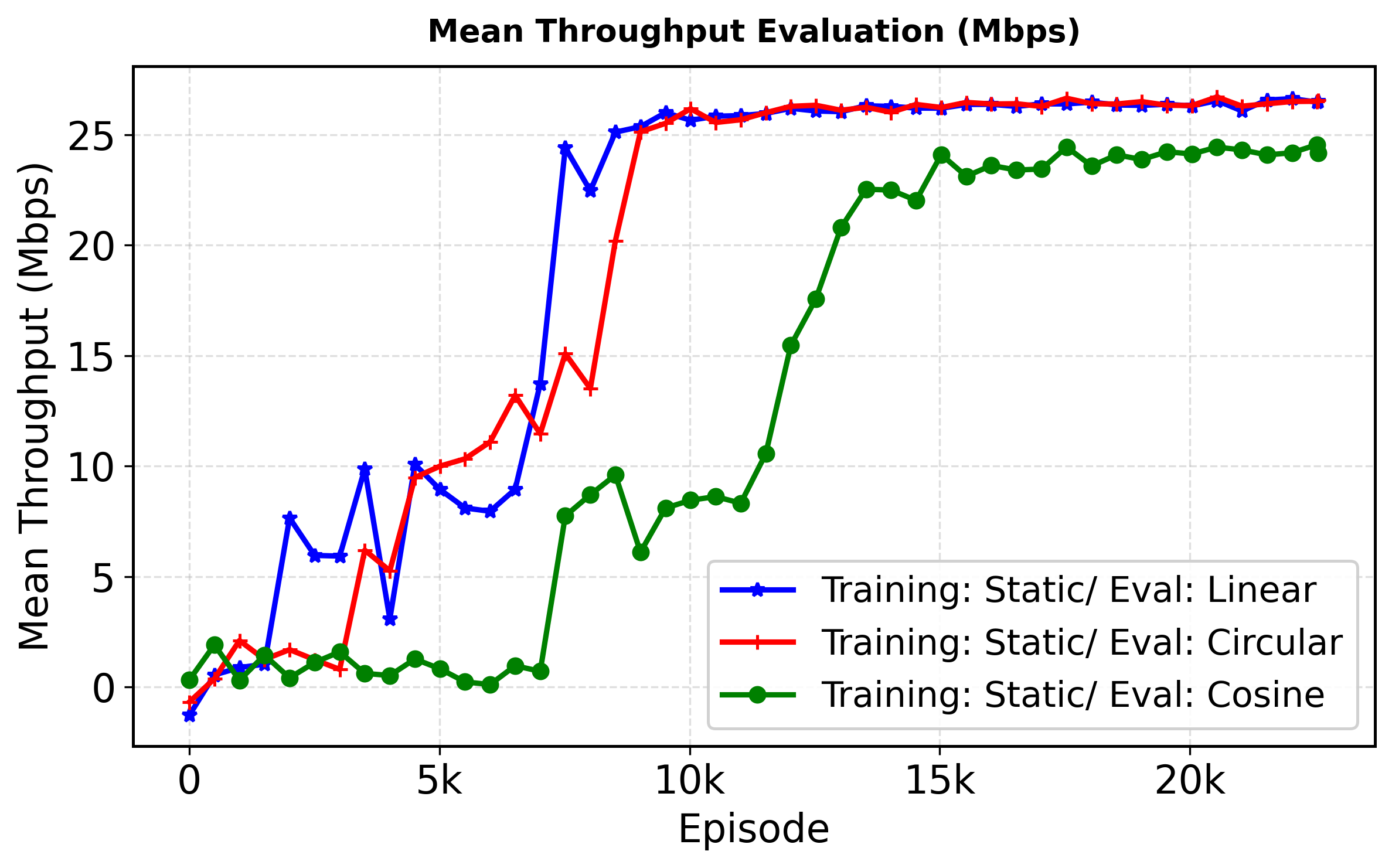}
\caption{Evaluation for UAV-BS Initial Deployment (Scenario~c)}
\label{static_eval3}
\end{subfigure}%
\hfill%
\begin{subfigure}[t]{0.48\textwidth}
\centering
\includegraphics[width=\linewidth,height=5.5cm]{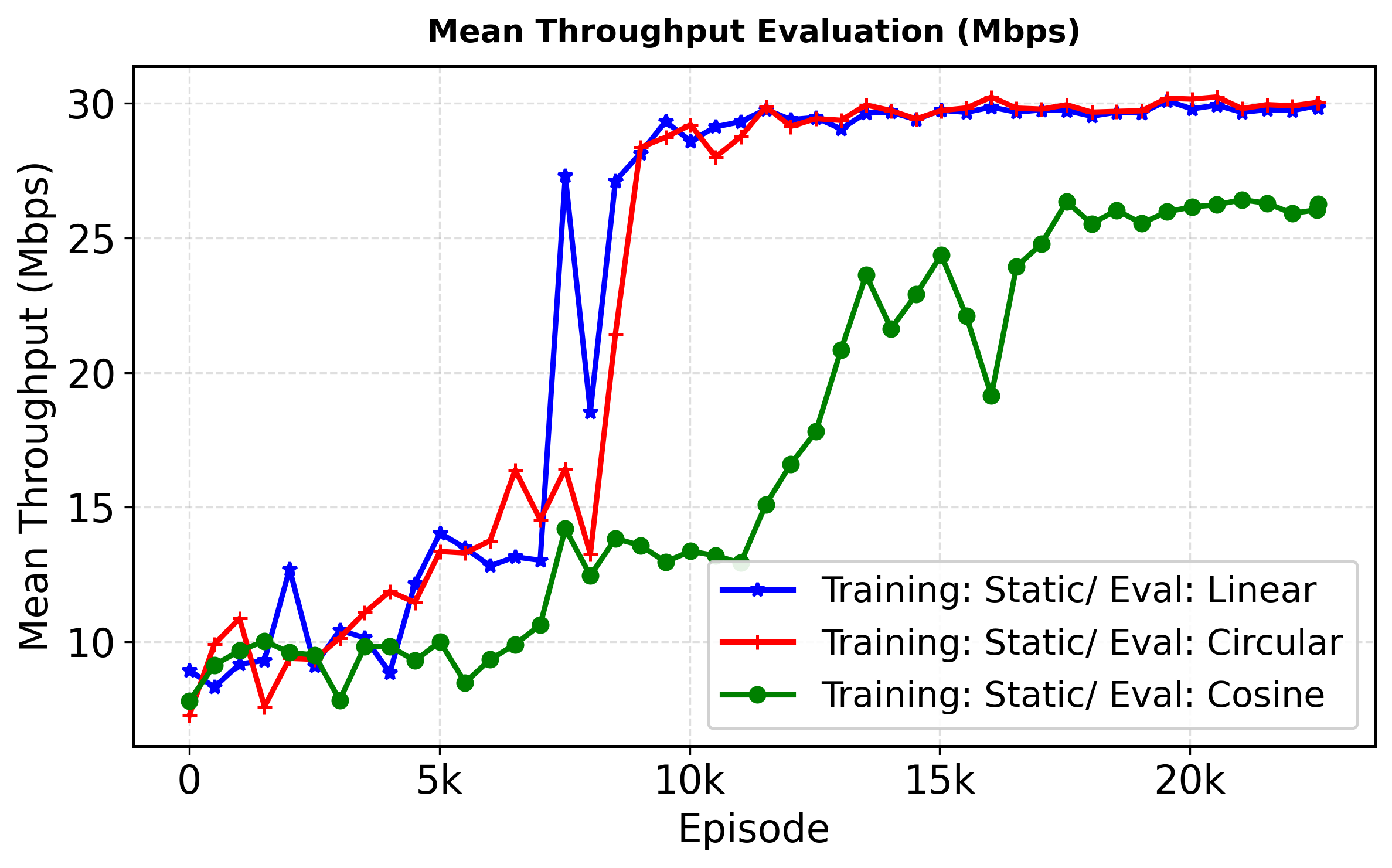}
\caption{Evaluation for UAV-BS Initial Deployment (Scenario~d)}
\label{static_eval4}
\end{subfigure}

\caption{Generalization performance under static training across different mobility patterns. Static-trained policies generalize effectively to linear and circular trajectories, achieving fast convergence and high throughput, whereas cosine-pattern evaluation shows slower convergence and lower throughput stability.}
\label{gen_static}
\end{figure*}

\begin{figure*}[!htb]
\centering
\textbf{\small PPO Generalization Performance Under Linear Motion Training Across Mobility Patterns}\\[6pt]

\begin{subfigure}[t]{0.48\textwidth}
\centering
\includegraphics[width=\linewidth,height=5.5cm]{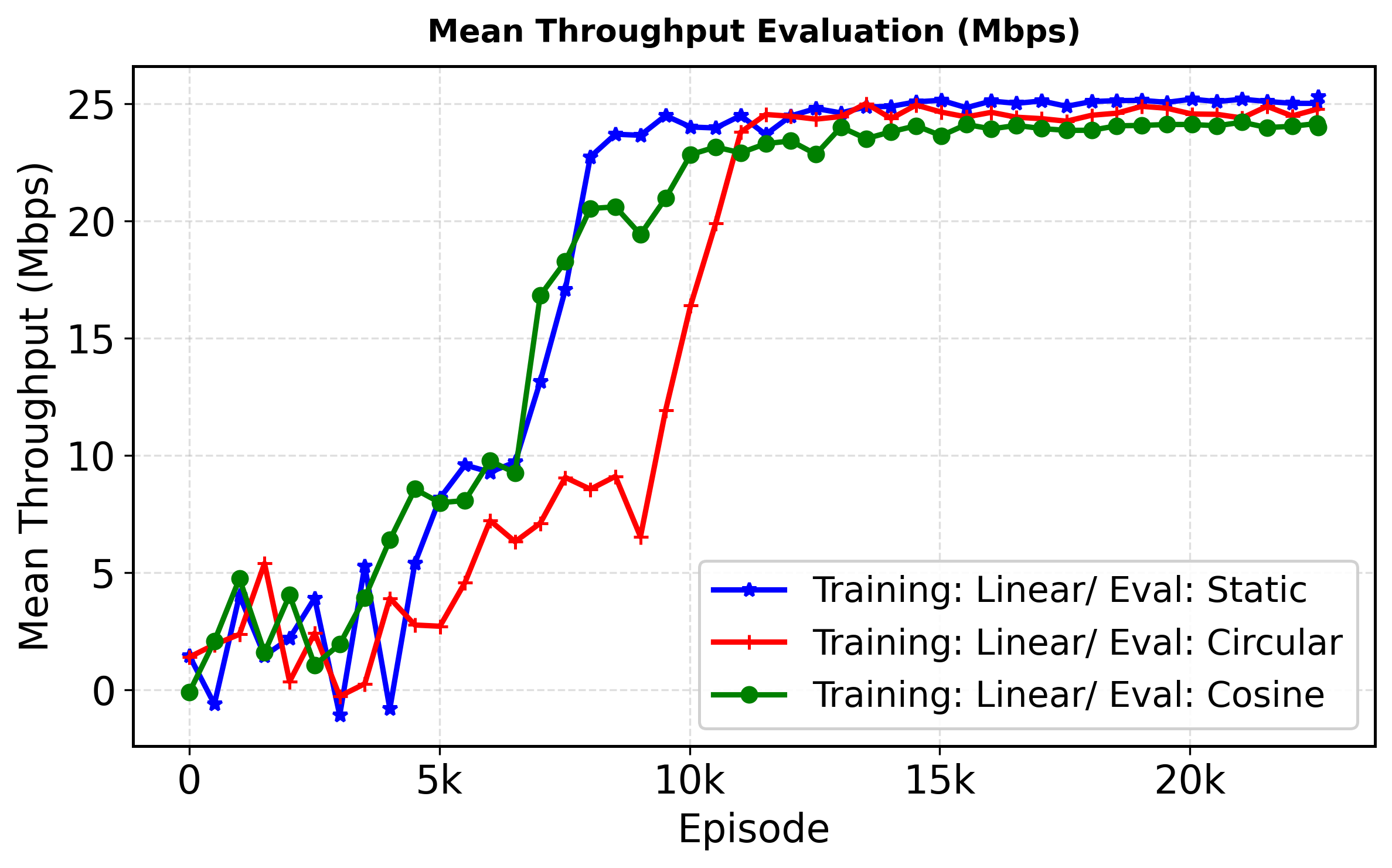}
\caption{Evaluation for UAV-BS Initial Deployment (Scenario~a)}
\label{straigh_eval1}
\end{subfigure}%
\hfill%
\begin{subfigure}[t]{0.48\textwidth}
\centering
\includegraphics[width=\linewidth,height=5.5cm]{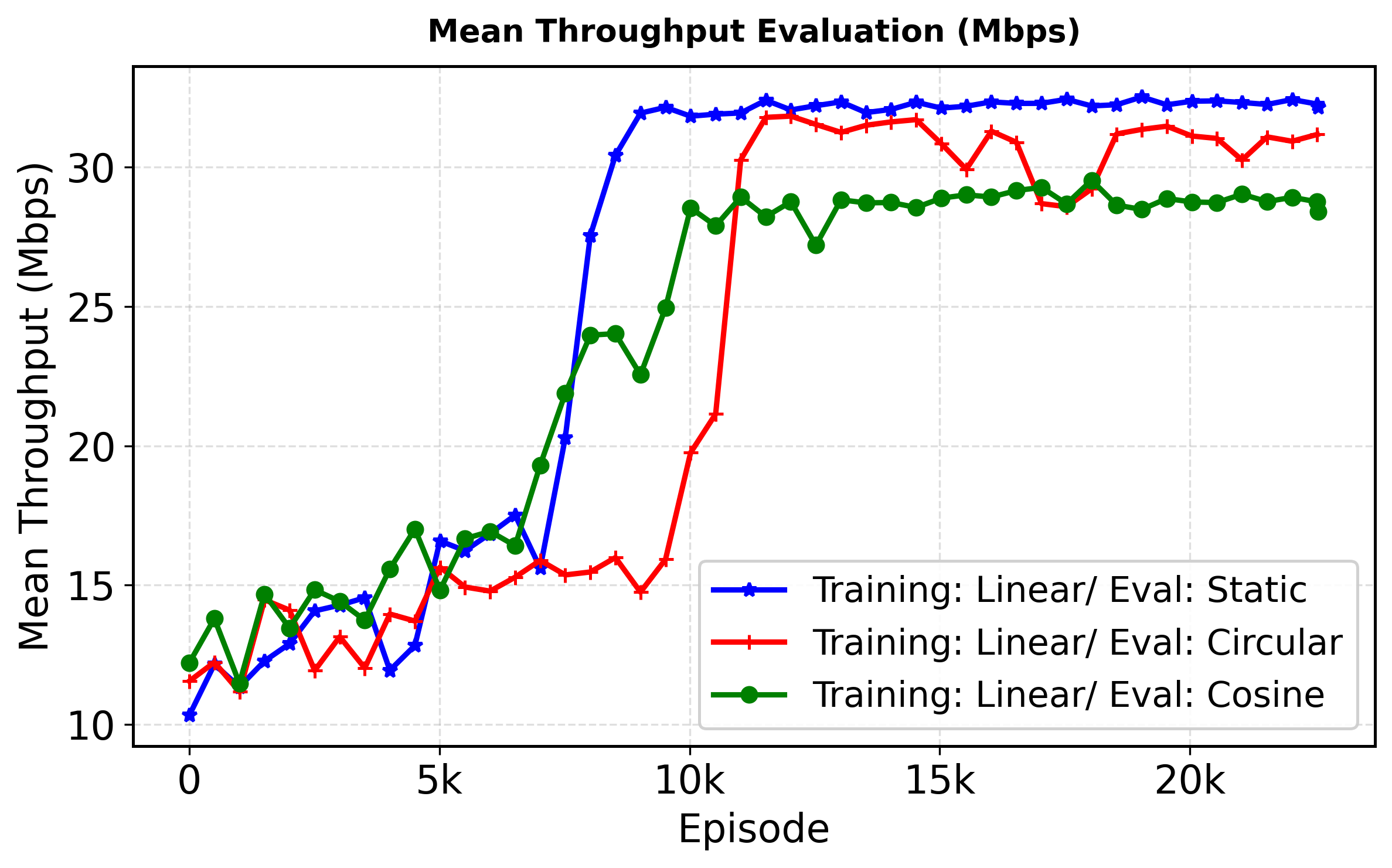}
\caption{Evaluation for UAV-BS Initial Deployment (Scenario~b)}
\label{straigh_eval2}
\end{subfigure}

\vspace{8pt}

\begin{subfigure}[t]{0.48\textwidth}
\centering
\includegraphics[width=\linewidth,height=5.5cm]{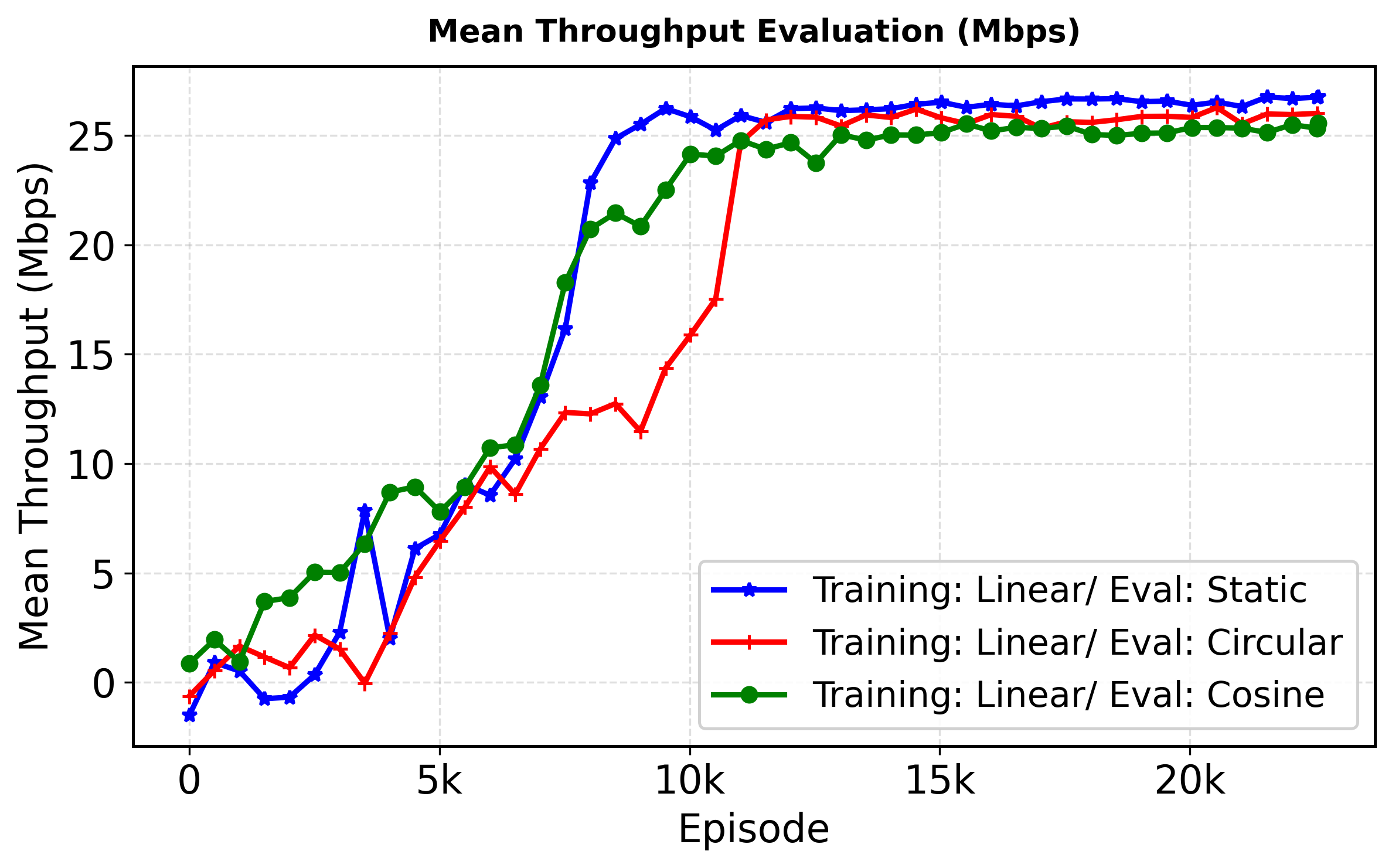}
\caption{Evaluation for UAV-BS Initial Deployment (Scenario~c)}
\label{straigh_eval3}
\end{subfigure}%
\hfill%
\begin{subfigure}[t]{0.48\textwidth}
\centering
\includegraphics[width=\linewidth,height=5.5cm]{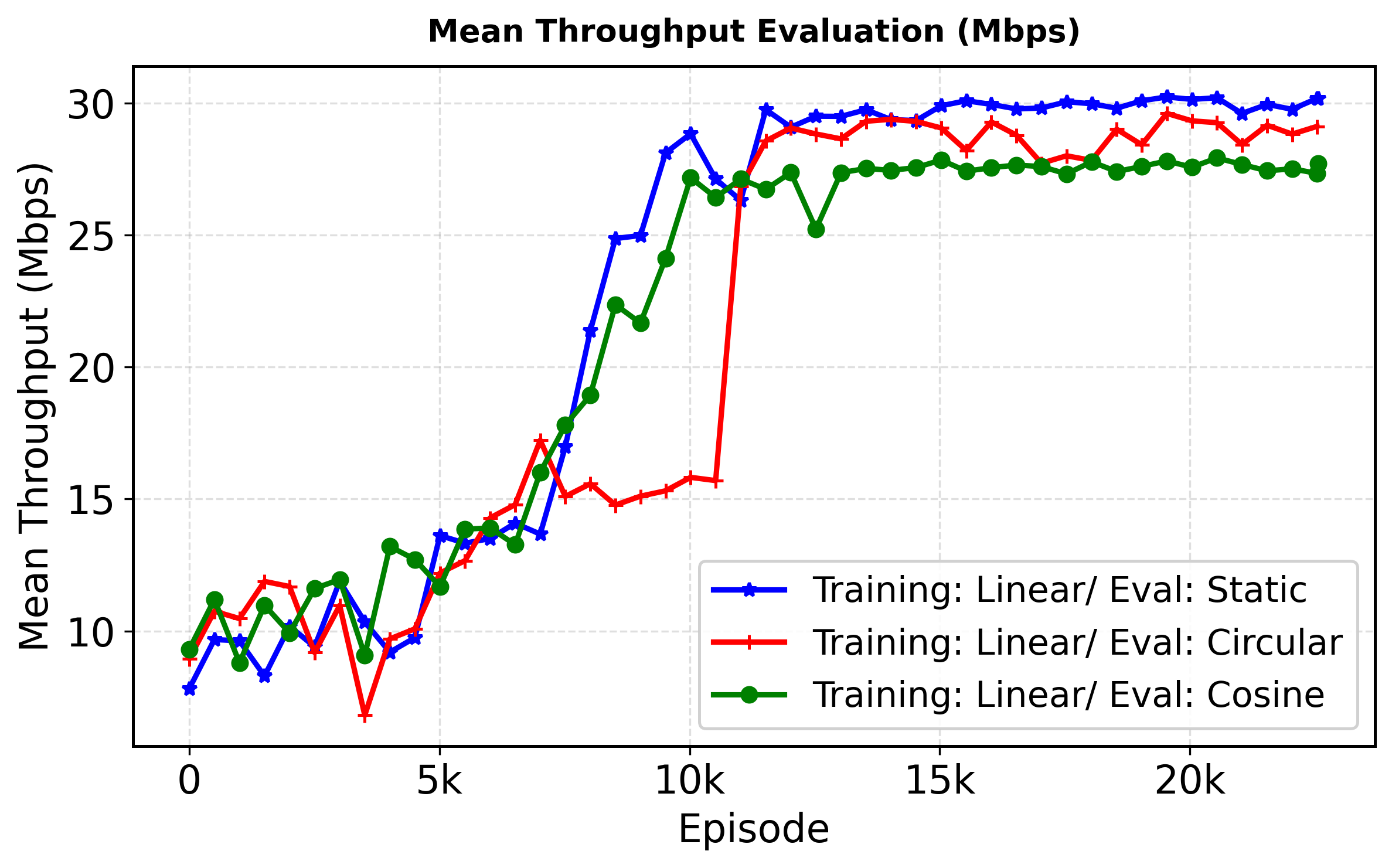}
\caption{Evaluation for UAV-BS Initial Deployment (Scenario~d)}
\label{straigh_eval4}
\end{subfigure}

\caption{Generalization performance under linear-motion training across different mobility patterns. Policies trained using linear trajectories generalize effectively to static and cosine-based motions, while circular-motion evaluation exhibits comparatively slower convergence and reduced throughput performance.}
\label{gen_straight}
\end{figure*}

\begin{figure*}[!htb]
\centering
\textbf{\small PPO Generalization Performance Under Circular Motion Training Across Mobility Patterns}\\[6pt]

\begin{subfigure}[t]{0.48\textwidth}
\centering
\includegraphics[width=\linewidth,height=5.5cm]{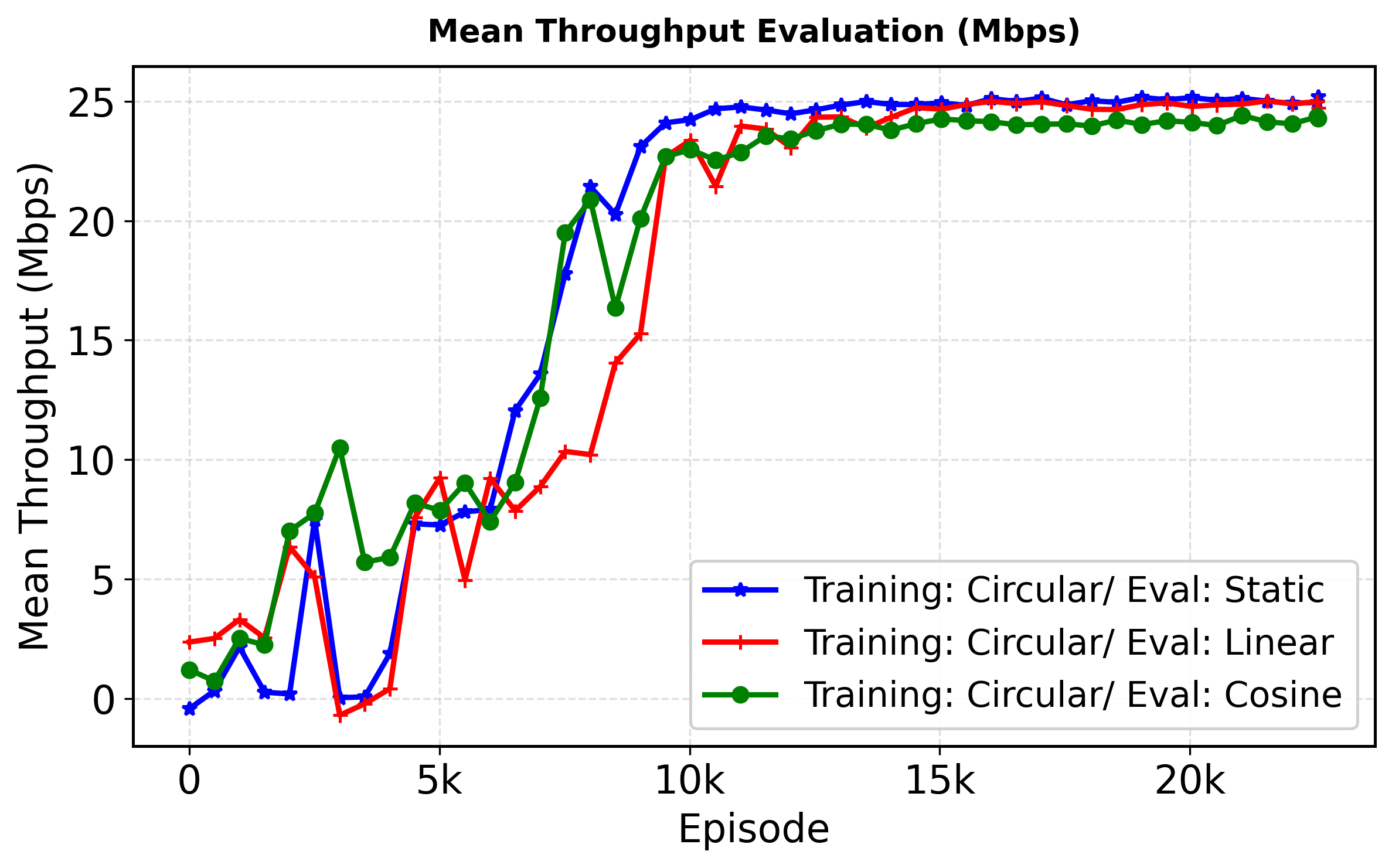}
\caption{Evaluation for UAV-BS Initial Deployment (Scenario~a)}
\label{circular_eval1}
\end{subfigure}%
\hfill%
\begin{subfigure}[t]{0.48\textwidth}
\centering
\includegraphics[width=\linewidth,height=5.5cm]{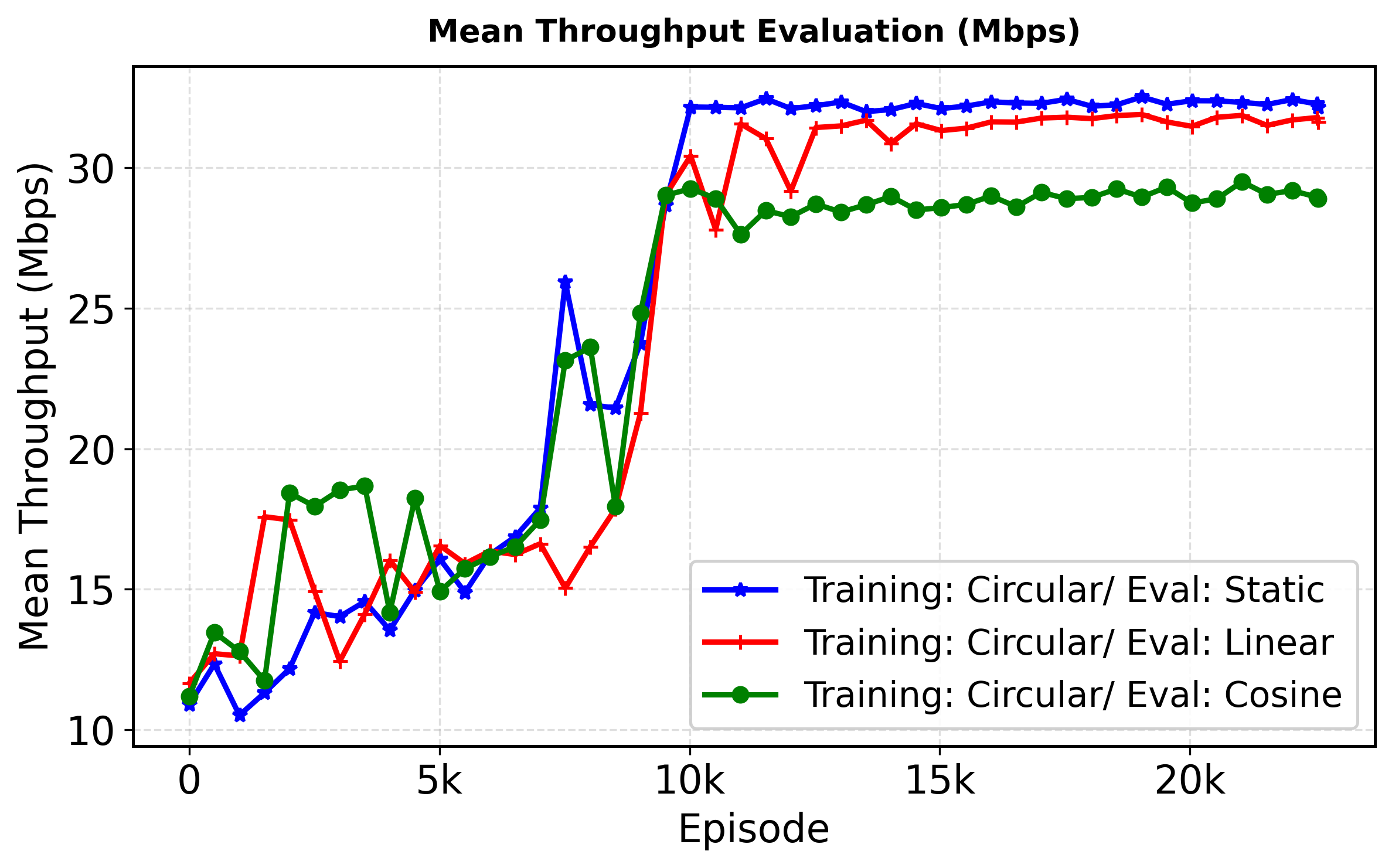}
\caption{Evaluation for UAV-BS Initial Deployment (Scenario~b)}
\label{circular_eval2}
\end{subfigure}

\vspace{8pt}

\begin{subfigure}[t]{0.48\textwidth}
\centering
\includegraphics[width=\linewidth,height=5.5cm]{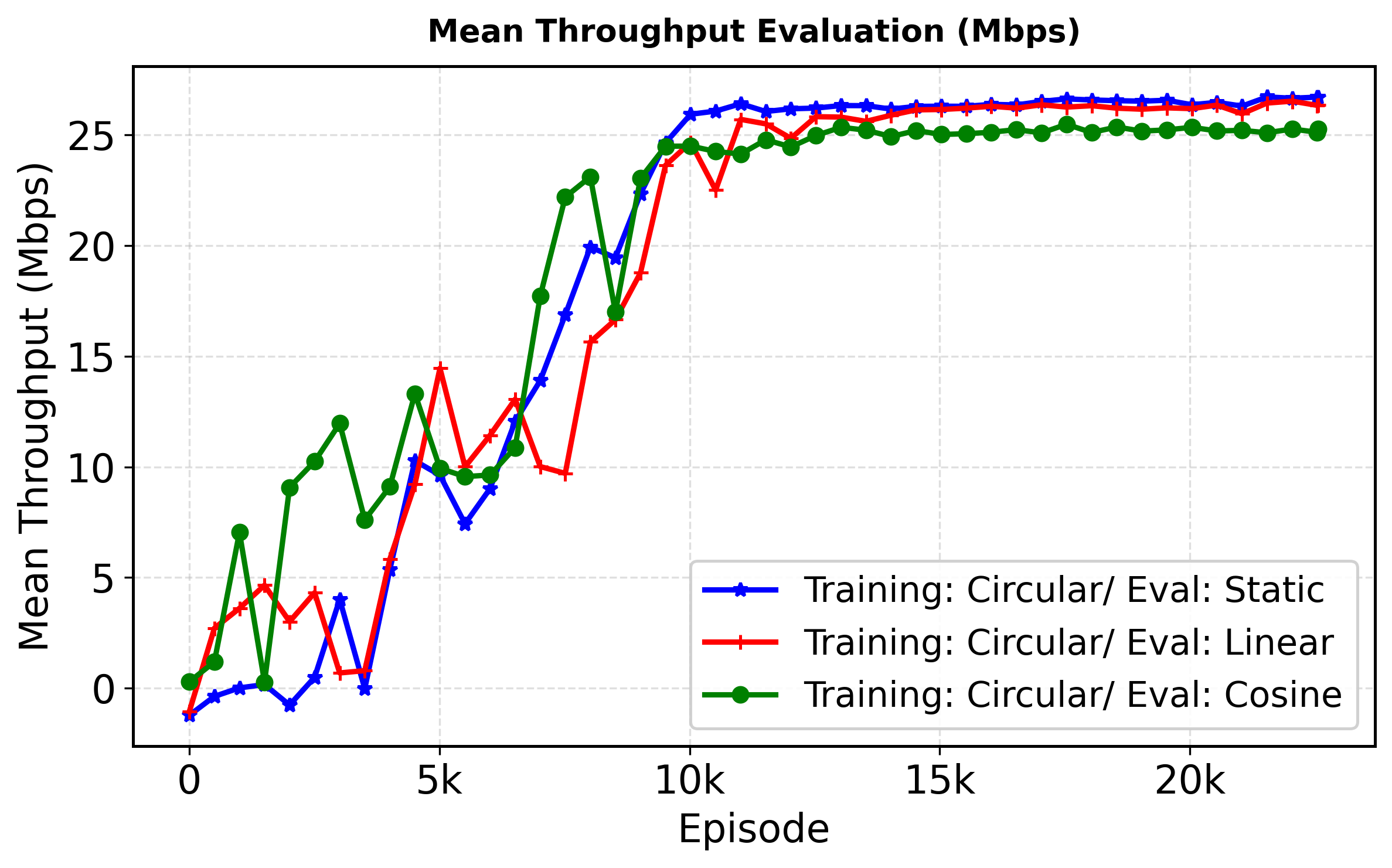}
\caption{Evaluation for UAV-BS Initial Deployment (Scenario~c)}
\label{circular_eval3}
\end{subfigure}%
\hfill%
\begin{subfigure}[t]{0.48\textwidth}
\centering
\includegraphics[width=\linewidth,height=5.5cm]{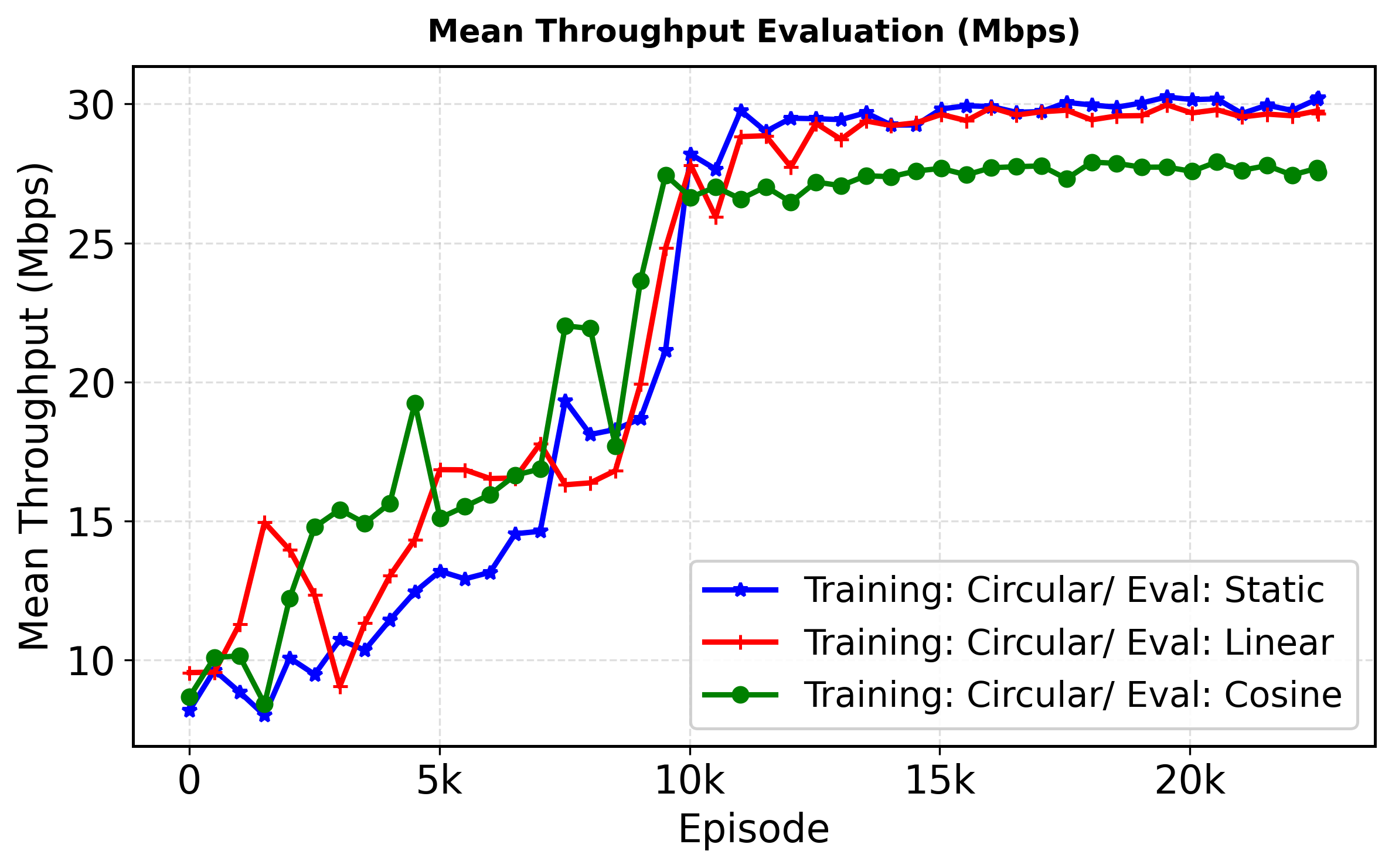}
\caption{Evaluation for UAV-BS Initial Deployment (Scenario~d)}
\label{circular_eval4}
\end{subfigure}

\caption{Generalization performance under circular-motion training across different mobility patterns. Policies trained using circular trajectories maintain fast convergence and high throughput across static, linear, and cosine-based evaluations, demonstrating strong transferability across mobility patterns.}
\label{gen_circular}
\end{figure*}

\begin{figure*}[!htb]
\centering

\textbf{\small PPO Generalization Performance Under Cosine Motion Training Across Mobility Patterns}\\[6pt]

\begin{subfigure}[t]{0.48\textwidth}
\centering
\includegraphics[width=\linewidth,height=5.5cm]{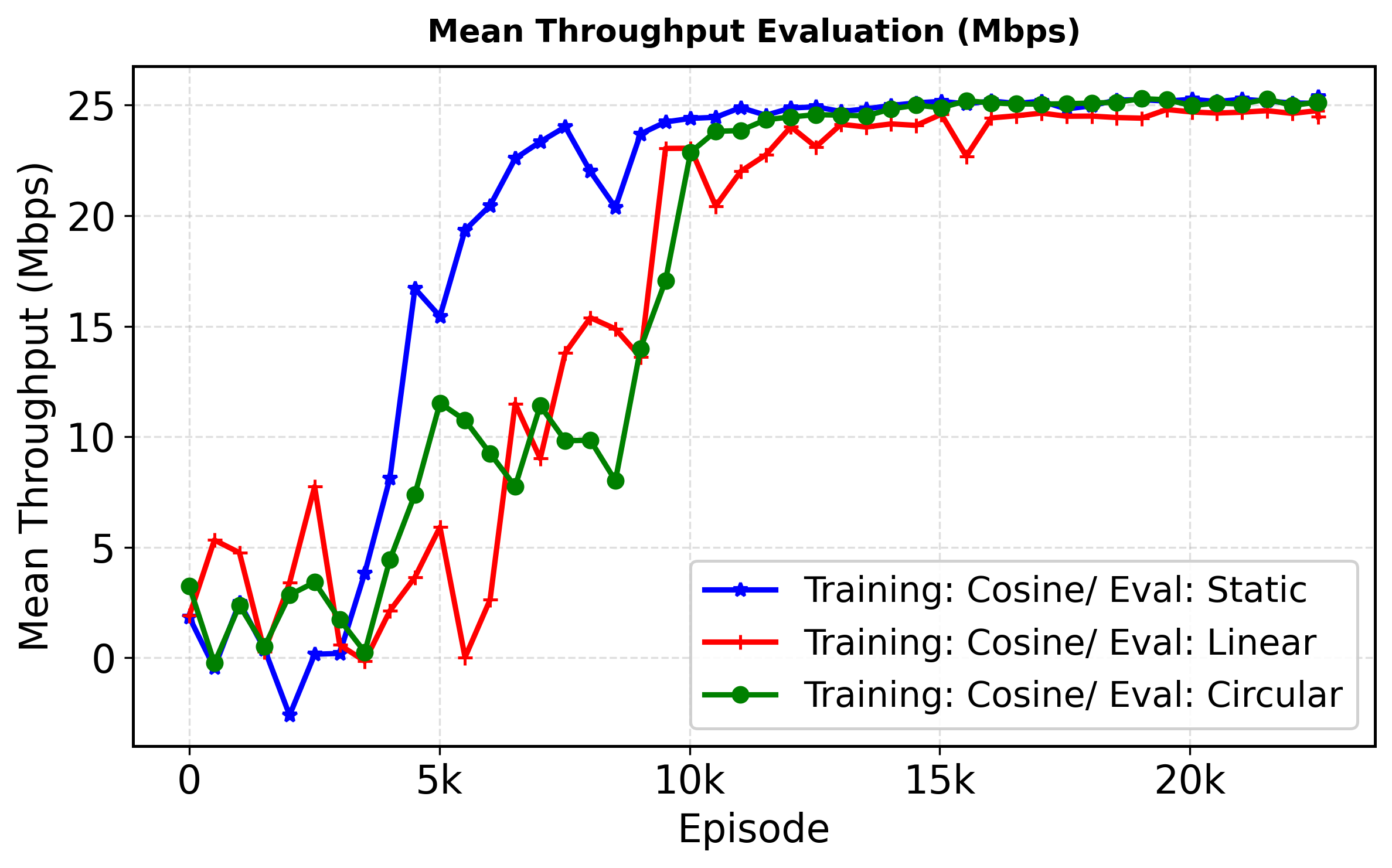}
\caption{Evaluation for UAV-BS Initial Deployment (Scenario~a)}
\label{Cosine_eval1}
\end{subfigure}%
\hfill%
\begin{subfigure}[t]{0.48\textwidth}
\centering
\includegraphics[width=\linewidth,height=5.5cm]{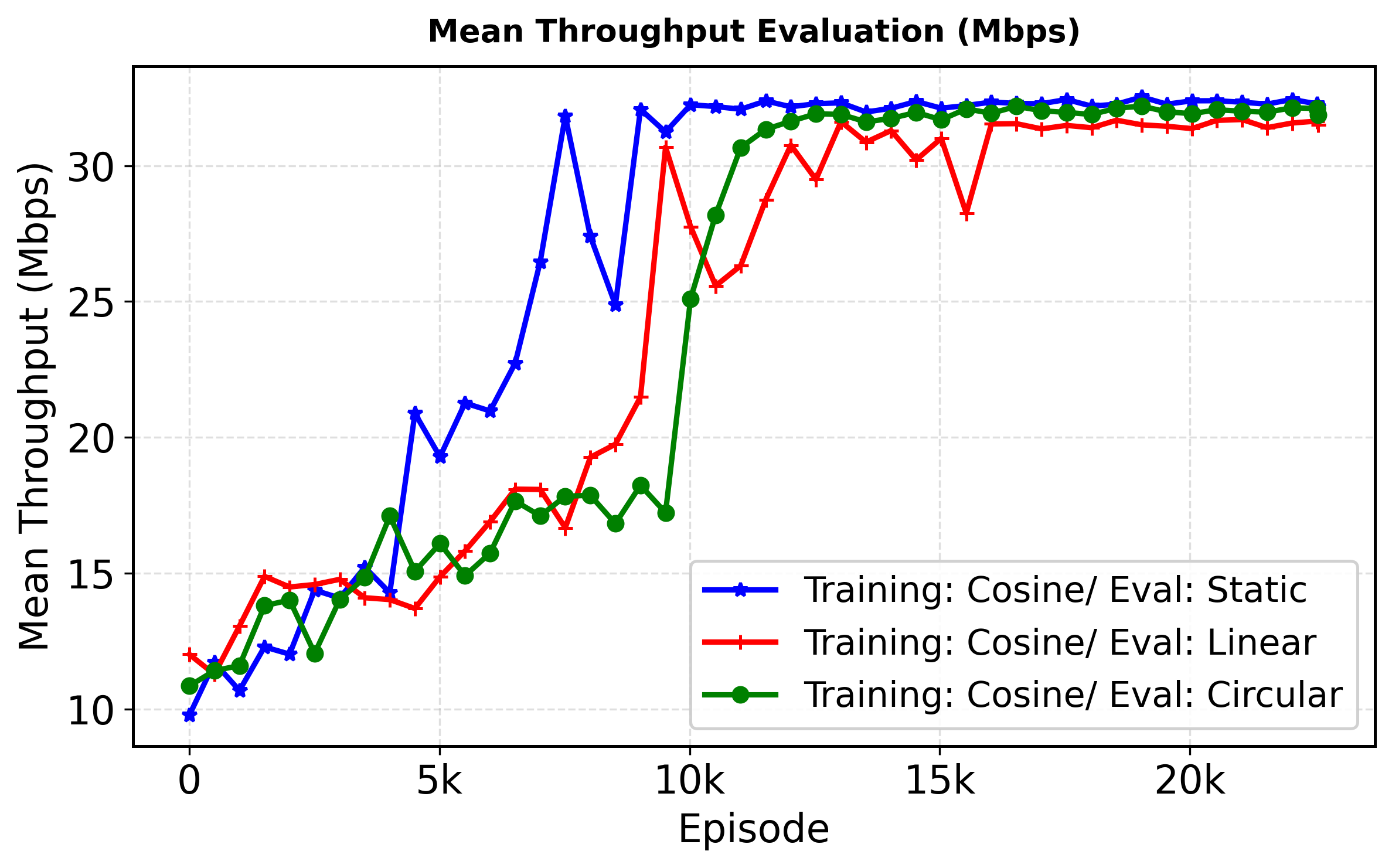}
\caption{Evaluation for UAV-BS Initial Deployment (Scenario~b)}
\label{Cosine_eval2}
\end{subfigure}

\vspace{8pt}

\begin{subfigure}[t]{0.48\textwidth}
\centering
\includegraphics[width=\linewidth,height=5.5cm]{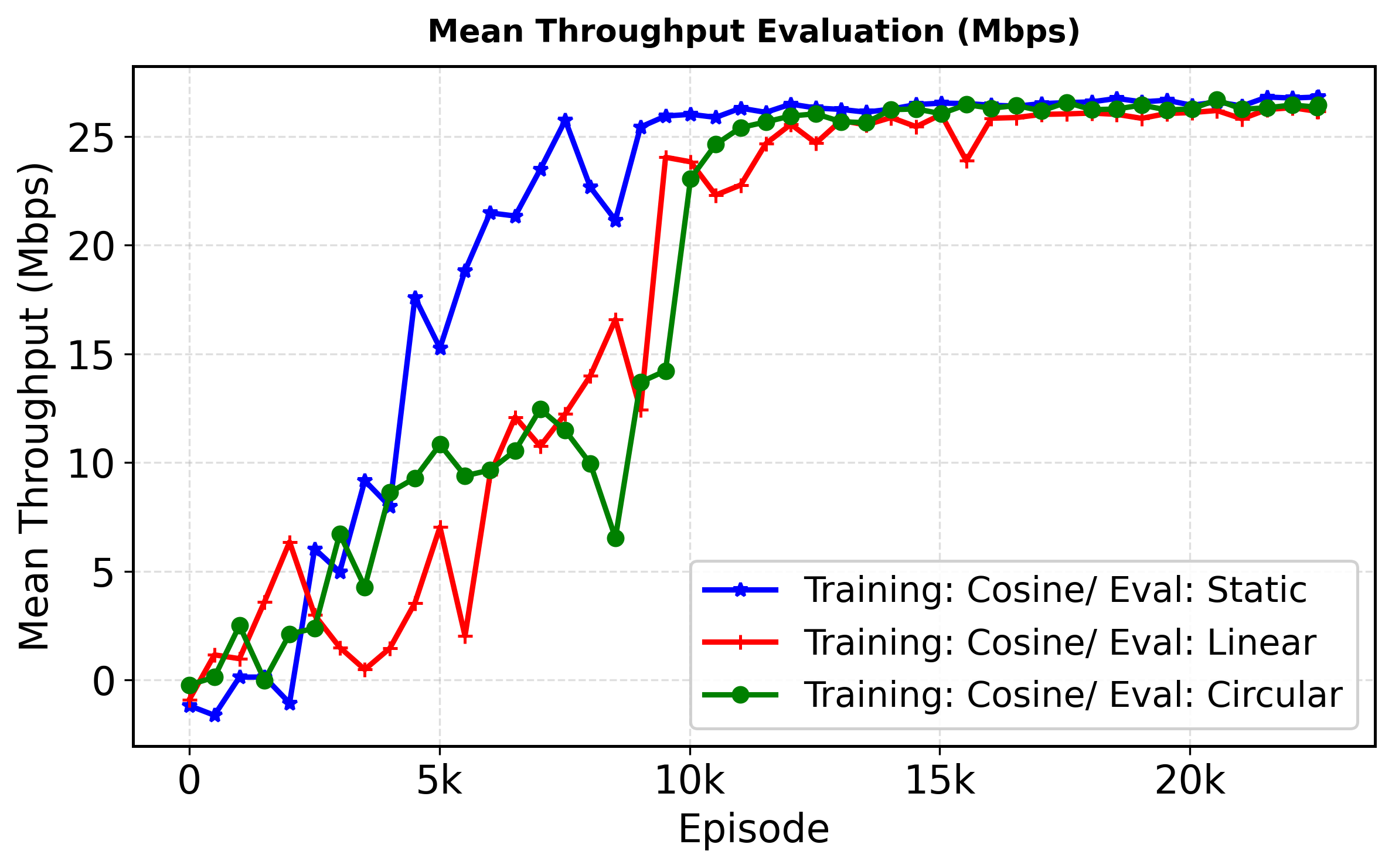}
\caption{Evaluation for UAV-BS Initial Deployment (Scenario~c)}
\label{Cosine_eval3}
\end{subfigure}%
\hfill%
\begin{subfigure}[t]{0.48\textwidth}
\centering
\includegraphics[width=\linewidth,height=5.5cm]{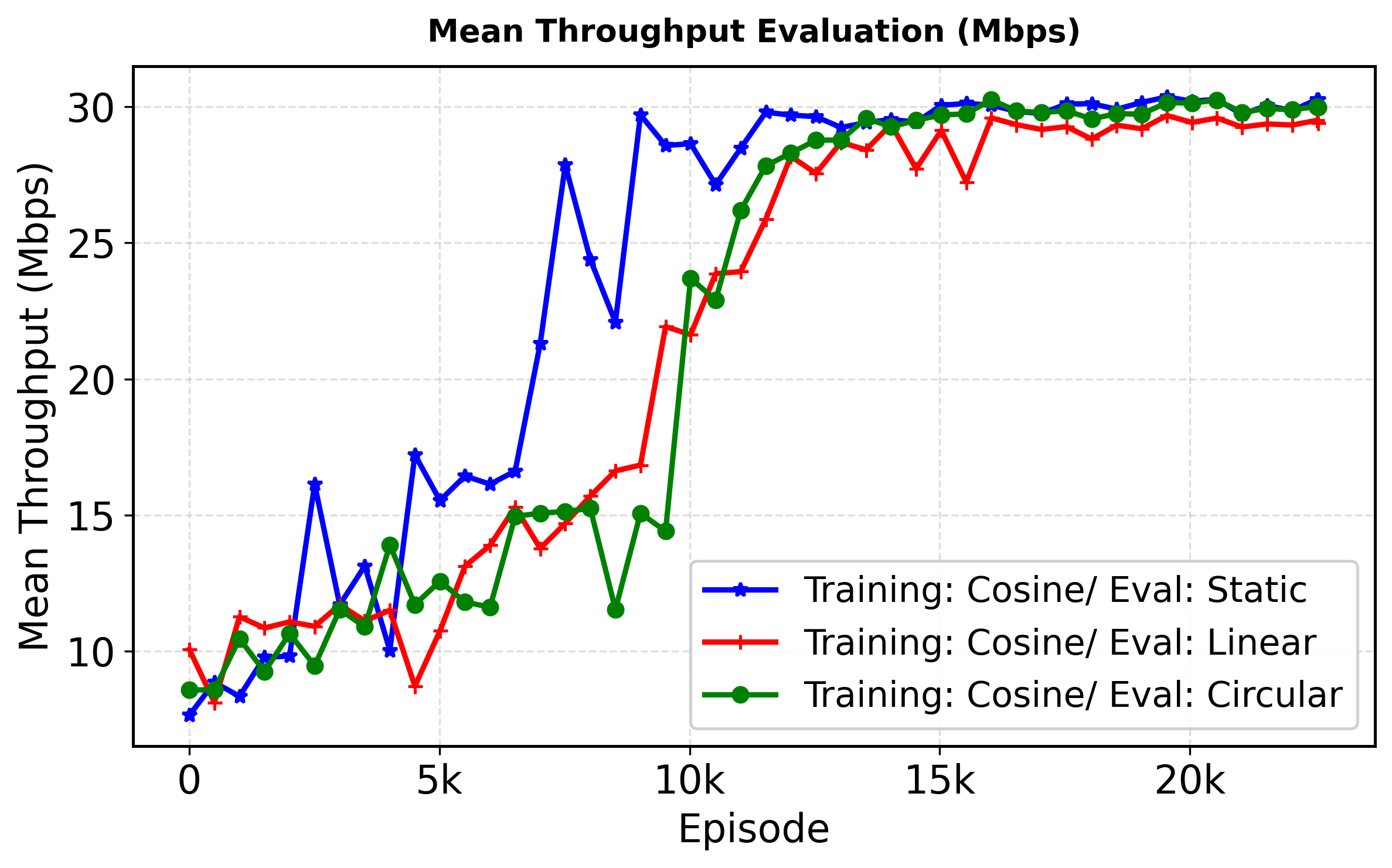}
\caption{Evaluation for UAV-BS Initial Deployment (Scenario~d)}
\label{Cosine_eval4}
\end{subfigure}

\caption{Generalization performance under cosine-pattern motion training across different mobility patterns. Policies trained using cosine trajectories generalize effectively to static, linear, and circular mobility patterns, although static evaluation consistently achieves faster convergence and slightly higher throughput stability.}
\label{gen_cosine}
\end{figure*}

\begin{figure*}[!htb]
\centering

\textbf{\small PPO Generalization Performance Under Composite Mobility Across Scenarios}\\[6pt]

\begin{subfigure}[t]{0.48\textwidth}
\centering
\includegraphics[width=\linewidth,height=5.5cm]{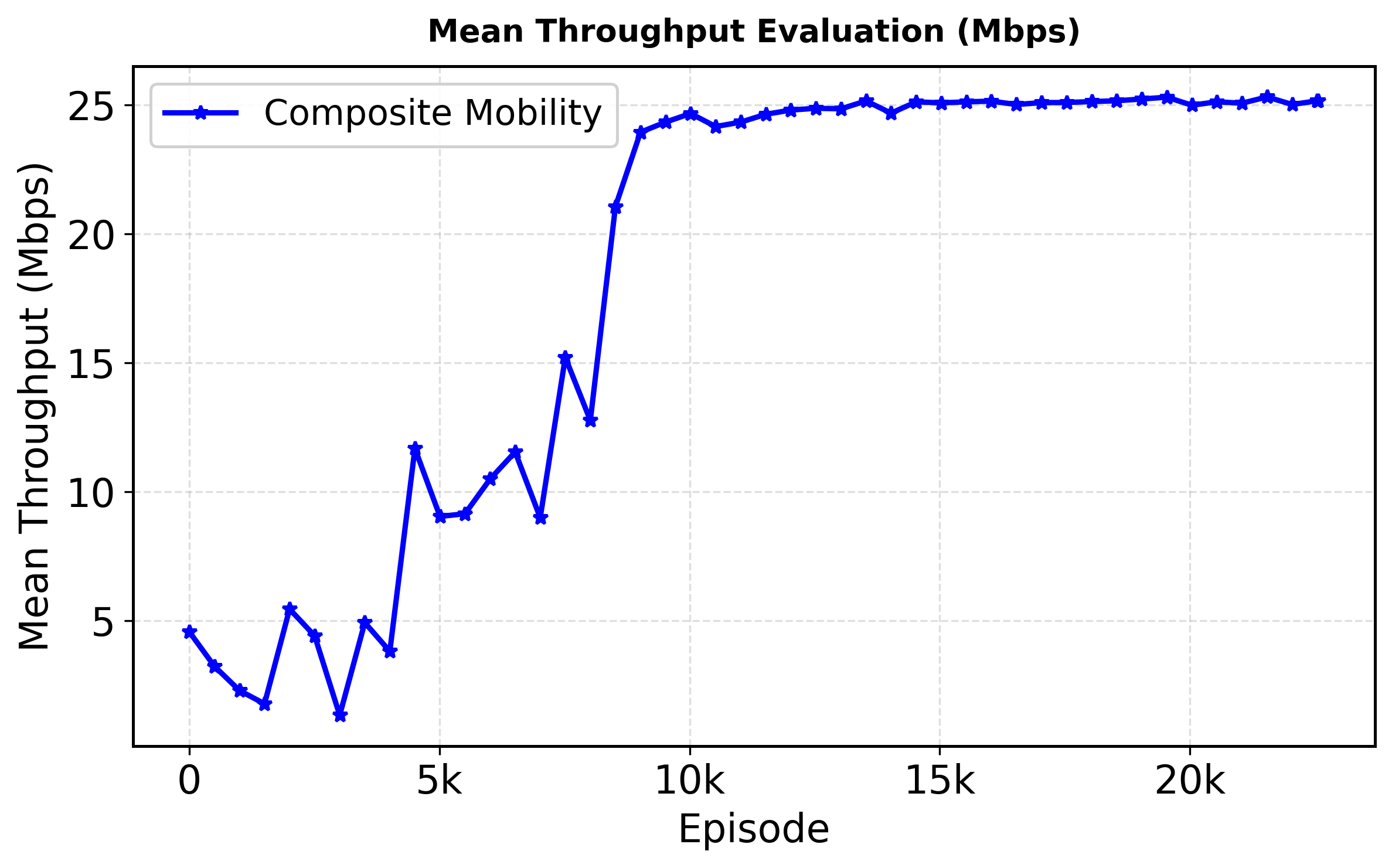}
\vspace{4pt}
\caption{Evaluation for UAV-BS Initial Deployment (Scenario~a)}
\label{compo_eval1}
\end{subfigure}%
\hfill%
\begin{subfigure}[t]{0.48\textwidth}
\centering
\includegraphics[width=\linewidth,height=5.5cm]{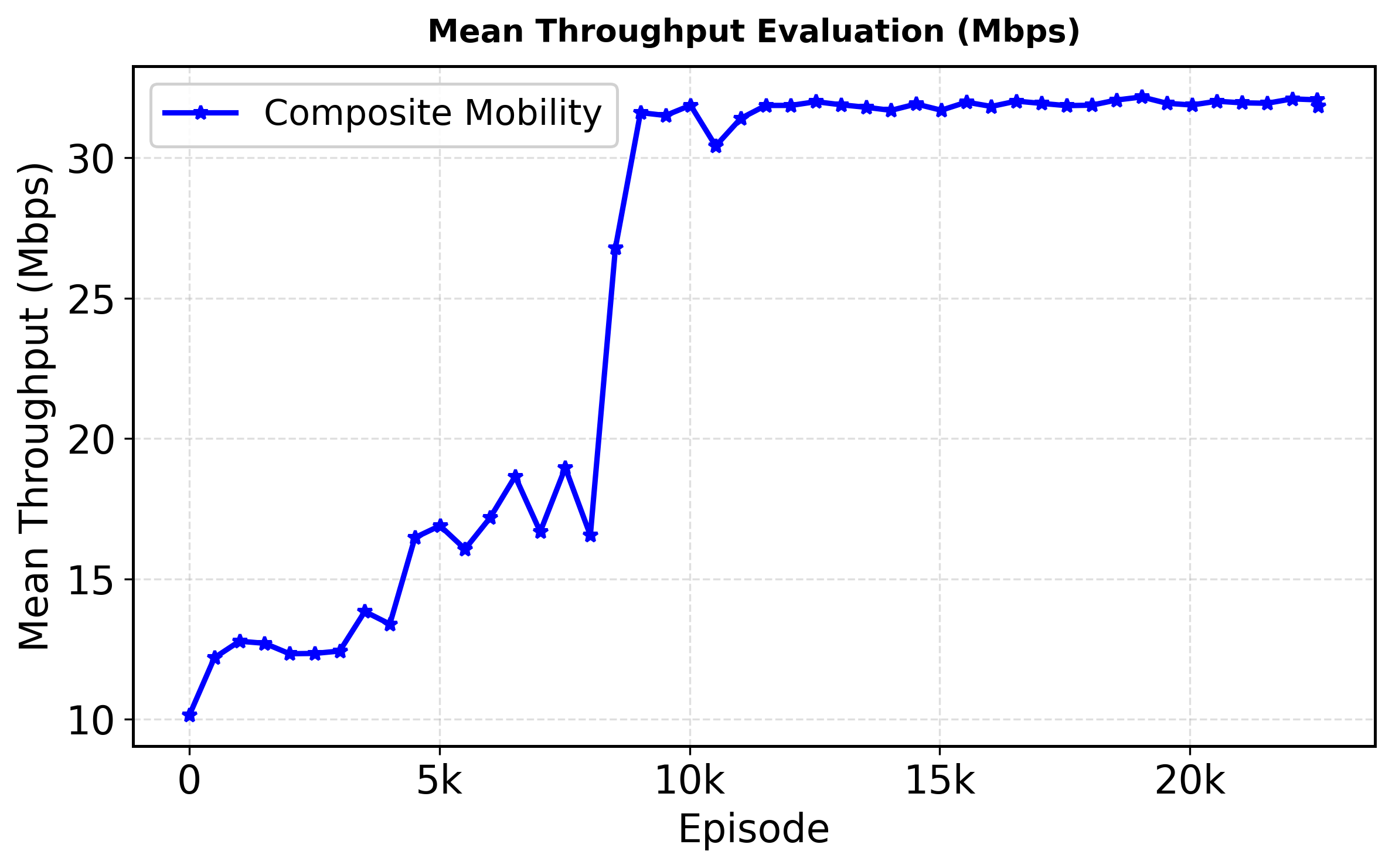}
\vspace{4pt}
\caption{Evaluation for UAV-BS Initial Deployment (Scenario~b)}
\label{compo_eval2}
\end{subfigure}

\vspace{8pt}

\begin{subfigure}[t]{0.48\textwidth}
\centering
\includegraphics[width=\linewidth,height=5.5cm]{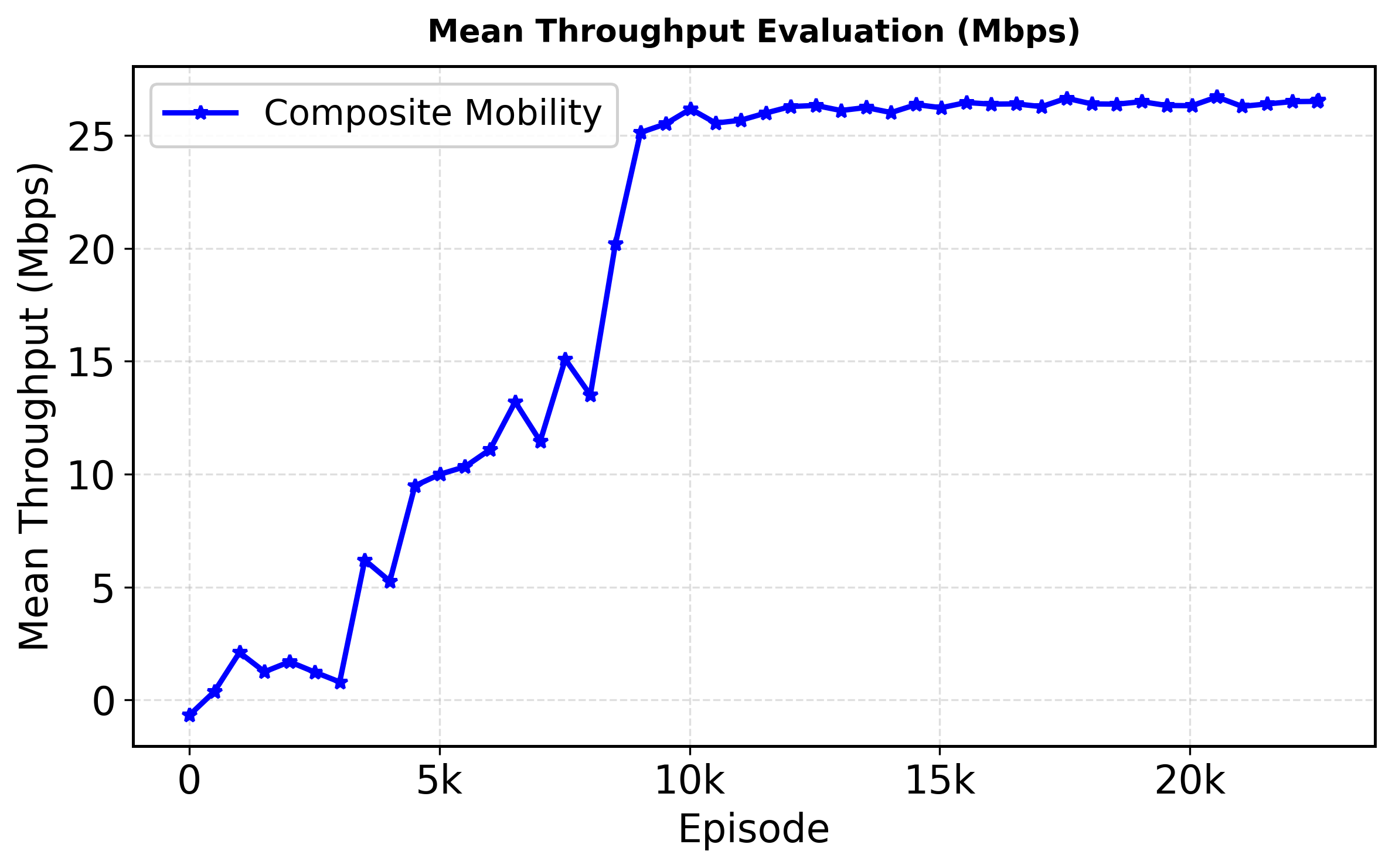}
\vspace{4pt}
\footnotesize
\caption{Evaluation for UAV-BS Initial Deployment (Scenario~c)}
\label{compo_eval3}
\end{subfigure}%
\hfill%
\begin{subfigure}[t]{0.48\textwidth}
\centering
\includegraphics[width=\linewidth,height=5.5cm]{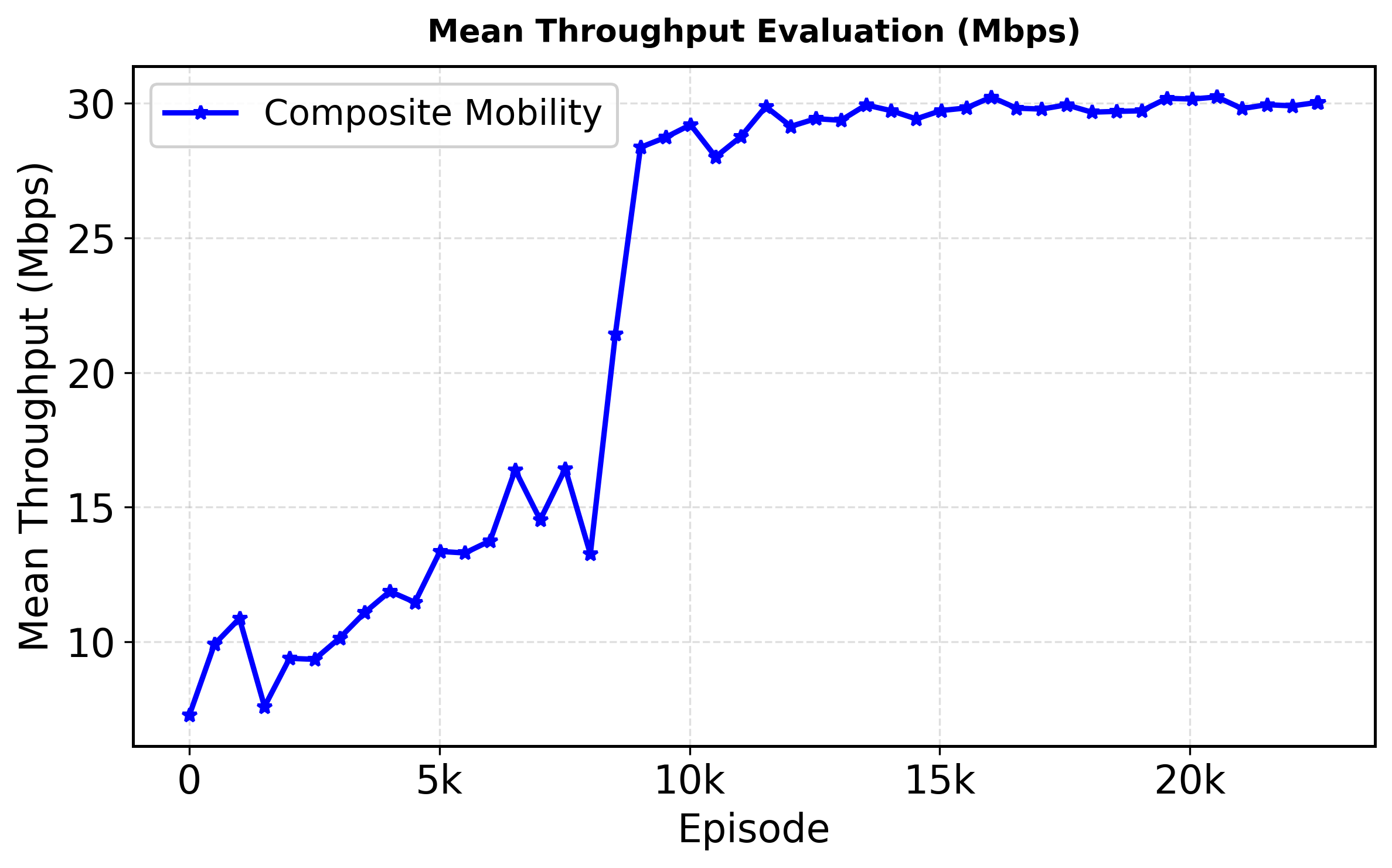}
\vspace{4pt}
\footnotesize
\caption{Evaluation for UAV-BS Initial Deployment (Scenario~d)}
\label{compo_eval4}
\end{subfigure}

\caption{PPO generalization performance under composite mobility across four predefined scenarios. Each plot shows throughput behavior during evaluation for a given initial UAV-BS configuration.}
\label{fig:comp_reward_comparison_eval}
\end{figure*}

\subsubsection{Simulation Results}

Fig.~\ref{fig:PPO_DQN_DDPG_training} compares the training performance of PPO, DQN, and DDPG in terms of mean reward and mean throughput. The shaded regions represent the variability around the smoothed mean curves and provide insight into the stability of each algorithm during training. PPO exhibits the fastest convergence, the smallest performance variability, and the highest final reward and throughput. Its clipped surrogate objective improves training stability by preventing excessively large policy updates.
Fig.~\ref{fig:DDPG_mean_reward} and Fig.~\ref{fig:DDPG_mean_throughput} show that DQN achieves the lowest overall performance among the evaluated algorithms. The training process exhibits slower convergence, noticeable fluctuations, and larger uncertainty regions during the early learning stages. Since DQN relies on a value-based formulation with discretized actions, its ability to model the continuous control requirements of UAV-BS trajectory and resource optimization is limited, resulting in lower final reward and throughput performance.
The DDPG results in Fig.~\ref{fig:DDPG_mean_reward} and Fig.~\ref{fig:DDPG_mean_throughput} indicate higher training variability, reflected by the wider shaded regions throughout much of the learning process. Although DDPG eventually converges, it stabilizes at lower reward and throughput levels than PPO. This behavior is mainly related to sensitivity to critic estimation errors, exploration noise, and the sensitivity of UAV-BS control environments to small action perturbations.
Overall, PPO provides the best balance between convergence speed, training stability, and final performance compared with DQN and DDPG.

\subsection{PPO Generalization Performance Results}
\label{subsec:generalization_mobility_ppo1}
Building on the refined reward design and state-space configuration obtained in the previous phases, as well as the statistical evaluation across multiple random seeds and comparisons with DQN and DDPG, this section investigates the generalization capability of PPO using the selected sigmoid-based reward mapping and the reduced state space composed of UAV-BS positions, SINR, and AoA statistics.
The objective is to evaluate PPO performance when the training and evaluation phases are conducted under different mobility patterns selected from \{static, linear, circular, cosine\}. To provide a systematic analysis, each mobility pattern is individually used during training, after which the learned policy is evaluated on the remaining mobility patterns. For example, a policy trained under static mobility is evaluated under linear, circular, and cosine-pattern motions. The same procedure is repeated for all mobility configurations.
Finally, we investigate PPO generalization under a composite mobility scenario, where multiple UE mobility patterns coexist simultaneously. In this setting, linear, circular, and cosine-pattern motions are jointly considered during both training and evaluation to better reflect realistic emergency communication environments with heterogeneous and dynamically evolving user movements.

\subsubsection{Simulation Results}

\paragraph{Training: Static; Evaluate: Linear, Circular, Cosine}

Figures~\ref{static_eval1}-\ref{static_eval4} illustrate the generalization performance of PPO policies trained under linear-motion mobility and evaluated across different motion patterns. Across all deployment scenarios, the linear-motion evaluation (blue curve) demonstrates the fastest convergence and achieves the highest throughput performance, typically reaching approximately $25$--$33$~Mbps with stable and consistent learning behavior.
The circular-motion evaluation (red curve) also achieves relatively high final throughput values comparable to the linear case; however, its convergence process is noticeably slower due to the increased mobility complexity introduced by rotational trajectory dynamics.
In contrast, the cosine-motion evaluation (green curve) exhibits slower convergence and converges to comparatively suboptimal throughput levels, generally stabilizing between approximately $23$ and $28$~Mbps. This behavior is primarily attributed to the higher complexity and continuously varying characteristics of cosine-based mobility patterns, which make policy adaptation more challenging.


\paragraph{Training: Linear Motion; Evaluate: Static, Circular, Cosine}

Figures~\ref{straigh_eval1}-\ref{straigh_eval4} illustrate the generalization performance of PPO policies trained under linear-motion trajectories and evaluated across different UAV-BS deployment scenarios and mobility patterns. Overall, the learned policies demonstrate effective transferability across heterogeneous mobility conditions while maintaining stable convergence and high throughput performance.
Across all deployment scenarios, the static-motion evaluation (blue curve) achieves the fastest convergence and the highest throughput performance, typically stabilizing between approximately $26$ and $33$~Mbps. The cosine-motion evaluation (green curve) also demonstrates strong generalization capability, converging relatively quickly and maintaining stable throughput levels ranging from approximately $24$ to $28$~Mbps.
In contrast, the circular-motion evaluation (red curve) exhibits noticeably slower convergence and slightly reduced throughput performance, generally stabilizing between approximately $25$ and $31.5$~Mbps depending on the deployment scenario. The delayed convergence behavior indicates that rotational mobility introduces more challenging environmental dynamics and adaptation requirements compared to static and cosine-based trajectories.
Although all mobility patterns eventually converge to stable high-throughput operating regions, the results indicate that linear-motion training generalizes more effectively to static and cosine-based mobility patterns than to circular-motion trajectories.




\paragraph{Training: Circular Motion; Evaluate: Static, Linear, Cosine}

Figures~\ref{circular_eval1}-\ref{circular_eval4} illustrate the generalization performance of PPO policies trained under circular-motion trajectories and evaluated across different UAV-BS deployment scenarios and mobility patterns. Overall, the learned policies demonstrate strong transferability across all evaluated mobility conditions while maintaining stable convergence and high throughput performance.
Across all deployment scenarios, the static-motion evaluation (blue curve) achieves the fastest convergence and the highest throughput performance, typically stabilizing between approximately $25$ and $33$~Mbps. The linear-motion evaluation (red curve) also demonstrates effective generalization capability, exhibiting convergence behavior comparable to the static case with only minor throughput differences across most scenarios.
The cosine-motion evaluation (green curve) likewise maintains stable convergence and competitive throughput performance across all deployment scenarios. However, in several cases, the achieved final throughput is slightly lower than that of the static and linear evaluations, generally stabilizing between approximately $24$ and $30$~Mbps. This behavior indicates that cosine-based mobility patterns introduce additional trajectory variations that slightly reduce the achievable long-term throughput performance.
Compared with other mobility-training configurations, circular-motion training provides relatively balanced generalization performance across static, linear, and cosine evaluation patterns. The performance gap among different evaluation scenarios remains comparatively small, indicating strong robustness to mobility variations and trajectory dynamics.

\paragraph{Training: Cosine Motion; Evaluate: Static, Linear, Circular}

Figures~\ref{Cosine_eval1}-\ref{Cosine_eval4} illustrate the generalization performance of PPO policies trained under cosine-pattern mobility across different UAV-BS deployment scenarios and evaluation mobility patterns. Overall, cosine-motion training demonstrates effective transferability across static, linear, and circular mobility conditions while maintaining stable convergence and high throughput performance.
Across all deployment scenarios, the static-motion evaluation (blue curve) achieves the fastest convergence and the highest throughput performance, typically stabilizing between approximately $25$ and $33$~Mbps. The linear-motion evaluation (red curve) also demonstrates strong generalization capability, converging to throughput levels comparable to the static case, generally ranging between approximately $25$ and $32$~Mbps, although with comparatively slower convergence in several deployment scenarios.
Similarly, the circular-motion evaluation (green curve) maintains stable convergence and competitive throughput performance across all deployment scenarios. However, its convergence process is also relatively slower than the static evaluation, and in some scenarios, the achieved final throughput is marginally lower than that of the static and linear evaluations, generally stabilizing between approximately $24.5$ and $32.5$~Mbps.
Overall, the results indicate that cosine-motion training enables robust cross-mobility generalization, while static evaluations consistently provide the fastest convergence and the highest achievable throughput performance.



\paragraph{Training: Composite Mobility; Evaluation: Composite Mobility}
Fig.~\ref{fig:comp_reward_comparison_eval} presents the mean throughput evaluation of PPO under composite mobility models for different predefined UAV-BS positions. Across all plots in~\ref{compo_eval1}-\ref{compo_eval4}, throughput increases rapidly during training and converges to stable values in the range of approximately 24-30~Mbps. This indicates consistent training dynamics. The findings demonstrate that the PPO's learnt policy generalizes well across varied spatial configurations and maintains strong throughput performance under composite UE mobility.
\par Overall, across all evaluated scenarios, including one-to-all and composite mobility models, PPO shows significant generalization even when evaluation mobility patterns change from training. 
This robustness is due to learning mobility-invariant, high-level spatial control strategies rather than trajectory-specific behaviors. Clustered UE distributions, in particular, allow PPO to use a centroid-tracking method, rendering the exact mobility model essentially unnecessary while maintaining SINR and throughput performance under heterogeneous and time-varying dynamics.



\subsection{Data Availability}

The dataset generated and analyzed during this study is publicly available on Zenodo at:

\url{https://doi.org/10.5281/zenodo.20720697}

The source code, plotting scripts, and reproducibility materials used to generate the figures and performance evaluations presented in this paper are publicly available through the accompanying GitHub repository:

\url{https://github.com/aakhtarshenas/centralized-ppo-multi-uav-bs}

Experimental tracking, logging, and visualization were conducted using the Weights \& Biases (W\&B) platform. The archived dataset and accompanying source code are provided to support transparency, reproducibility, and future research on DRL-based UAV-BS positioning and trajectory optimization in disaster response networks.

\section{Conclusion}

This paper presented a centralized PPO-based DRL framework for joint positioning and trajectory optimization of multiple UAV-BSs in GPS-free emergency communication scenarios. The problem was formulated as a fairness-aware sum-throughput maximization task in continuous state and action spaces, where UE mobility was modeled through hotspot-based representations.

The proposed framework enables autonomous UAV-BS coordination using radio-level observations, including SINR, received power, and AoA measurements, without requiring explicit UE location information. In addition, circular-statistics-based AoA processing and sigmoid reward shaping significantly improved convergence stability and learning efficiency.

Extensive simulations under static, linear, circular, cosine, and composite mobility patterns demonstrated stable convergence, strong throughput performance, and robust generalization across different deployment scenarios. The results further showed that richer state representations combining spatial, directional, and radio-level information consistently improved PPO performance. Statistical evaluation across multiple random seeds confirmed the robustness and reliability of the proposed approach. Comparative analysis also demonstrated that PPO consistently outperformed DQN and DDPG in convergence stability and overall throughput performance.

Future work will investigate decentralized multi-agent reinforcement learning, dynamic UAV-BS and hotspot association, multi-connectivity mechanisms, and energy-aware trajectory optimization under practical flight constraints.

\vspace{12pt}

\bibliography{bibliography}

\appendix
\section{Role of Angle-of-Arrival Averaging in UAV-BS Stability}
\label{app:aoa_uav_stability}

\subsection{Circular Mean for Angular Averaging}

This appendix motivates the use of the circular mean $\psi$ instead of the arithmetic mean $\mu$ for averaging angular quantities such as UAV-BS headings and angle-of-arrival (AoA). Since angular variables are periodic, averaging must account for circular geometry to avoid incorrect directional estimates and unstable UAV-BS motion.

\subsubsection{Circular Mean and Resultant Vector}

Let $\theta_1,\ldots,\theta_n \in (-\pi,\pi]$ denote angular samples. Each angle is represented on the unit circle as $e^{j\theta_i}$. The complex resultant is defined as
\begin{equation}
\mathcal{R} = \sum_{i=1}^{n} e^{j\theta_i} = C + jS,
\end{equation}
where
\begin{equation}
C = \sum_{i=1}^{n} \cos\theta_i,
\qquad
S = \sum_{i=1}^{n} \sin\theta_i.
\end{equation}

The normalized resultant length is
\begin{equation}
R = \frac{1}{n}\sqrt{C^2 + S^2},
\end{equation}
which measures angular concentration. The circular mean is then given by
\begin{equation}
\boldsymbol{\psi} = \operatorname{atan2}(S,C),
\end{equation}
corresponding to the mean direction of the angular samples \cite{mardia2009directional}.

\subsubsection{Circular Standard Deviation}

Directional dispersion is quantified using the circular standard deviation
\begin{equation}
\boldsymbol{\omega} = \sqrt{-2\ln R},
\end{equation}
which, for concentrated samples ($R \approx 1$), can be approximated as
\begin{equation}
\boldsymbol{\omega} \approx \sqrt{2(1-R)}.
\end{equation}

Thus, $R$ and $\omega$ jointly characterize directional concentration and uncertainty.

\subsubsection{Failure of the Arithmetic Mean}

The arithmetic mean
\begin{equation}
\mu = \frac{1}{n}\sum_{i=1}^{n}\theta_i
\end{equation}
ignores angular periodicity and may produce incorrect directional estimates near wrap boundaries.

Consider
\[
\theta_1 = \pi - \varepsilon,
\qquad
\theta_2 = -\pi + \varepsilon,
\]
where $\varepsilon > 0$ is small. Although both angles correspond to nearly the same direction, the arithmetic mean becomes
\begin{equation}
\mu = 0,
\end{equation}
which points in the opposite direction. In contrast, the circular mean correctly yields
\begin{equation}
\boldsymbol{\psi} \approx \pi.
\end{equation}

\subsubsection{Implications for UAV-BS Stability}

In the proposed UAV-BS framework, AoA measurements guide trajectory adaptation and hotspot tracking. Using arithmetic averaging may introduce incorrect directional estimates near $\pm\pi$, leading to oscillatory motion and unstable convergence.

By contrast, the circular mean $\psi$ preserves angular consistency, while the circular standard deviation $\omega$ quantifies directional uncertainty. For concentrated AoA distributions, $\psi$ varies smoothly and $\omega$ remains small, enabling stable and reliable UAV-BS trajectory updates. Therefore, circular statistics provide a robust foundation for AoA-based UAV-BS control and DRL optimization.

\end{document}